\documentclass[sigplan,screen]{acmart}
\acmConference[PL'18]{ACM SIGPLAN Conference on Programming Languages}{January 01--03, 2018}{New York, NY, USA}
\acmYear{2018}
\acmISBN{} 
\acmDOI{} 
\startPage{1}

\setcopyright{rightsretained}
\acmPrice{}
\acmDOI{10.1145/3519939.3523450}
\acmYear{2022}
\copyrightyear{2022}
\acmSubmissionID{pldi22main-p187-p}
\acmISBN{978-1-4503-9265-5/22/06}
\acmConference[PLDI '22]{Proceedings of the 43rd ACM SIGPLAN International Conference on Programming Language Design and Implementation}{June 13--17, 2022}{San Diego, CA, USA}
\acmBooktitle{Proceedings of the 43rd ACM SIGPLAN International Conference on Programming Language Design and Implementation (PLDI '22), June 13--17, 2022, San Diego, CA, USA}

\ccsdesc[500]{Software and its engineering~Automatic programming}
  
\keywords{Program Synthesis, RESTful API, Type Inference}
  
\usepackage{booktabs}   
\usepackage{subcaption} 

\usepackage{amssymb}

\usepackage{amsthm,mathtools,stmaryrd}
\usepackage{amsfonts}
\usepackage{graphicx}
\usepackage{listings}
\usepackage{xspace}
\usepackage[T1]{fontenc}
\usepackage[scaled=0.75]{beramono}
\usepackage{paralist}
\usepackage[inference]{semantic}
\usepackage{algorithmicx}
\usepackage[noend]{algpseudocode}
\usepackage{wrapfig}
\usepackage{pdfpages}
\usepackage{multirow}
\usepackage[inline]{enumitem}

\hypersetup{colorlinks,
  linkcolor=ACMDarkBlue,
  citecolor=ACMPurple,
  urlcolor=ACMDarkBlue,
  filecolor=ACMDarkBlue}

\makeatletter
\g@addto@macro{\UrlBreaks}{%
  \do\/\do\a\do\b\do\c\do\d\do\e\do\f%
  \do\g\do\h\do\i\do\j\do\k\do\l\do\m%
  \do\n\do\o\do\p\do\q\do\r\do\s\do\t%
  \do\u\do\v\do\w\do\x\do\y\do\z%
  \do\A\do\B\do\C\do\D\do\E\do\F\do\G%
  \do\H\do\I\do\J\do\K\do\L\do\M\do\N%
  \do\O\do\P\do\Q\do\R\do\S\do\T\do\U%
  \do\V\do\W\do\X\do\Y\do\Z%
  \do\0\do\1\do\2\do\3\do\4%
  \do\5\do\6\do\7\do\8\do\9}
\makeatother

\usepackage[version=0.96]{pgf}
\usepackage{tikz}
\usetikzlibrary{arrows,shapes,automata,backgrounds,petri,positioning}
\usepackage{todonotes}

\usepackage{ltablex,array}

\newcommand{\ie}{\emph{i.e.}\xspace}

\newcommand{\eg}{\emph{e.g.}\xspace}

\newcommand{\tygar}{\tname{TYGAR}}
\newcommand{\emphbf}[1]{\emph{\textbf{#1}\xspace}}
\newcommand{\mypara}[1]{\smallskip\noindent\emphbf{#1.}\xspace}

\newif\ifdraft
\drafttrue

\ifdraft
\newcommand\authorrnote[3]{\textcolor{#1}{({#2}: {#3})}\xspace}
\newcommand{\TODO}[1]{{\color{orange!80!black}[\textsl{#1}]}}
\else
\newcommand\authorrnote[2]{}
\newcommand{\TODO}[1]{}
\fi

\newif\iflong
\longtrue

\newcommand{\tname}[1]{\textsc{#1}\xspace}
\newcommand{\tool}{\tname{APIphany}}

\newcommand{\sover}{\tname{StackOverflow}}

\newcommand{\sypet}{\tname{SyPet}}
\newcommand{\insynth}{\tname{InSynth}}
\newcommand{\prospector}{\tname{Prospector}}

\newcommand{\corelang}{$\lambda_A$\xspace}
\newcommand{\hplus}{\tname{Hoogle+}}
\newcommand{\github}{\tname{GitHub}}
\newcommand{\slack}{\tname{Slack}}
\newcommand{\stripe}{\tname{Stripe}}
\newcommand{\squareapi}{\tname{Square}}

\definecolor{commentgreen}{rgb}{0.25,0.5,0.35}

\lstdefinestyle{numbers}
{
  numbers=left,
  numberstyle=\sf\tiny,
  xleftmargin=15pt
}

\lstdefinelanguage{dsl}{
    numbers=none,
    escapeinside={(*@}{@*)},
    basicstyle=\ttfamily\small,
    morekeywords={if,return,let}
}

\lstdefinestyle{dsl}
{
  language=dsl,
  keywordstyle=[1]{\bfseries\ttfamily},
  literate=%
    {<-}{$\leftarrow$}2
    {->}{$\rightarrow$}2      
    {-->}{$\rightarrow$\space}2      
    {==}{$=$}1
}

\lstdefinestyle{dslnum}
{
  language=dsl,
  keywordstyle=[1]{\bfseries\ttfamily},
  numbers=left,
  numberstyle=\sf\tiny,
  numbersep=0pt,
  literate=%
    {<-}{$\leftarrow$}2
    {->}{$\rightarrow$}2      
    {-->}{$\rightarrow$\space}2      
    {==}{$=$}1
}

\newcommand{\T}[1]{\mbox{\lstinline[style=dsl]^#1^}}

\definecolor{light-gray}{gray}{0.87}
\newcommand{\graybox}[1]{\colorbox{light-gray}{#1}}

\tikzstyle{place}=[circle,thick,draw=blue!75,fill=blue!20,minimum size=6mm]
\tikzstyle{final}=[place,double]
\tikzstyle{other}=[place,draw=red!75,fill=red!20]
\tikzstyle{blank}=[minimum size=6mm]
\tikzstyle{transition}=[rectangle,thick,draw=black!75,fill=black!20,minimum size=4mm]
\tikzstyle{sol}=[very thick]
\tikzstyle{mismatch}=[thick,red!75]

\algrenewcommand\algorithmicindent{.7em}
\newcommand*\Let[2]{\State #1 $\gets$ #2}
\algrenewcommand\algorithmicrequire{\textbf{Input:}}
\algrenewcommand\algorithmicensure{\textbf{Output:}}
\algnewcommand{\LineFor}[2]{
  \State\algorithmicfor\ {#1}\ \algorithmicdo\ {#2}}
\algnewcommand\algorithmicswitch{\textbf{match}}
\algnewcommand\algorithmiccase{\textbf{case}}
\algdef{SE}[SWITCH]{Switch}{EndSwitch}[1]{\algorithmicswitch\ #1}{\algorithmicend\ \algorithmicswitch}%
\algdef{SE}[CASE]{Case}{EndCase}[1]{\algorithmiccase\ #1:}{\algorithmicend\ \algorithmiccase}%
\algtext*{EndSwitch}%
\algtext*{EndCase}%

\newcommand{\pre}{\mathsf{pre}}
\newcommand{\post}{\mathsf{post}}

\newcommand{\many}[1]{\overline{#1}}

\newcommand{\net}{\mathcal{N}}

\newcommand{\proj}{\mathsf{proj}}
\newcommand{\guard}{\mathsf{filter}}

\newcommand{\dsu}{\ensuremath{\mathit{DS}}}
\newcommand{\prog}{\ensuremath{\mathcal{E}}\xspace}
\newcommand{\bank}{\ensuremath{\mathcal{V}}}
\newcommand{\tok}{\ensuremath{\mathsf{tok}}}
\newcommand{\fire}{\ensuremath{\mathsf{fire}}}

\newcommand{\lloc}{\mathit{loc}}
\newcommand{\lin}{\mathsf{in}}
\newcommand{\lout}{\mathsf{out}}
\newcommand{\lidx}{\mathsf{0}}

\newcommand{\rorig}{$r_{\mathit{orig}}$\xspace}
\newcommand{\rre}{$r_{\mathit{RE}}$\xspace}
\newcommand{\rreto}{$r_{\mathit{RE}}^{\mathit{TO}}$\xspace}

\newcommand{\sema}[1]{\ensuremath{\hat{#1}}}

\newcommand{\elam}[2]{\lambda {#1}.{#2}}

\newcommand{\synarrow}{\Longrightarrow}

\newcommand{\jtyping}[3]{\ensuremath{\sema{\Lambda};#1 \vdash #2 :: #3}}
\newcommand{\jtypinglib}[3]{\ensuremath{#1 \vdash #2 :: #3}}
\newcommand{\jtinfer}[2]{\ensuremath{\Lambda \vdash #1 \synarrow #2}}
\newcommand{\jtinferacc}[3]{\ensuremath{\Lambda, #1 \vdash #2 \synarrow #3}}
\newcommand{\witness}[3]{\ensuremath{\langle #1, #2, #3 \rangle}}
\newcommand{\construct}[4]{\ensuremath{#1 \vdash #2 \xrightarrow{#3} #4}}
\newcommand{\downgrade}[1]{\ensuremath{\lfloor{#1}\rfloor}}

\newcommand{\jlift}[4]{\ensuremath{ #1 \vdash #2  \;\leadsto\; #3 \dashv #4}}
\newcommand{\jliftexp}[5]{\ensuremath{ #1 \vdash #2 \uparrow #3   \;\leadsto\; #4 \dashv #5}}

\newcommand{\witnesses}{\mathcal{W}}

\newcommand{\elet}[2]{\texttt{\textbf{let }} #1 = #2}

\newcommand{\eif}[2]{\texttt{\textbf{if }}#1=#2}

\newcommand{\eret}[1]{\texttt{\textbf{return }}#1}
\newcommand{\ebind}[2]{\ensuremath{#1 \leftarrow #2}}
\newcommand{\eseq}[2]{\ensuremath{#1; #2}}

\newcommand{\step}[5]{\ensuremath{\langle #1 ; #2 ; #3 \mid #4\rangle \Rightarrow #5}}
\newcommand{\stepenv}[3]{\step{\witnesses}{\Gamma}{#1}{#2}{#3}}

\newcommand{\nosem}{\tname{\tool-Syn}}
\newcommand{\nomerge}{\tname{\tool-Loc}}
\newcommand{\nAPI}{three\xspace}
\newcommand{\nBench}{32\xspace}
\newcommand{\nTyQueries}{32\xspace}

\newcommand{\synTO}{150 seconds\xspace}
\newcommand{\repeatTimes}{three\xspace}
\newcommand{\nCorrect}{29\xspace}

\newcommand{\nMultiplicity}{15\xspace}
\newcommand{\nFiveSec}{19\xspace}
\newcommand{\avgTime}{17.8\xspace}
\newcommand{\medianTime}{1.3\xspace}

\newcommand{\nTopFive}{8\xspace}
\newcommand{\nTopTen}{12\xspace}

\newcommand{\nTopFiveRE}{19\xspace}
\newcommand{\nTopTenRE}{23\xspace}

\begin{document}

\title{Type-Directed Program Synthesis for RESTful APIs}

\author{Zheng Guo}
\affiliation{
  \institution{UC San Diego}
  \country{USA}
}
\email{zhg069@ucsd.edu}

\author{David Cao}
\affiliation{
  \institution{UC San Diego}
  \country{USA}
}
\email{dmcao@ucsd.edu}

\author{Davin Tjong}
\affiliation{
  \institution{UC San Diego}
  \country{USA}
}
\email{dtjong@ucsd.edu}

\author{Jean Yang}
\affiliation{
  \institution{Akita Software}
  \country{USA}
}
\email{jean@akitasoftware.com}

\author{Cole Schlesinger}
\affiliation{
  \institution{Akita Software}
  \country{USA}
}
\email{cole@akitasoftware.com}

\author{Nadia Polikarpova}
\affiliation{
  \institution{UC San Diego}
  \country{USA}
}
\email{npolikarpova@ucsd.edu}

\begin{abstract}
With the rise of software-as-a-service and microservice architectures, 
RESTful APIs are now ubiquitous in mobile and web applications.
A service can have tens or hundreds of API methods, 
making it a challenge for programmers to 
find the right combination of methods to solve their task.

We present \tool, a component-based synthesizer 
for programs that compose calls to RESTful APIs.
The main innovation behind \tool is the use of precise \emph{semantic types},
both to specify user intent and to direct the search.
\tool contributes three novel mechanisms to overcome challenges in adapting
component-based synthesis to the REST domain:
\begin{inparaenum}[(1)]
  \item a type inference algorithm 
  for augmenting REST specifications with semantic types; 
  \item an efficient synthesis technique for ``wrangling'' semi-structured data,
  which is commonly required in working with RESTful APIs; 
  and 
  \item a new form of simulated execution to avoid executing APIs calls during synthesis.
\end{inparaenum}
We evaluate \tool on \nAPI real-world APIs 
and \nTyQueries tasks extracted from \github repositories and \sover. 
In our experiments, \tool found correct solutions to \nCorrect tasks, 
with \nTopTenRE of them reported among top ten synthesis results.

\end{abstract}


\maketitle

\begin{figure*}
  \centering
  \includegraphics[width=.7\linewidth]{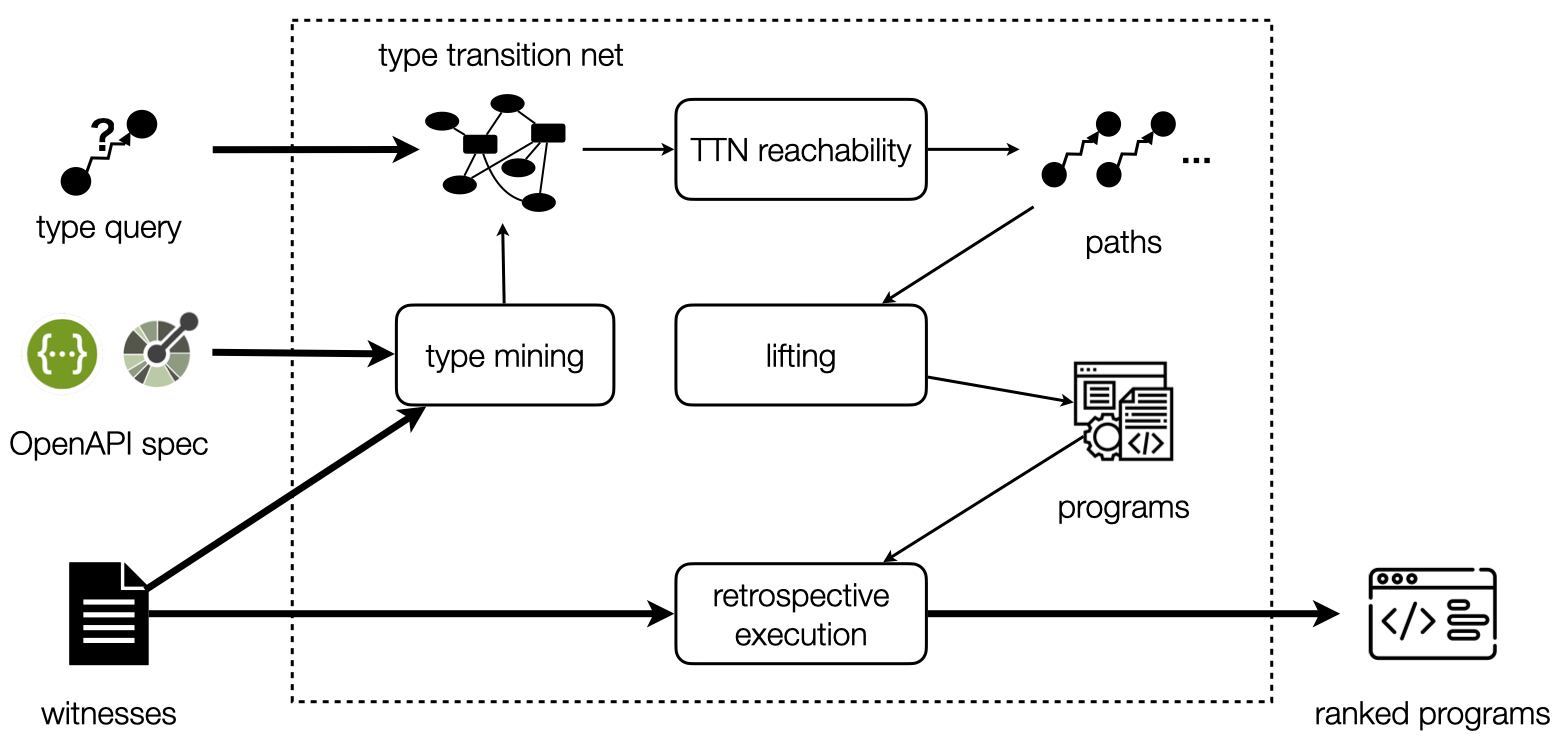}
  \caption{Overview of \tool}\label{fig:overview}
\end{figure*}

\section{Introduction}\label{sec:intro}

Software-as-a-service has emerged as a widely-used means for developers to leverage third-party software.  
Developers might send requests to \stripe to handle payments 
or integrate with \slack to publish notifications, 
all while making use of cloud providers to provision
various form of storage and compute.
According to recent industry surveys, more than 80\% of respondents' services
offer RESTful APIs~\cite{smartbear-state-of-api, postman-state-of-api}, and
these APIs are extensive.  
\slack, for example, has 174 API methods as of version~1.5.0.  
Amazon Web Services offers over two hundred products and
services, each with tens or hundreds of API methods.  
Even with comprehensive
documentation---which is by no means guaranteed---using a new service can be a
daunting proposition.

As an example, consider a question posed on \sover about the \slack API:
\emph{How do I retrieve all member emails from a \slack channel with a given name?}  
The answer is surprisingly complicated:
\begin{enumerate}

  \item First, call \T{conversations_list}%
  \footnote{
    We shorten method names for brevity
    and elide the distinction between REST \emph{methods} and \emph{endpoints},
    irrelevant in this context.
  }  
  to retrieve the array of all channel
  objects, and then search for a channel object with a given name and get its ID;

  \item Next, call \T{conversations_members} on the channel ID to get all user IDs of its members;

  \item Finally, for each user ID, call \T{users_info} to retrieve a user
  object \T{u}, and then access the user's email via \T{u.profile.email}.

\end{enumerate}
To come up with this solution,
one must be familiar with \T{channel} objects, \T{user} objects, 
and three different API methods.

Component-based program
synthesis~\cite{Mandelin05,GveroKKP13,FengM0DR17,JamesGWDPJP20} has been
previously used to help programmers navigate APIs in Java, Scala, and Haskell.
Component-based synthesizers take as input a type signature and (in most cases)
a set of input-output examples, and return a list of program snippets that
compose API calls and have the desired type and input-output
behavior.  This is a powerful approach for navigating APIs, because it
allows developers to start with information easily at hand---%
the types of inputs they have and the outputs they desire---%
and requires no knowledge of which API methods to apply.


\mypara{Challenges}
Unfortunately, there are three significant challenges in applying component-based
synthesis to RESTful APIs.  
First, component-based synthesis relies on \emph{types} both for expressing user
intent and for efficient search, but types in REST APIs are quite shallow.  
For example, in the \slack API specification, 
both channel names and emails have type \T{String},
so our example, which transforms a channel name into an array of emails,
would have a very imprecise type signature $\T{String} \to \T{[String]}$.

Second, RESTful APIs commonly transmit \emph{semi-structured data},
\ie arrays of objects, 
which may themselves contain nested objects and arrays.  
As a result, using an API is often not as simple as sequencing together a handful of method calls;
instead, the calls must be interleaved with ``data wrangling'' operations
such as projections, maps, and filters.
%
These data wrangling operations are challenging for component-based synthesis:
they are extremely generic, and hence significantly expand the search space.

Finally, to compensate for the inherent ambiguity of types,
component-based synthesis typically relies on \emph{executing} candidate
program snippets and matching them against user-provided input-output examples.  
In a software-as-a-service environment, this is a complete non-starter:
not only is the user generally unaware of the internal state of the service 
and hence unable to provide accurate examples,
but executing API calls during synthesis 
can also be prohibitively expensive due to rate limits imposed by the services
and, even more importantly, can have unrecoverable side effects, 
such as deleting accounts or publishing messages.

\mypara{\tool: synthesis with semantic types}
Our core insight is that type-based specifications are actually a good fit for REST APIs,
as long as the types are more fine-grained.
In our example, if the \slack API had dedicated types for \T{Channel.name} and \T{Profile.email},
the programmer could specify their intent as the type $\T{Channel.name} \to \T{[Profile.email]}$.
Although this specification is still somewhat ambiguous,
intuitively it has enough information to narrow down the synthesis results 
to a manageable number such that the programmer can manually inspect
the remaining solutions.
We refer to such fine-grained types as \emph{semantic types}.

In this paper, we present \tool, a component-based synthesizer for REST APIs
guided by semantic types.
\autoref{fig:overview} shows a high-level overview of our approach,
which is structured into two phases:
\begin{inparaenum}[(1)]
  \item the \emph{analysis} phase infers semantic type annotations for a given API;
  \item the \emph{synthesis} phase uses these type annotations to perform component-based synthesis.
\end{inparaenum} 
For the \slack API,
\tool is able to infer, for example, that the method \T{conversations_members}
has the semantic type $\T{Channel.id} \to \T{[User.id]}$.
At synthesis time, given the \emph{type query}
$\T{Channel.name} \to \T{[Profile.email]}$,
\tool returns a ranked list of programs of this type,
where the desired solution (shown in \autoref{fig:slack-solution}) appears among the top ten.
\tool's output is expressed in a compact DSL inspired by Haskell's monadic \T{do}-notation and Scala's \T{for}-comprehensions,
which, however, can be easily translated into the user's language of choice for communicating with the API.

\mypara{Contributions}  We present the design, implementation, and evaluation
of \tool, including:
\begin{enumerate}

  \item \emph{Type mining} (\autoref{sec:mining}),
  a technique that infers semantic types from a set of \emph{witnesses}
  (observed invocations of API methods).
  Witnesses can be generated in a sandbox or by tapping live production traffic;
  in either case, they are collected ahead of time, once per API, 
  which avoids inducing side effects during synthesis.

  \item Efficient synthesis of wrangling operations for semi-structured data 
  via \emph{array-oblivious search} (\autoref{sec:synthesis}), 
  which omits challenging array operations during search, 
  and recovers them later via type-directed lifting.


  \item Ranking synthesis results with the help of \emph{retrospective execution}
  (\autoref{sec:ranking}), 
  a type of simulated execution using previously collected witnesses.
  Retrospective execution helps \tool weed out uninteresting programs 
  (\eg programs that always return an empty array),
  reducing the number of synthesis results the user has to inspect to find their expected solution.

\end{enumerate}

\begin{figure}
      \begin{lstlisting}[style=dslnum]
  \channel_name -> { 
    c   <- conversations_list()
    if c.name == channel_name
    uid <- conversations_members(channel=c.id)
    let u = users_info(user=uid)
    return u.profile.email      
  }
      \end{lstlisting}
      \caption{Solution for retrieving all member emails from a \slack channel in \tool DSL.}
      \label{fig:slack-solution}  
\end{figure}

We evaluate \tool on \nAPI real-world APIs,
and \nTyQueries tasks extracted from \github repositories and \sover (\autoref{sec:eval}).
Our evaluation shows that \tool can find solutions to the majority of tasks (\nCorrect/\nTyQueries)
within \synTO.
Moreover, semantic types are crucial to its effectiveness:
without type mining, \tool can only solve four tasks.
Finally, ranking significantly improves the quality of reported solutions,
increasing the number of correct solutions appearing in top ten results 
from \nTopTen/\nCorrect to \nTopTenRE/\nCorrect.

\section{\tool by Example}\label{sec:overview}

In this section we use the task of retrieving all member emails in a \slack channel
as a running example
to illustrate the \tool workflow depicted in \autoref{fig:overview}.
%

\begin{figure*}
  \begin{minipage}{.48\linewidth}
    \includegraphics[width=.9\textwidth]{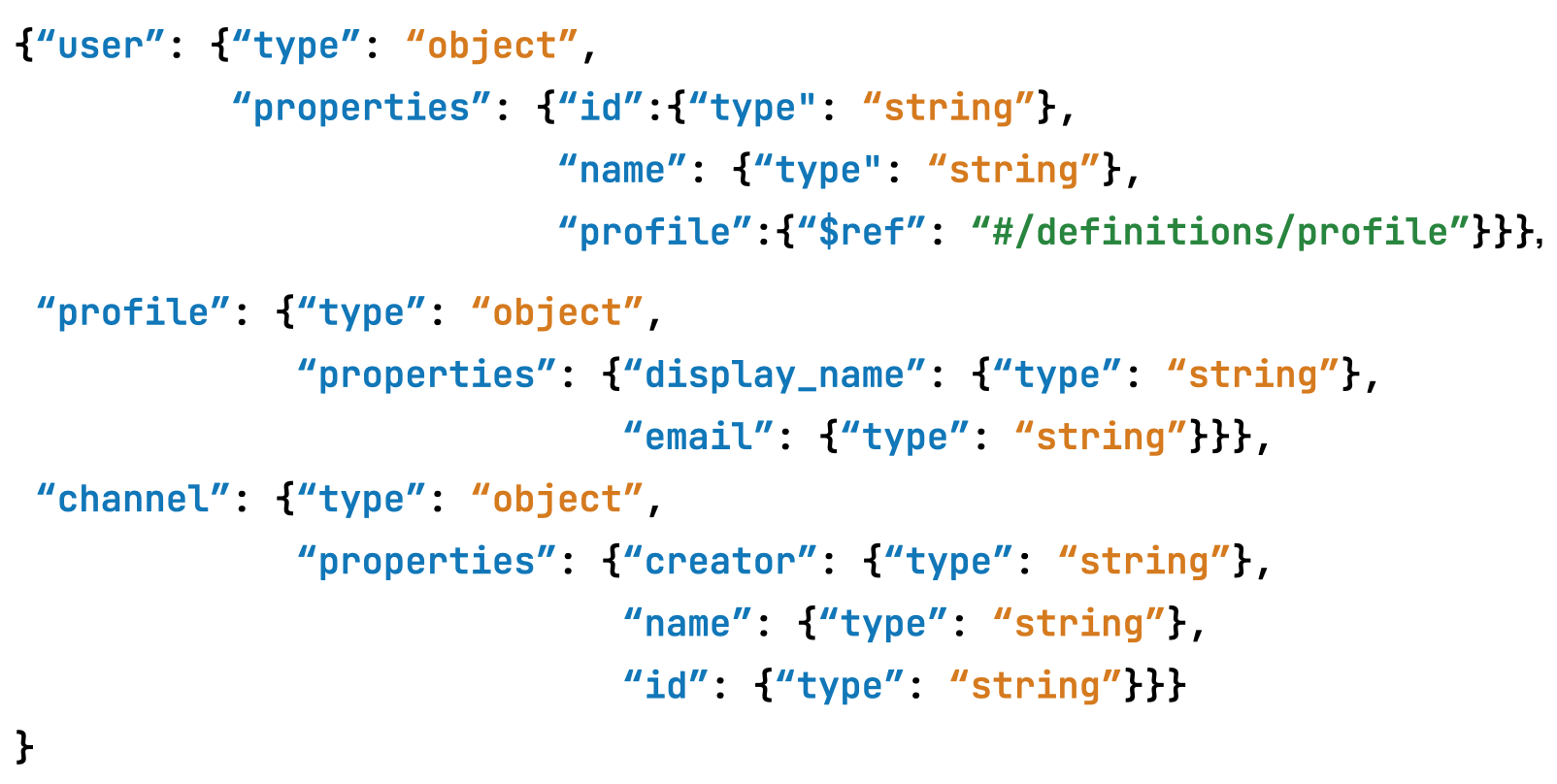}
  \end{minipage}
  \hfill\vline\hfill
  \begin{minipage}{.4\linewidth}
    \includegraphics[width=.9\textwidth]{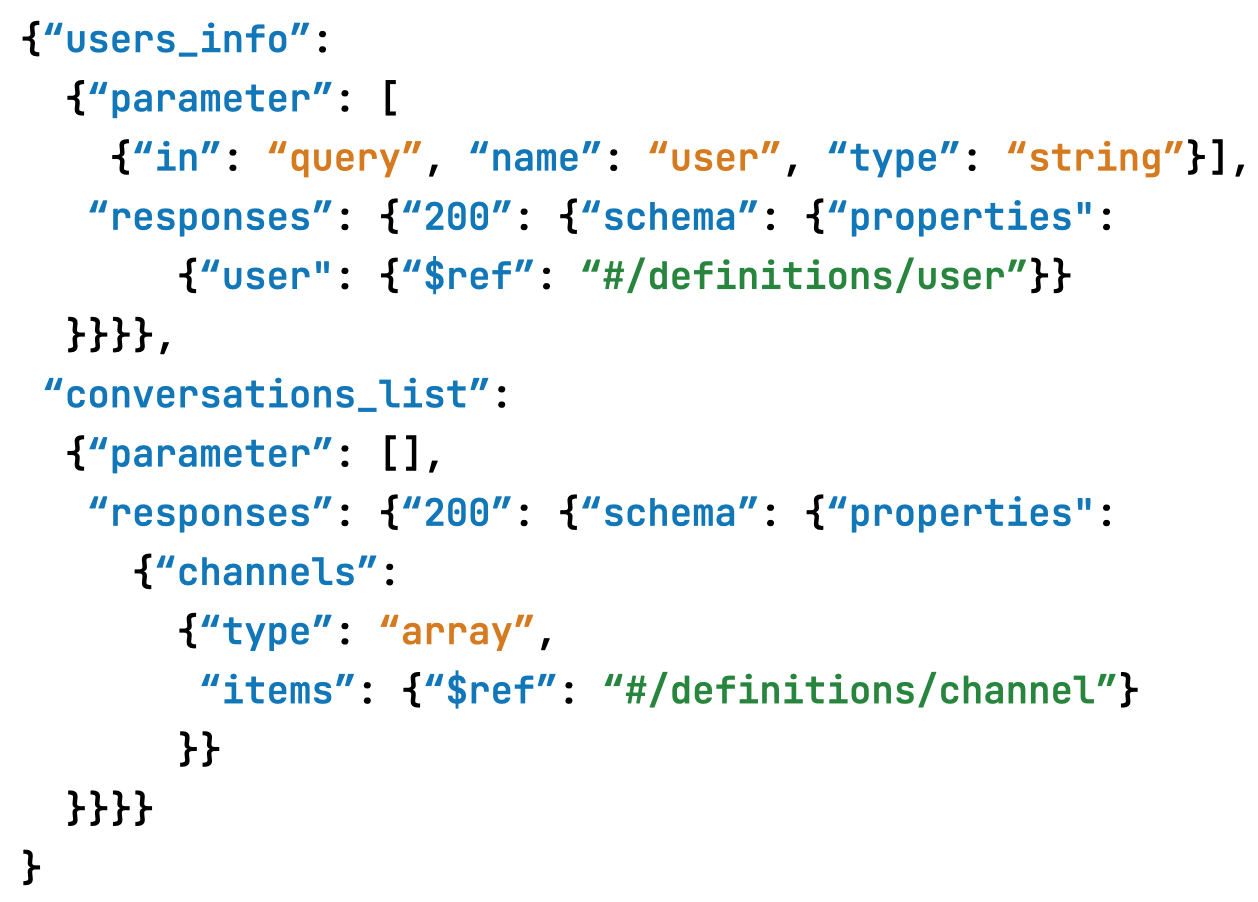}
  \end{minipage}
  \caption[Fragment of the Slack API's OpenAPI specification]{Fragment of the Slack API's OpenAPI specification. 
  (left) Definitions of \T{user}, \T{profile} and \T{channel} objects.
  (right) Parameters and responses of the methods \T{users_info} and \T{conversations_list}.}
  \label{fig:objs}
\end{figure*}

\subsection{API Analysis by Example}\label{sec:overview:analysis}

API analysis is performed once per API.
It takes as input a \emph{spec} in the popular OpenAPI format%
\footnote{\url{https://swagger.io/}. \tool supports both OpenAPI v2 and v3.}
and a set of \emph{witnesses} (successful API method calls);
it produces a spec annotated with semantic types.
OpenAPI specs are publicly available for most popular APIs.%
\footnote{\slack OpenAPI spec is available at: \url{https://raw.githubusercontent.com/slackapi/slack-api-specs/master/web-api/slack\_web\_openapi\_v2.json}}
Witnesses can be generated in a number of ways,
for example, by running an integration test suite in a sandbox
or by passively listening to production API traffic.
We envision witness collection and API analysis being performed by the API maintainer
(or another interested party),
not by regular users of the \tool synthesizer.


\mypara{OpenAPI specs}
\autoref{fig:objs} shows a fragment of the OpenAPI spec provided by \slack.  
An OpenAPI spec consists of object definitions and method definitions.
We show definitions of three objects, \T{user}, \T{profile} and \T{channel},
and two methods, \T{users_info} and \T{conversations_list},
relevant to our example.
As you can see, the spec does provide precise type information for some of the locations:
for example, the response of \T{users_info} clearly has type \T{User}
(it is annotated with a reference to the corresponding object definition).
The bulk of the locations, however,
such as the field \T{user.id} or the parameter of \T{users_info},
are simply annotated with \T{String},
which is not very helpful for the purposes of type-directed synthesis.
Our goal is to replace these \T{String} annotations with more fine-grained types.

\begin{figure}
  \includegraphics[width=.9\linewidth]{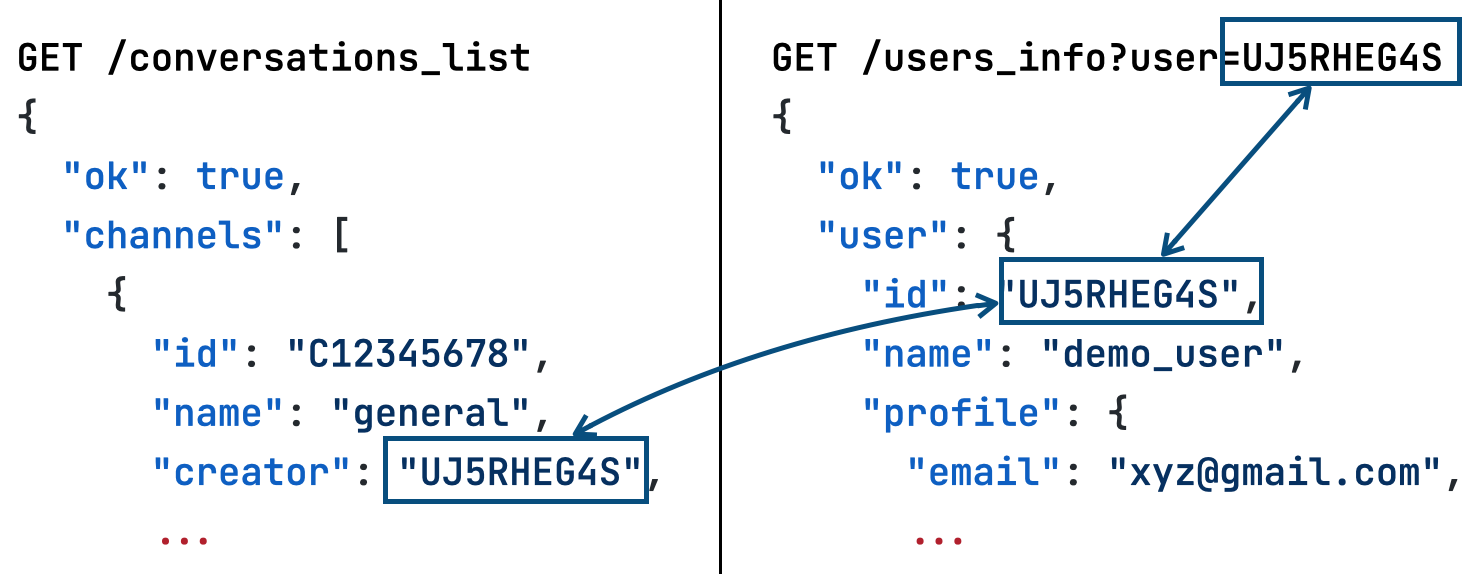}
  \caption{Witnesses for two \slack API methods.
  Arrows connect equal values observed at different locations.
  Type mining ascribes the type \T{User.id} to all the boxed locations.}
  \label{fig:witnesses}
\end{figure}

\mypara{Mining types from witnesses}
To this end, we build upon an algorithm first proposed
in~\cite{specmining} that infers types by \emph{mining} them from execution traces,
based on the insight
that \emph{equal values observed at different locations likely have the same type}.
More specifically, 
our type mining algorithm starts by ascribing a unique semantic type to each \T{String} location
and then merges locations that share a value anywhere in the witness set.
As an illustration,
consider \autoref{fig:witnesses},
which lists two witnesses for the API methods from our running example.
%
In this witness set we observe the same value \T{"UJ5RHEG4S"} in three locations:
\begin{inparaenum}[(1)]
  \item the \emph{parameter} of \T{users_info},
  \item the \T{id} \emph{field} of a \T{User} object (we know from the spec that \T{users_info} returns a \T{User}), and
  \item the \T{creator} \emph{field} of a \T{Channel} object (we know from the spec that \T{conversations_list} returns an array of \T{Channel}s).
\end{inparaenum}
Hence we merge all three locations into the same semantic type.
For presentation purposes, we assign the name \T{User.id} to this type,
which is derived from location (2) above.
The choice of name is not important, however:
the user is free to refer to this semantic type via any of its representative locations;
for example, \T{Channel.creator} also denotes the same type.

\subsection{Program Synthesis by Example}\label{sec:overview:synthesis}
The program synthesis phase of \tool is meant to be used by regular programmers,
any time they need help accomplishing a task with one of the supported APIs.
The programmer queries \tool
with a type signature built from semantic types.
Although the UI for constructing queries is beyond the scope of this paper,
we envision the programmer browsing object definitions and selecting relevant fields as semantic types.
For our running example, 
the programmer knows that they need to go from a channel name to an array of user emails;
they might first look through the \T{channel} object definition and find the \T{name} field;
they might then search globally for a field called \T{email} and find it inside the \T{profile} object;
hence they settle on the type query $\T{Channel.name} \to [\T{Profile.email}]$.

The program synthesis phase itself comprises two steps, 
beginning with a program \emph{search} step to generate a list of candidate programs with a given type, 
followed by a \emph{ranking} step to identify promising candidates
(described in \autoref{sec:overview:ranking}).

\mypara{Challenge: components meet control flow}
Given the type query  $\T{Channel.name} \to [\T{Profile.email}]$,
how would \tool go about enumerating all programs of this type?
This task presents a challenge to existing synthesis techniques
because our candidate programs have \emph{both}
a large component library to choose from---%
from dozens to hundreds of methods---%
\emph{and} non-trivial control flow---%
\eg the solution to our running example has to \emph{loop} over the members of a channel.
One line of prior work that scales to large component libraries
is graph-based search using \emph{type-transition nets}~(TTNs)~\cite{FengM0DR17,GuoJJZWJP20};
unfortunately, this approach can only generate sequences of method calls,
and does not support loops.


\mypara{The \tool DSL}
We observe that the loops we need for manipulating semi-structured data
are restricted to iterating over (possibly nested) arrays of objects.
To capture this restricted class of programs we have designed a DSL
%
inspired by Scala's \T{for}-comprehensions,
Haskell's monadic \T{do}-notation,
and \T{LINQ}~\cite{linq}.
The solution to our running example in this DSL is given in \autoref{fig:slack-solution}.
In this language, iteration over an array is expressed using the monadic \emph{bind} operation (written \T{<-}).
For example, the second bind in \autoref{fig:slack-solution}
has the effect of performing the subsequent computation for every element \T{uid} 
of the array returned in line 4:
\begin{lstlisting}[style=dslnum,firstnumber=4]
  uid  <- conversations_members(channel=c.id);
  let u = users_info(user=uid);
  return u.profile.email      
\end{lstlisting}

\mypara{Array-oblivious search}
The main idea behind \tool's search is that
although we cannot directly synthesize the program above using existing TTN-based techniques,
we can synthesize an \emph{array-oblivious} version of this program,
where we pretend that \T{conversations_members} returns a single \T{User.id} instead of an array,
and hence we can simply sequence the two method calls, without monadic binding:
\begin{lstlisting}[style=dslnum,firstnumber=4]
  let uid = conversations_members(channel=c.id);
  let u   = users_info(user=uid);
  u.profile.email      
\end{lstlisting}

To transform an array-oblivious program into the final solution,
\tool \emph{lifts} it into a comprehension
by replacing each \T{let} binding that causes a type mismatch with a monadic bind.
In our example, the \T{let} in line 4 causes a type error
(because \T{conversations_members} returns \T{[User.id]}, while \T{users_info} expects a single \T{User.id}),
while the \T{let} in line 5 does not
(since \T{users_info} returns a single \T{User});
hence lifting replaces the first \T{let}-binding with \T{<-} but not the second.

\subsection{Ranking via Retrospective Execution}\label{sec:overview:ranking}

\begin{figure}
  \small
  \centering
  \begin{minipage}{.5\columnwidth}
    \centering
  \begin{lstlisting}[style=dsl, basicstyle=\ttfamily\footnotesize]
  \channel_name -> { 
    c <- conversations_list()
    if c.name == channel_name
    let uid = (*@\graybox{\texttt{c.creator}}@*)
    let u = users_info(user=uid)
    return u.profile.email      
  }
  \end{lstlisting}
\end{minipage}%
\begin{minipage}{.5\columnwidth}
  \centering
\begin{lstlisting}[style=dsl, basicstyle=\ttfamily\footnotesize]
\channel_name -> { 
  c <- (*@\graybox{\texttt{conversations\_open()}}@*)
  if c.name == channel_name
  let uid = (*@\graybox{\texttt{c.creator}}@*)
  let u = users_info(user=uid)
  return u.profile.email      
}
\end{lstlisting}
\end{minipage}
\vspace{-10pt}
    \caption{A sample of incorrect candidate solutions.}
    \label{fig:ranking}
\end{figure}

Although semantic types are less ambiguous than primitive types 
for expressing user intent,
they are still not precise enough to exactly identify the desired program.
For example, our synthesizer generates more than 1000 candidates
for the type signature $\T{Channel.name} \to \T{[Profile.email]}$;
clearly, it is infeasible for the user to manually go through all of them.
Hence, \tool must be able to rank the candidates in
order to show the user a small number of likely solutions.

Fortunately, most of the 1000 candidates are easy to weed out
because they produce uninteresting results.
Consider two of the candidates depicted in \autoref{fig:ranking},
which differ from our desired solution (\autoref{fig:slack-solution})
in the highlighted fragments:
the first program returns the email of the channel's \emph{creator} (as opposed to all of its members),
and the second one gets the list of channels from \T{conversations_open},
which is intended for opening a direct message channel.
It turns out that the second program \emph{always fails} at run time,
because a successful call to \T{conversations_open} 
requires providing exactly one of its two optional arguments (a channel ID or a list of users).
The first program executes successfully,
but it always returns a single email, while the user asked for an array of emails.
%
For these reasons, both of these programs are less likely to be the intended solution
than the program in \autoref{fig:slack-solution},
which successfully returns multiple emails at least sometimes.

A natural idea is to test all candidate programs on random inputs
and rank them based on the results they produce.
%
Unfortunately, as we have hinted above,
there are several barriers to systematically executing many candidate programs 
that make calls to REST APIs.
First, most REST APIs set a rate limit on
how frequently a user can make method calls or
how many calls a user can make in a day.
%
Second, many REST API methods are side-effecting.  Unlike a self-contained
binary, a remotely-hosted service cannot be restarted from a clean state for
each execution.  

\mypara{Retrospective execution}
We propose \emph{retrospective execution} (RE) as an efficient, 
non-side effecting alternative to program execution.
The main idea is to simulate execution
by ``replaying'' witnesses collected for the API analysis phase.  
When evaluating a candidate program, rather than executing an API call,
RE instead searches for a matching witness
and substitutes its response at the call site.
If done naively, however, 
this process almost always yields failure or an empty array;
so making RE useful for ranking purposes
requires explicitly \emph{biasing} execution towards meaningful results.

As an illustration,
consider executing the program in \autoref{fig:slack-solution}
using the witnesses in~\autoref{fig:witnesses}.
As the first step, we simulate the call to \T{conversations_list} 
using the first witness;
the response is an array of channels with names \T{"general"}, \T{"private-test"}, and \T{"team"}.
The second step is to filter this array,
retaining only those channels whose name is equal to the input parameter \T{channel_name}.
If we had sampled the value for \T{channel_name} eagerly, before running the program,
we could scarcely have chosen one of the three names actually present in the array,
so the filtering step (and hence the whole program) would almost always return an empty array.
Instead we sample the value for \T{channel_name} \emph{lazily}, once we encounter the filter,
picking one of the names present in the array.

Assume that we picked \T{channel_name = "general"},
and hence the filter returns the first channel.
%
Next, we simulate the call to \T{conversations_members} on this channel's ID.
Because our witness set is sparse,
we may or may not find an exact match for this call;
in the latter case,
we sample the response from the set of \emph{approximate matches}, 
\ie witnesses with the same method names and argument names,%
\footnote{Because in REST some arguments are optional, the same method can be called with different subsets of arguments.}
but not necessarily the same argument values.
Due to approximate matching,
RE results do not always equal the results of a real execution,
but they are still useful for estimating whether a program candidate is able to produce meaningful outputs.
For each candidate, we run RE multiple times (with different random seeds)
and use the outputs to assign a rank to each candidate.

\section{The Core Language}\label{sec:lang}

In this section, we formalize the core of \tool's DSL as \corelang, 
a functional language specialized for manipulating semi-structured data.
The syntax of \corelang is summarized in~\autoref{fig:syntax}.

\begin{figure}[t]
  \small
  \centering
  $$
  \begin{array}{rll}
    o ::=& \T{User} \mid \T{Channel} \mid \ldots & \text{object names}\\
    f ::=& \T{u_info} \mid \ldots & \text{method names}\\
    l ::=& \lin \mid \lout \mid \lidx \mid \texttt{id} \mid \texttt{name} \mid \ldots & \text{field labels}\\    
    \ell ::=& l \mid ?l & \text{record fields}\\
    \lloc ::=& o\many{.l} \mid f\many{.l} & \text{locations}
  \end{array}
  $$
  \textbf{Terms}
  $$
  \begin{array}{lll}
  e ::=& & \text{\emph{Expressions}}\\
  & \mid x \mid e.l & \text{variable, projection} \\  
  & \mid f(\many{l_i=e_i}) \mid \eseq{\elet{x}{e}}{e} & \text{method call, pure binding} \\
  & \mid \eseq{\eif{e}{e}}{e} \mid \eseq{\ebind{x}{e}}{e} & \text{guard, monadic binding} \\
  & \mid \eret{e} & \text{pure value lifting} \\
  \prog ::= & \elam{\many{x}}{e} & \text{\emph{Top Level Programs}}
  \end{array}
  $$
  \textbf{Values} 
  $$
  v ::= \T{"..."} \mid [\many{v}]  \mid \{\many{l_i = v_i}\} \quad \text{strings, arrays, objects}   
  $$
  \textbf{Types}
  $$
  \begin{array}{lll}
    t ::= & & \text{\emph{Syntactic types}} \\
    & \mid \T{String} & \text{strings} \\
    & \mid o \mid [t]  \mid \{\many{\ell_i: t_i}\} & \text{named objects, arrays, records}\\
    s ::= & t \to t & \text{function types}\\
    \sema{t} ::= & & \text{\emph{Semantic types}} \\
    & \mid \{\many{\lloc\}} & \text{loc-sets} \\
    & \mid o \mid [\sema{t}]  \mid \{\many{\ell_i: \sema{t_i}}\} & \text{named objects, arrays, records}\\
    \sema{s} ::= & \sema{t} \to \sema{t} & \text{function types}    
  \end{array}
  $$
  \textbf{Libraries}
  $$
  \begin{array}{lll}
  \Lambda ::= & \many{o : t} ; \many{f : s} & \text{object and method definitions}\\
  \sema{\Lambda} ::= & \many{o : \sema{t}} ; \many{f : \sema{s}} & \text{semantic definitions}
  \end{array}
  $$
  \caption{Syntax of the language \corelang}
  \label{fig:syntax}
\end{figure}

\mypara{Types}
The types of \corelang include \emph{syntactic types} $t$ (those used in the OpenAPI spec)
and \emph{semantic types} $\sema{t}$, which we infer.
Both categories of types have named objects $o$,
arrays $[t]$, and records $\{\many{\ell_i:t_i}\}$.%
\footnote{We write $\many{X}$ to denote zero or more occurrences of a syntactic element $X$.}
Records are mappings from field labels to types;
some fields are optional, indicated with a $?$ before its label.
For example, the record type $\{\T{id}: \T{String}, ?\T{time_zone}: \T{String}\}$,
has a required field \T{id} and an optional field \T{time_zone}.
The two categories of types differ in their base types:
the sole primitive syntactic type is \T{String},%
\footnote{In practice, REST APIs also include integers and booleans;
these types are handled slightly differently in \tool, as discussed in \autoref{sec:discussion}.}
while the sole primitive semantic type is a \emph{loc-set}, \ie a set of locations.

A \emph{location} is an object or method name followed by a sequence of labels,
such as \T{User.id}.
Apart from field labels that correspond to object fields in the OpenAPI spec,
we introduce three reserved labels---$\lin$, $\lout$, and $\lidx$---%
for addressing method parameters and responses, and array elements, respectively.
For example, $\T{c_list}.\lout.\lidx$ refers to an element type of the response array
of the method \T{c_list}.

%
%
Function types are written $t \to t$,
and multiple arguments are represented as a record
whose fields encode argument names
(with optional fields encoding optional arguments).

A library $\Lambda$ models an OpenAPI spec.
It contains object definitions,
which bind object identifiers to (record) types,
and method definitions,
which bind method names to function types.
A semantic library $\sema{\Lambda}$, which is the output of type mining, 
binds object identifiers and method names to semantic types.
As an example, \autoref{tab:library} shows $\Lambda$ definitions
that correspond to a portion of the \slack OpenAPI spec (with method names shortened for brevity),
and their corresponding definitions in the semantic library $\sema{\Lambda}$.

\begin{figure*}[t]
  \begin{minipage}{.65\textwidth}
\small
\begin{tabular}{|c|p{.39\textwidth}|p{.46\textwidth}|}
\toprule
& Syntactic library $\Lambda$ & Semantic library $\sema{\Lambda}$\\
\midrule
\multirow{6}{*}{\rotatebox{90}{Objects}} &
\vspace{-1em}
\begin{lstlisting}[belowskip=-1em]
Channel: { id: String, 
           name: String, 
           creator: String }
User: { id: String, 
        name: String, 
        profile: Profile }
\end{lstlisting}
&
\vspace{-1em}
\begin{lstlisting}[belowskip=-1em]
Channel: { id: (*@\graybox{$\texttt{Channel.id}$}@*), 
           name: (*@\graybox{$\texttt{Channel.name}$}@*), 
           creator:(*@\graybox{$\texttt{User.id}$}@*) }
User: { id: (*@\graybox{$\texttt{User.id}$}@*), 
        name: (*@\graybox{$\texttt{User.name}$}@*), 
        profile: Profile }
\end{lstlisting}
\\
\midrule
\multirow{6}{*}{\rotatebox{90}{Methods}} & 
\vspace{-1em}
\begin{lstlisting}[belowskip=-1em]
c_list: 
  {} -> [Channel]
u_info: 
  {user: String} -> User
c_members: 
  {channel: String} -> [String]
\end{lstlisting}
&
\vspace{-1em}
\begin{lstlisting}[belowskip=-1em]
c_list: 
  {} -> [Channel]
u_info: 
  {user: (*@\graybox{$\texttt{User.id}$}@*)} -> User
c_members: 
  {channel: (*@\graybox{$\texttt{Channel.id}$}@*)} -> [(*@\graybox{$\texttt{User.id}$}@*)] 
\end{lstlisting} \vspace{-1em}\\
\bottomrule
\end{tabular}
\caption{Library $\Lambda$ that models a portion of the \slack OpenAPI spec
and the corresponding semantic library $\sema{\Lambda}$. 
Each gray box is a loc-set type inferred by type mining,
depicted for brevity using a single representative location from the set.
}\label{tab:library}
\end{minipage}\hfill\vline\hfill
\begin{minipage}{.33\textwidth}
  \small
    \begin{algorithmic}[1]
        \Require{A library $\Lambda$ and witnesses $\witnesses$}
        \Ensure{A semantic library $\sema{\Lambda}$}
        \Function{\textsc{MineTypes}}{$\Lambda, \witnesses$}
            \Let{$\dsu$}{empty disjoint-set}
            \For{$\witness{f}{v_{in}}{v_{out}} \in \witnesses$}
                \State{\Call{AddWitness}{$\dsu, f, \mathsf{in}, v_{in}$}}
                \State{\Call{AddWitness}{$\dsu, f, \mathsf{out}, v_{out}$}}
            \EndFor
            \Let{\sema{\Lambda}}{\Call{AddDefinitions}{$\Lambda, \dsu$}}
            \State \Return{$\sema{\Lambda}$}
            \Statex
        \EndFunction
      \Function{\textsc{AddWitness}}{$\dsu, \lloc, v$}
        \Switch{$v$}
        \Case{\T{"..."}}
          \State $\jtinfer{\lloc}{\{\lloc'\}}$
          \Let{$\dsu$}{\Call{$\mathsf{insert}$}{$\dsu, \lloc', v$}}
        \EndCase
        \Case{$[\many{v_i}]$}
        \State \textbf{forall} $i:$ \Call{AddWitness}{$\dsu, \lloc.\lidx, v_i$}
        \EndCase
        \Case{$\{\many{l_i=v_i}\}$}
          \State \textbf{forall} $i:$ \Call{AddWitness}{$\dsu, \lloc.l_i, v_i$}
        \EndCase
        \EndSwitch      
      \EndFunction
    \end{algorithmic}
    \caption{Type mining algorithm.}
    \label{fig:comp-inference}
\end{minipage}
\end{figure*}

\mypara{Terms}
Values of \corelang include string literals,
arrays, and objects;
objects are mappings from field labels to values.
Similarly to Haskell's \T{do}-notation,
\eret{e} returns an array with a single element $e$,
and the monadic binding \eseq{\ebind{x}{e_1}}{e_2}
evaluates $e_2$ for each element $x$ of the array $e_1$,
and concatenates all resulting arrays.
In contrast,
the pure binding \eseq{\elet{x}{e_1}}{e_2}
binds $x$ to the entire result of $e_1$
and then evaluates $e_2$.
The guard expression \eseq{\eif{e_1}{e_2}}{e}
evaluates $e$ if the guard holds and returns an empty array otherwise;
guards are restricted to equalities,
since these are the only guards generated by \tool.
At the top level, a program \prog~is an abstraction with a list of arguments $\many{x}$
and body $e$.
%

\section{Type Mining}\label{sec:mining}

In this section we detail \tool's type mining algorithm,
using the library $\Lambda$ in \autoref{tab:library}
and the witnesses in \autoref{fig:witnesses} as a running example.
Informally, the idea is to first assign every \T{String} location
$\lloc$ in $\Lambda$
a unique type $\{\lloc\}$,
and then merge the types of some locations based on the witnesses.

\mypara{Assigning location-based types}
We formalize the first step
as a judgement $\jtinfer{\lloc}{\sema{t}}$,
which assigns a semantic type $\sema{t}$ to location $\lloc$
based only on the information present in the syntactic library $\Lambda$.
The reader might be wondering why isn't the assigned type $\sema{t}$ always simply $\{\lloc\}$.
This is indeed the case for \T{String}-annotated locations
explicitly present in $\Lambda$,
such as \T{User.id} or $\T{u_info}.\lin.\T{user}$.
But in other cases, location-based type assignment is more involved;
for example:
\begin{compactitem}
\item $\jtinfer{\T{u_info}.\lout}{\T{User}}$
because this location is annotated with a named object type.
\item $\jtinfer{\T{c_members}.\lout}{[\{\T{c_members}.\lout.\lidx\}]}$
because array types do not themselves get replaced with loc-sets; 
instead, we recursively assign a location-based type to an array's element.
\item $\jtinfer{\T{u_info}.\lout.\T{id}}{\{\T{User.id}\}}$
because type assignment \emph{canonicalizes} locations inside types
to make sure they explicitly appear in $\Lambda$;
to this end, we recursively assign a type to location's prefix, $\jtinfer{\T{u_info}.\lout}{\T{User}}$,
and then follow the field \T{id} of the \T{User} object.
\end{compactitem}
The formalization of location-based type assignment is mostly straightforward
and relegated to 
\iflong
\autoref{appendix:mining} (\autoref{fig:type-inference}).
\else
the technical report~\cite{techreport}.
\fi

\mypara{Merging types via a disjoint-set}
Type mining relies on a variant of the \emph{disjoint-set} data structure
(also known as \emph{union-find}~\cite{union-find}).
Our disjoint-set $DS$ stores disjoint groups of pairs $(\lloc, v)$,
where $\lloc$ is a location and $v$ is a string value.
When two pairs are in the same group,
their corresponding locations have the same semantic type.

$\dsu$ supports two efficient operations: 
$\mathsf{insert}$ and $\mathsf{find}$.
$\mathsf{insert}$ takes a pair $(\lloc, v)$ and checks whether either of its components already appears in $\dsu$;
if so, it merges the new pair into the corresponding group,
and otherwise puts it into a new group.
$\mathsf{find}$ takes a location $\lloc$ and returns a semantic type $\sema{t}$;
internally, $\mathsf{find}$ locates the group to which the pair $(\lloc, \_)$ belongs in $\dsu$
and returns the loc-set $\{\lloc, \lloc_1, \ldots\}$ that contains all locations in that group.

\mypara{Type mining algorithm}
\autoref{fig:comp-inference} presents the top-level algorithm \textsc{MineTypes},
which takes as input a syntactic library $\Lambda$ and a set of witnesses $\witnesses$,
and returns a semantic library $\sema{\Lambda}$.
A \emph{witness} $W$ is a triple \witness{f}{v_{in}}{v_{out}},
where $f$ is a method name
and $v_{in}$, $v_{out}$ are its argument and response value
(multiple arguments are represented as an object).
\textsc{MineTypes} operates in two phases:
in lines 2--5 it builds the disjoint-set $\dsu$ from $\witnesses$
and in line 6 it build $\sema{\Lambda}$ from $\dsu$.

In the first phase, the algorithm iterates over the witnesses,
registering the input value $v_{in}$ at the location $f.\lin$ 
and the output value $v_{out}$ at the location $f.\lout$.
To this end, we call a helper function \textsc{AddWitness},
which drills down into composite values (arrays and objects) to get to string literals,
and then $\mathsf{insert}$s each string into $\dsu$ with its location-based type.
For example, when processing the response from the first witness in \autoref{fig:witnesses},
\textsc{AddWitness} iterates over all channel objects in the array,
and over all fields of each channel object;
once it reaches the value \T{"UJ5RHEG4S"}, 
it computes the type of its location as $\jtinfer{\T{c_list}.\lout.\lidx.\T{creator}}{\{\T{Channel.creator}\}}$,
and inserts $(\T{Channel.creator}, \T{"UJ..."})$ into $\dsu$.
Processing the second witness results in inserting the pairs $(\T{u_info}.\lin.\T{user}, \T{"UJ..."})$
and $(\T{User.id}, \T{"UJ..."})$, which share the same string value,
and hence all three pairs get merged into the same group.
Once all the witnesses are added to $\dsu$,
its groups represent the final set of semantic types.

In the second phase, 
the algorithm calls \textsc{AddDefinitions}
to iterate over all object and method definitions in $\Lambda$,
and add corresponding definitions to $\sema{\Lambda}$,
relying on $\mathsf{find}$ to retrieve the semantic type for each location.
For example, when adding the method \T{u_info},
we query $\mathsf{find}(\dsu, \T{u_info}.\lin.\T{user})$,
which finds the group mentioned above and returns its loc-set:
$\{\T{User.id}, \T{Channel.creator}, \ldots\}$.
If the requested location is not in $\dsu$---%
because $\witnesses$ has no witnesses for the enclosing method or object---%
it is annotated with the unmerged location-based type.

\section{Type-Directed Synthesis}\label{sec:synthesis}
In this section, we discuss how \tool generates a set of well-typed programs given a query type,
using the same running example as in previous sections.
%

\mypara{Synthesis problem}
Formally, our synthesis problem is defined by a semantic library $\sema{\Lambda}$ and 
a semantic query type $\sema{s}$.
For our running example, we use the semantic library from \autoref{tab:library}
and the query type $\T{Channel.name} \to [\T{Profile.email}]$.%
\footnote{Here and throughout this section, we write loc-set types using an arbitrarily chosen representative;
the user can query \tool using any locations of their choosing,
and the tool interprets them as the loc-sets they belong to.}
A \emph{candidate solution} is any program $\prog$ that type-checks against $\sema{s}$.
To formalize this notion, we introduce the program typing judgment $\jtypinglib{\sema{\Lambda}}{\prog}{\sema{s}}$,
which is mostly straightforward.
We note only that in a monadic binding $\eseq{\ebind{x}{e_1}}{e_2}$, both $e_1$ and $e_2$ must have array types;
in a guard $\eseq{\eif{e_1}{e_2}}{e}$, $e$ must have an array type,
while $e_1$ and $e_2$ must have (the same) loc-set type,
since equality is only supported over string values.
Full definition can be found in 
\iflong
\autoref{appendix:synthesis} (\autoref{fig:lang}).
\else
the technical report~\cite{techreport}.
\fi

\begin{figure*}
  \begin{minipage}{.65\textwidth}
  \small
      \centering
      \includegraphics[width=.9\textwidth]{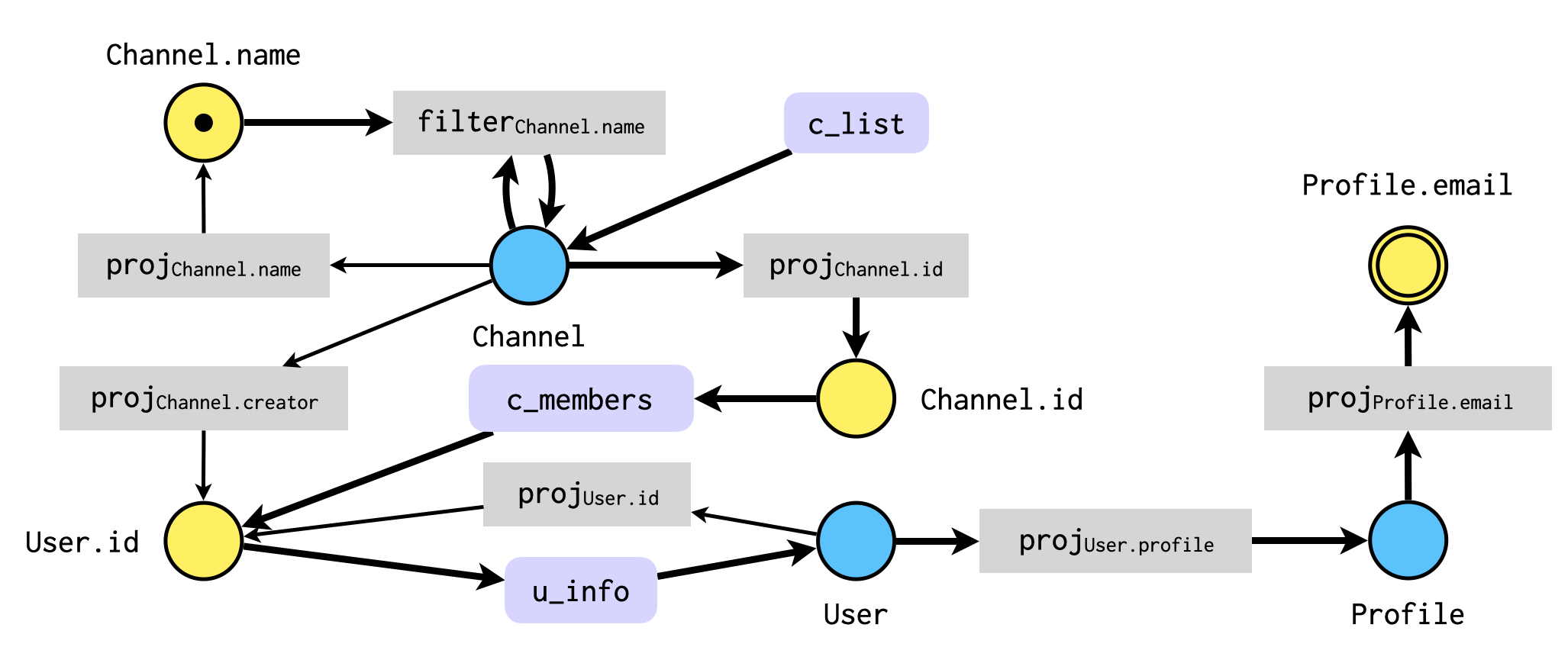}    
  \caption{A fragment of the type-transition net (TTN) for \slack.
  Places (circles) are semantic types;
  transitions (rectangles) are API methods and data transformations.
  The bold path represents the solution to our running example.
  }
  \label{fig:ttn}
  \end{minipage}\hfill\vline\hfill%
  \begin{minipage}{.3\textwidth}
    \small
    \begin{algorithmic}[1]
        \Require{Semantic library $\sema{\Lambda}$ query type $\sema{s}$}
        \Ensure{Set of candidate solutions $\many{\prog}$}
        \Statex
        \Function{Synthesize}{$\sema{\Lambda}, \sema{s}$}
            \Let{$\net$}{\Call{BuildTTN}{$\sema{\Lambda}$}}
            \Let{$I, F$}{\Call{PlaceTokens}{$\sema{s}$}}
            \For{$\pi \in \textsc{Paths}(\net, I, F)$}
                \For{$\prog \in \textsc{Progs}(\pi)$}
                  \State \textbf{yield} $\textsc{Lift}(\sema{\Lambda}, \sema{s},\prog)$
                \EndFor
            \EndFor
        \EndFunction
    \end{algorithmic}
    \caption{Synthesis algorithm}\label{alg:synthesis}
  \end{minipage}
\end{figure*}

\mypara{Type transition nets}
To efficiently enumerate well-typed programs 
we follow prior work~\cite{FengM0DR17,GuoJJZWJP20}
and encode the search space as a special kind of Petri net, called \emph{type-transition net}~(TTN).
Intuitively, a TTN encodes how each API method transforms values of one semantic type into another;
\eg \T{u_info} transforms a \T{User.id} into a \T{User}.
\autoref{fig:ttn} shows a TTN for our running example.
%
Places (circles) correspond to semantic types, 
transitions (rectangles) correspond to methods, 
and edges connect methods with their input and output types.
In addition to API methods,
the TTN contains transitions that correspond to \corelang projections
(\eg $\proj_{\texttt{User.profile}}$ and $\proj_{\texttt{Profile.email}}$)
and guards
(\eg $\guard_{\texttt{Channel.name}}$).

\mypara{Array-oblivious search}
For our search space encoding to be useful, 
we need to make sure that every well-typed \corelang program
corresponds to a path in the TTN.
This is where we encounter a challenge:
there is no straightforward way to encode \corelang's monadic bind operation into the TTN.
Although prior work on \hplus~\cite{GuoJJZWJP20} supports higher-order functions,
the arguments to those functions are syntactically restricted to variables
(\ie inner lambda abstractions are not supported),
which is insufficient for our purposes.
To address this problem, 
we introduce a new, \emph{array-oblivious} TTN encoding,
which does not distinguish between array types and types of their elements,
and hence does not require monadic binds.
For example, in \autoref{fig:ttn} \T{c_members} returns \T{User} instead of \T{[User]},
and hence its output can be passed directly to \T{u_info},
without iterating over it.

\mypara{Search in the TTN}
Once the TTN is built, 
we enumerate paths from the input to the output type
(or rather, array-oblivious versions thereof).
In our example, we place a \emph{token} in the input type \T{Channel.name}
and search for a path (a sequence of transitions)
that would get this token to the output type \T{Profile.email},
possibly generating and consuming extra tokens along the way.
The bold path in~\autoref{fig:ttn} corresponds to our desired solution
from \autoref{fig:slack-solution}.
On this path, we first fire the transition \T{c_list}
(which does not consume any tokens)
to produce an extra token in \T{Channel}.
Next, we fire $\guard_{\texttt{Channel.name}}$,
which consumes the two tokens in \T{Channel} and \T{Channel.name},
and produces a single token in \T{Channel}.
The remaining five transitions on the bold path simply move this one token along
until it reaches \T{Profile.email}.

Like in prior work~\cite{FengM0DR17,GuoJJZWJP20}, 
a path is only considered valid
if the final state contains exactly one token in the output type
(and no tokens in any other types);
this condition ensures that the generated programs use all their inputs.

\mypara{Synthesis algorithm}
\tool's top-level synthesis algorithm is depicted in \autoref{alg:synthesis}.
The algorithm first constructs a TTN $\net$
and encodes the query type $\sema{s}$ as an initial and final token placement, $I$ and $F$; 
it then enumerates all paths from $I$ to $F$ in $\net$ in the order of length (until timeout).
For each path $\pi$, 
the algorithm iterates over the corresponding array-oblivious programs \prog
and \emph{lifts} them into well-typed \corelang programs. 
%
The reason $\pi$ might yield multiple programs is that
the TTN does not distinguish different arguments of the same type,
and hence we must try all their combinations.

Because TTN construction and search for valid paths is similar to prior work, 
we omit their detailed description
and refer an interested reader to 
\iflong
\autoref{appendix:synthesis}.
\else
our technical report~\cite{techreport}.
\fi

One difference worth mentioning, however, is that we use an \emph{integer linear programming} (ILP) solver to find paths in the TTN,
unlike prior approaches, which relied on SAT/SMT solvers.
We found that although both solvers are equally quick at finding \emph{one valid path},
when it comes to computing \emph{all valid paths} of a given length, 
the ILP solver is much more efficient,
as it has native support for enumerating multiple solutions.
%

\begin{figure}
  \small
  \begin{minipage}{.53\columnwidth}
    \begin{lstlisting}[style=dsl,basicstyle=\ttfamily\footnotesize,numbers=left,numberstyle=\sf\tiny,numbersep=5pt,xleftmargin=10pt]
\channel_name ->
  let x1 = c_list({});

  let x2 = x1.name;
  if x2 == channel_name;
  let x3 = x1.id;
  let x4 = c_members(channel=x3);

  let x5 = u_info(user=x4);
  let x6 = x5.profile;
  let x7 = x6.email;

  x7
    \end{lstlisting}
  \end{minipage}%
  \begin{minipage}{.47\columnwidth}
    \begin{lstlisting}[style=dsl,basicstyle=\ttfamily\footnotesize]
\channel_name ->
  let x1 = c_list({});
  x1' <- x1;
  let x2 = x1'.name;
  if x2 == channel_name;
  let x3 = x1'.id;
  let x4 = c_members(channel=x3);
  x4' <- x4;
  let x5 = u_info(user=x4');
  let x6 = x5.profile;
  let x7 = x6.email;
  let x7' = return x7
  x7'
    \end{lstlisting}
  \end{minipage}
  \caption{Array-oblivious program built from the bold path in \autoref{fig:ttn} (left)
  and its lifted version (right).}
  \label{fig:lift-example}
\end{figure}

\mypara{Lifting array-oblivious programs}
The function $\textsc{Progs}(\pi)$ (line 5 in \autoref{alg:synthesis}) converts a TTN path $\pi$
into a set of array-oblivious programs in A-Normal Form (ANF).
\autoref{fig:lift-example} (left) shows the full array-oblivious program extracted from the bold path in \autoref{fig:ttn}.
As you can see from this example, array-oblivious programs can be ill-typed:
for example, the projection $x_1.\T{name}$ in line 4
does not type-check since $x_1$ actually has an array type $[\T{Channel}]$.
What we really want this program to do is to project \T{name} 
(and execute the remaining steps in the program) 
\emph{for each} channel in $x_1$.
This can be accomplished by inserting a monadic binding $\ebind{x_1'}{x_1}$
and using $x_1'$ instead of $x_1$ in line 4
(and elsewhere in the program where a non-array version of $x_1$ is required, such as line 6).
We refer to this process of repairing type errors by inserting monadic bindings and \T{return}s
as \emph{lifting}.%
\footnote{A reader familiar with monads might think of the array-oblivious program as written in the identity monad instead of the list monad,
and lifting as lifting the program back into the list monad.}

The function \textsc{Lift} (line 6 in \autoref{alg:synthesis})
takes as input a semantic library $\sema{\Lambda}$, a query type $\sema{s}$, and an array-oblivious program $\prog$,
and produces a program $\prog'$ that is well-typed at $\sema{s}$.
\autoref{fig:lift-example} (right) depicts the result of lifting the program in \autoref{fig:lift-example} (left)
to the query type $\T{Channel.name} \to [\T{Profile.email}]$ with $\sema{\Lambda}$ from \autoref{tab:library}.
The full definition of lifting can be found in 
\iflong
\autoref{appendix:synthesis} (\autoref{fig:lifting}).
\else
the technical report~\cite{techreport}.
\fi
Informally, lifting type-checks the program ``line by line'',
and whenever it encounters a type mismatch (in a projection, guard, or a method argument),
it inserts the appropriate number of monadic bindings or \T{return}s
in order to fix the mismatch.
This is always possible because the only kind of type mismatch we can encounter
is between an actual type $[..[\sema{t}]..]$ and the expected type $\sema{t}$,
or vice versa.
One thing worth noting is that we assume that the top-level return type of the program is an array type:
since the lifted programs have top-level monadic bindings, 
they can only return arrays.
If the user requests a scalar return type,
we take this into account at the ranking stage 
by prioritizing programs that always return singleton arrays.

\mypara{Completeness}
Strictly speaking, array-oblivious search is incomplete:
there are multiple programs that map to the same array-oblivious program, 
but lifting only returns a single, canonical representative.
For example, consider the program in \autoref{fig:lift-example} (right),
where we iterate over the array \T{x1} only once (line 3), 
and reuse the same ``iterator" variable \T{x1'} in lines 4 and 6. 
An alternative would be to iterate over \T{x1} the second time before line 6,
effectively retrieving names and IDs from all \emph{pairs} of channels
(instead of the name and the ID belonging to the same channel).
We consider this a benign incompleteness because it is much less likely 
that the user intended to loop twice over the same array. 
If they did, we believe they would be able to repair the program by hand, 
as we discuss in \autoref{sec:discussion}.

\section{Ranking}\label{sec:ranking}

As we mentioned in \autoref{sec:overview}, the algorithm \textsc{Synthesize}
may generate hundreds or even thousands of well-typed candidate solutions,
most of which, however, are uninteresting.
We now formalize how \tool ranks these candidates 
with the help of \emph{retrospective execution} (RE).


\mypara{Cost computation}
To rank the programs, we assign them a positive cost,
and then order them from lowest to highest cost.
To compute the cost of a program \prog, 
we retrospectively execute it multiple times,
accumulating execution results in a set $\mathit{res}$;
retrospective execution is non-deterministic,
and executing a program more times lead to more precise cost estimates.
We then compute the cost of \prog based on its result set $\mathit{res}$
and the return type \sema{t} of the query as follows:
\begin{enumerate}
  \item The base cost is the size of \prog\ in AST nodes.
  \item If $\mathit{res} = \emptyset$ (all executions have failed), 
  the candidate receives a large penalty.
  \item If $\mathit{res} = \{[]\}$ (all executions return an empty array), 
  the candidate receives a medium penalty.
  \item Finally, we compare the values $v\in \mathit{res}$ with the desired result type $\sema{t}$;
  recall that \corelang programs always return an array, 
  while $\sema{t}$ might or might not be an array type.
  We assign a small penalty for a \emph{multiplicity mismatch}, \ie
  if either $\sema{t}$ is a scalar type and \emph{any} value $v$ has more than one element,
  or $\sema{t}$ is an array type and \emph{all} values $v$ have a single element.
\end{enumerate}

\begin{figure}
    \small
    \textbf{Retrospective Execution}\quad$\boxed{\stepenv{\Sigma}{e}{v}}$
    \begin{gather*}
    %
      \inference[\textsc{E-If-True-L}]
      {
        x_1 \in \Sigma &
        x_2 \not\in \Sigma &
        \Sigma(x_1) = v_1 \\
        \stepenv{x_2 \mapsto v_1, \Sigma}{e}{v}
      }
      {
        \stepenv{\Sigma}{\eseq{\eif{x_1}{x_2}}{e}}{v}
      }
      \\
      \inference[\textsc{E-If-True-R}]
      {
        x_1 \not\in \Sigma &
        \stepenv{\Sigma}{x_2}{v_2} \\
        \stepenv{x_1 \mapsto v_2, x_2 \mapsto v_2, \Sigma}{e}{v}
      }
      {
        \stepenv{\Sigma}{\eseq{\eif{x_1}{x_2}}{e}}{v}
      }
      \\
    %
    \inference[\textsc{E-Method-val}]
    {
      (f, \many{l_i=v_i}, v_{out}) \in \witnesses
    }
    {
      \stepenv{\Sigma}{f(\many{l_i=v_i})}{v_{out}}
    }
    \\
    \inference[\textsc{E-Method-name}]
    {
      \forall (f, \many{l_i=v'_i}, v_{out}) \in \witnesses.~\exists i : v'_i \neq v_i \\
      (f, \many{l_i=v'_i}, v_{out}) \in \witnesses
    }
    {
      \stepenv{\Sigma}{f(\many{l_i=v_i})}{v_{out}}
    }
    \end{gather*}
  \caption{Retrospective execution.}
  \label{fig:re-rules}
\end{figure}

\mypara{Retrospective execution}
We formalize RE as a judgement \stepenv{\Sigma}{e}{v},
stating that $v$ is a valid result for executing the expression $e$
in the environment $\Sigma$ (which maps variables to values). 
The judgment is also parameterized by a type context $\Gamma$
and witness set $\witnesses$,
used to replay method calls and sample program inputs.
To run a candidate solution $\prog$,
we execute its body in an \emph{empty environment} $\Sigma = \cdot$
and with $\Gamma$ storing the types of $\prog$'s arguments.
As we explain in more detail below,
program inputs are selected lazily, during execution, in order to maximize its chances of producing meaningful results.

\mypara{Replaying method calls}
Most of the rules for the RE judgement describe standard big-step operational semantics
(they can be found in 
\iflong
\autoref{fig:re-full} in \autoref{appendix:re}%
\else
the technical report~\cite{techreport}%
\fi
),
but two groups of rules, shown in \autoref{fig:re-rules}, deserve more attention.
The first group of interest includes \textsc{E-Method-Val} and \textsc{E-Method-Name},
which replay a method call by looking it up in $\witnesses$.
The rule \textsc{E-Method-Val} applies when $\witnesses$ contains an exact match for the current call,
\ie we have previously observed a call to the same method, with the same parameter names and parameter values.
The rule \textsc{E-Method-Name} applies when an exact match cannot be found (see first premise);
in this case we pick an approximate match, where only the method name and parameter names match.
Matching parameter names is important because many REST API methods admit optional parameters,
and behave very differently based on which pattern of optional parameters is provided.
If an approximate match cannot be found either, RE fails.
Note that for a given call $f(\many{l_i=v_i})$,
there might be multiple approximate matches in $\witnesses$,
which makes RE non-deterministic
(in fact, there can even be multiple precise matches because services are stateful).
Due to hidden state and approximate matches,
the results of RE are not guaranteed to match actual execution,
but our experiments show that they are precise enough 
for the purposes of ranking.

\mypara{Lazy sampling of program inputs}
%
The remaining two rules in \autoref{fig:re-rules}
are responsible for choosing program inputs
so as to bias guard expressions to evaluate to true.
We observe that when inputs are sampled eagerly ahead of time,
guard expressions almost always evaluate to false,
causing RE to return an empty array;
as a result, our ranking heuristic cannot distinguish
meaningful candidates from those that return an empty array regardless of the input.
%
%
To address this issue, we postpone adding program inputs to the environment $\Sigma$
until they are used.
If the first usage of a program input is in a guard,
the rules \textsc{E-If-True-L} and \textsc{E-If-True-R} pick its value to make the guard true:
\textsc{E-If-True-L} applies when only the right-hand side of a guard is undefined,
and \textsc{E-If-True-R} applies when the left-hand side or both are undefined.
If the first usage of an input is in a method call or a projection,
we instead randomly sample from all values of the same type observed in $\witnesses$.

\section{Evaluation}\label{sec:eval}

We implemented \tool in Python, 
except for retrospective execution, 
where we used Rust for performance reasons.
We used the Gurobi ILP solver~\cite{gurobi} v9.1
as the back-end for TTN search.
We ran all the experiments on a machine with an Intel Core i9-10850K CPU and 32GB of memory.

We designed our empirical evaluation to answer the following research questions:
\begin{enumerate}[label=(\bfseries RQ\arabic*)]
  \item Can \tool find solutions for a wide range of realistic tasks across multiple popular APIs?
  \item Is type mining effective and necessary for enabling type-directed synthesis?
  \item Is retrospective execution effective and necessary for prioritizing relevant synthesis results?
\end{enumerate}

\begin{table} 
  \centering
  \caption{
    APIs used in our experiments.
    For each API we report
    the number of methods $|\Lambda.f|$,
    min/max number of arguments per method $n_\mathit{arg}$,
    the number of objects $|\Lambda.o|$, 
    and min/max size of the objects $s_\mathit{obj}$.
    We also report the number of witnesses $|\witnesses|$ we collected for type mining 
    and the number of methods covered by those witnesses $n_{\mathit{cov}}$.
  }\label{tab:api}
  \small\begin{tabular}{l|rrrr|rr}
  \toprule
  & \multicolumn{4}{c|}{API size} & \multicolumn{2}{c}{API Analysis} \\
  \cmidrule(lr){2-5} \cmidrule(lr){6-7} 
  API & $|\Lambda.f|$ & $n_{\mathit{arg}}$ & $|\Lambda.o|$ & $s_{\mathit{obj}}$ & $|\witnesses|$ & $n_{\mathit{cov}}$ \\
  \midrule
  
    \slack & 174 & 0 - 15 & 79 & 1 - 70 & 3834 & 60  \\
    \stripe & 300 & 0 - 145 & 399 & 1 - 66 & 25402 & 124  \\
    \squareapi & 175 & 0 - 20 & 716 & 1 - 34 & 1749 & 67  \\
  \bottomrule\end{tabular}
\end{table}

\mypara{API selection}
For our evaluation, we selected \nAPI popular REST APIs:
the \slack communication platform
and two online payment platforms, \stripe and \squareapi. 
We selected these APIs because they are widely used and 
have both an OpenAPI specification and a web interface, 
which allowed us to set up the test environment and collect witnesses easily.
As shown in \autoref{tab:api}, these APIs are quite complex:
each has over a hundred methods with up to 145 arguments;
all three feature optional arguments.
The three APIs also contain a large number of object definitions,
with up to 70 fields.

\mypara{Experiment setup: type mining}
Recall that type mining relies on a witness set $\witnesses$.
Witnesses are straightforward to collect for API owners,
or when an integration test suite is publicly available;
neither was the case in our setting.
Instead, we collected witnesses by observing traffic from the services's web interface,
and then enhancing this initial (very sparse) witness set via random testing;
this process is described in more detail in 
\iflong
\autoref{sec:appendix:witness-collection}%
\else
our technical report~\cite{techreport}%
\fi
.
As shown in \autoref{tab:api},
we collected between 1.7K and 25K witnesses per API,
which covered 30--40\% of all methods.
It is hard to obtain full coverage for these closed source APIs as an outsider,
for instance, because many methods are only available to paid accounts;
our experiments show, however, that \tool performs well with this witness set.

%

\mypara{Benchmark selection}
For each API, we extracted programming tasks
from \sover questions that mention this API 
as well as \github repositories that use the API.
After excluding the tasks that were out of scope of our DSL,
we manually translated each of the remaining tasks
from a natural-language description or a code snippet into a type query,
resulting in \nBench benchmarks (see \autoref{tab:bench}).
Apart from our running example (benchmark 1.1),
these include, for instance:
``Send a message to a user given their email'' in \slack (1.2),
``Create a product and invoice a customer'' in \stripe (2.3),
and ``Delete catalog items with given names'' in \squareapi (3.10).
As noted in \autoref{tab:bench},
many of these tasks are \emph{effectful}:
they require creating, modifying, or deleting objects.
%
%

Each benchmark comes with a ``gold standard'' solution:
the accepted solution on \sover or the snippet we found on \github.
We manually translated these solutions into \tool's DSL.
As shown in the ``Solution Size'' portion of \autoref{tab:bench},
these solutions range in complexity from 7 to 22 AST nodes,
containing up to three method calls and guards and up to seven projections,
which makes them non-trivial for programmers to solve manually.
A complete list of tasks, type queries, and solutions
can be found in
\iflong
\autoref{appendix:solutions}%
\else
\cite{techreport}%
\fi
.

\begin{table}[t]
  \caption{
    Synthesis benchmarks and results.
    Benchmarks marked with $\dagger$ are effectful.
    For each benchmark we report the size of the desired solution:
    AST, $n_f$, $n_p$ and $n_g$ correspond to
    number of AST nodes, method calls,
    projections and guards, respectively.
    We also report the time to find the correct solution (in seconds),
    its rank without RE (\rorig),
    and the lower and upper bound on its rank with RE (\rre and \rreto).    
    `-' means no solution is found in \synTO.
  }\label{tab:bench}
  \small
\begin{tabular}{c|l|rrrr|r|rrr}
        \toprule
        \multirow{2}{*}{API} & \multirow{2}{*}{ID} & \multicolumn{4}{c|}{Solution Size} & \multicolumn{1}{c|}{Time} & \multicolumn{3}{c}{Rank} \\
        \cmidrule(lr){3-6} \cmidrule(lr){8-10} 
         &  & AST & $n_{f}$ & $n_{p}$ & $n_{g}$ & (sec) & \rorig & \rre & \rreto \\
        \midrule
        \multirow{8}{*}{\rotatebox{90}{\slack}} & 1.1 & 17 & 3 & 6 & 1 & 83.5 & 25230 & 5 & 5  \\
        & 1.2$^{\dagger}$ & 12 & 3 & 5 & 0 & 5.6 & 2224 & 10 & 10  \\
        & 1.3 & 16 & 3 & 7 & 0 & - & - & - & -  \\
        & 1.4 & 14 & 2 & 4 & 1 & 1.3 & 489 & 24 & 31  \\
        & 1.5$^{\dagger}$ & 10 & 2 & 3 & 0 & 3.4 & 788 & 5 & 5  \\
        & 1.6$^{\dagger}$ & 9 & 2 & 2 & 0 & 1.7 & 573 & 8 & 19  \\
        & 1.7$^{\dagger}$ & 12 & 2 & 4 & 1 & 1.3 & 757 & 8 & 9  \\
        & 1.8 & 9 & 2 & 3 & 0 & 42.0 & 16438 & 29 & 30  \\
        \hline
        \multirow{13}{*}{\rotatebox{90}{\stripe}} & 2.1$^{\dagger}$ & 9 & 2 & 2 & 0 & 95.4 & 4952 & 3 & 3  \\
        & 2.2$^{\dagger}$ & 10 & 2 & 2 & 0 & 92.4 & 4854 & 4 & 4  \\
        & 2.3$^{\dagger}$ & 12 & 3 & 2 & 0 & 121.2 & 6363 & 1 & 1  \\
        & 2.4 & 8 & 1 & 2 & 1 & 0.5 & 3 & 1 & 1  \\
        & 2.5 & 8 & 2 & 2 & 0 & 1.0 & 10 & 4 & 4  \\
        & 2.6$^{\dagger}$ & 9 & 3 & 2 & 0 & 12.2 & 270 & 3 & 3  \\
        & 2.7 & 5 & 1 & 2 & 0 & 0.6 & 4 & 2 & 2  \\
        & 2.8 & 16 & 2 & 7 & 1 & 20.2 & 679 & 17 & 17  \\
        & 2.9 & 6 & 1 & 2 & 0 & 0.5 & 2 & 1 & 1  \\
        & 2.10$^{\dagger}$ & 10 & 2 & 3 & 0 & 7.8 & 187 & 6 & 6  \\
        & 2.11$^{\dagger}$ & 7 & 2 & 1 & 0 & 17.2 & 490 & 6 & 6  \\
        & 2.12$^{\dagger}$ & 11 & 3 & 2 & 0 & - & - & - & -  \\
        & 2.13$^{\dagger}$ & 10 & 3 & 2 & 0 & - & - & - & -  \\
        \hline
        \multirow{11}{*}{\rotatebox{90}{\squareapi}} & 3.1 & 4 & 1 & 1 & 0 & 0.2 & 2 & 1 & 1  \\
        & 3.2 & 16 & 1 & 4 & 3 & 0.5 & 10 & 4 & 4  \\
        & 3.3 & 10 & 1 & 3 & 1 & 0.4 & 6 & 1 & 1  \\
        & 3.4 & 5 & 1 & 2 & 0 & 0.7 & 2 & 1 & 1  \\
        & 3.5$^{\dagger}$ & 14 & 2 & 3 & 0 & 2.2 & 99 & 2 & 2  \\
        & 3.6 & 5 & 1 & 2 & 0 & 0.2 & 1 & 1 & 1  \\
        & 3.7 & 6 & 1 & 2 & 0 & 0.3 & 7 & 4 & 4  \\
        & 3.8 & 9 & 1 & 3 & 0 & 0.7 & 1 & 1 & 1  \\
        & 3.9 & 8 & 1 & 2 & 1 & 0.2 & 3 & 2 & 2  \\
        & 3.10$^{\dagger}$ & 16 & 2 & 5 & 1 & 1.9 & 174 & 10 & 12  \\
        & 3.11$^{\dagger}$ & 8 & 2 & 3 & 0 & 1.0 & 68 & 16 & 16  \\
        \bottomrule\end{tabular}
\end{table}

\mypara{Experiment setup: program synthesis}
For each of the \nTyQueries benchmarks,
we ran the synthesizer with a timeout of \synTO.
For each new candidate generated,
we estimated its cost using \nMultiplicity rounds of RE
and recorded the synthesis time
(including both TTN search and RE time).  
After the timeout,
we checked whether the gold standard solution appears among the generated candidates
and compared its RE-based rank vs the original rank at which it was generated 
(based on path length).
Below we report average time and median rank over \repeatTimes runs
to reduce the impact of randomness.

\begin{figure}
  \centering
  \includegraphics[width=\linewidth]{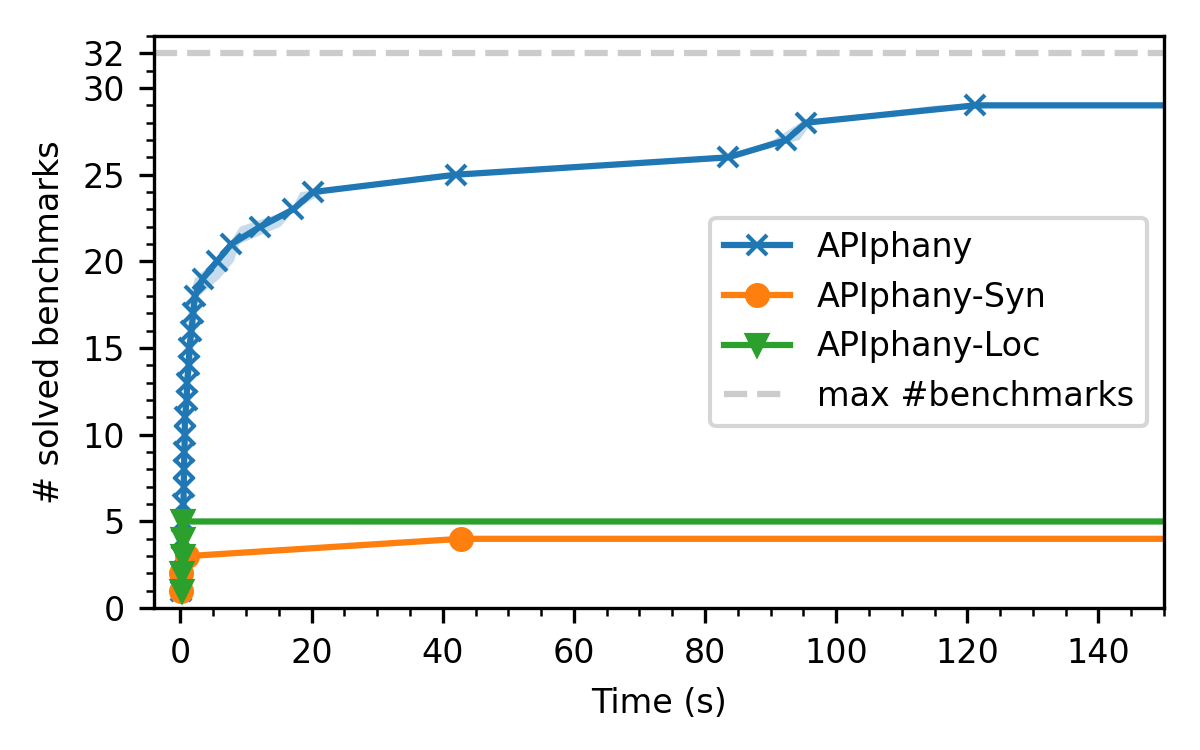}
  \vspace{-10pt}
  \caption{Comparison of synthesis performance between \tool and its two variants that do not use type mining.}
  \label{fig:ablations}
  \vspace{-5pt}
\end{figure}

\subsection{RQ1: Overall Effectiveness}\label{sec:rq1}
The last four columns of \autoref{tab:bench} detail \tool's performance on the \nTyQueries synthesis benchmarks.
\tool finds the correct solution for \nCorrect benchmarks.
The remaining three benchmarks fail with a timeout because their type queries are too ambiguous;
for example, in benchmark 1.3 (``Get unread messages of a user'') 
the type query has no means to specify that we are only interested in \emph{unread} messages;
as a result, the solution is drowned among thousands of other programs that map a user ID to messages.

We plot the number of benchmarks solved as a function of time (including RE) in \autoref{fig:ablations}.
As the plot shows, majority of benchmarks (\nFiveSec/\nTyQueries) can be solved within five seconds.
On average \tool takes \avgTime seconds to find the desired solution
(median time \medianTime seconds).

\medskip
\noindent
\framebox[\linewidth]{
  \parbox{.95\linewidth}{
    \centering
    \textbf{Takeaway:}
    \tool is able to solve 91\% of tasks from three real-world APIs.
    }    
}

\subsection{RQ2: Type Mining}\label{sec:rq2}

Recall that type mining involves replacing primitive \emph{syntactic types} in the spec with unique \emph{location-based types},
and then merging those based on the witness set to obtain \emph{semantic types}.
The merging process is not perfect:
it might \emph{fail to merge} two location that should have the same type
because the witness set lacks evidence to justify the merge;
or it might \emph{spuriously merge} two locations
if they share a value by chance.
It is hard to measure the accuracy of inferred types directly,
since we do not have an oracle for semantic types.
Instead, we evaluate type mining indirectly in two ways:
\begin{inparaenum}[1)]
  \item we run an \emph{ablation study} to measure its impact on the overall performance of the synthesizer, and
  \item we perform a small-scale \emph{qualitative analysis} of inferred types.
\end{inparaenum}

\mypara{Ablation study}
For this experiment, we compare the performance of \tool
and its two variants:
\begin{inparaenum}[(a)]
  \item \nosem, which builds the TTN directly from syntactic types, and
  \item \nomerge, which builds the TTN from (unmerged) location-based types.
\end{inparaenum}
We plot the number of benchmarks solved by each variant as the function of time in \autoref{fig:ablations}.

As expected, both variants perform poorly:
\nosem only solves 4/\nBench benchmarks and \nomerge solves 5.
All these benchmarks are ``easy'' (solved by \tool in under a second).
Intuitively, the two variants represent two extremes in terms of type \emph{granularity}.
Syntactic types are \emph{too coarse-grained} (all \T{String} locations have the same type),
which leads TTN search to return too many well-typed candidates.
As a result, \nosem struggles to solve all but the simplest tasks,
with many benchmarks running out of memory.
Location-based types, on the other hand, are \emph{too fine-grained} (each \T{String} location has a unique type),
which leads to most desired solutions simply being ill-typed,
because there is no way for one method to use values returned by another.
The solutions to all of the five benchmarks solved by \nomerge
have only one method call with no parameters,
followed by several projections or filters.

As you can see from \autoref{fig:ablations},
\tool drastically outperforms both variants.
This result indicates that type mining strikes a good balance
between coarse- and fine-grained types:
all \nBench benchmarks have a well-typed solution in terms of the mined types,
and \tool is able to find most of them within a reasonable time.

\mypara{Qualitative analysis}
To give a more direct account of the quality of inferred semantic types,
we randomly sampled five methods from each API
(among the methods covered by the collected witnesses),
and manually inspected the inferred types to check if they match our expectations.
More specifically, for each \T{String} location in a method spec,
we pick a location type $\lloc^*$, 
which we deem most natural for a programmer to use in a type query
(for example, for the parameter to \T{users_info}, $\lloc^* = \T{User.id}$);
we consider the inferred loc-set type sufficient if it contains $\lloc^*$.
The detailed results appear in
\iflong
\autoref{appendix:solutions} (\autoref{tab:qualitative})%
\else
the technical report~\cite{techreport}%
\fi
.

In the methods we examined,
type mining was able to infer a sufficient semantic type for all responses, required parameters,
and about half of optional parameters.
The remaining optional parameters were assigned unmerged location types,
because they were never used in our witness set.
This is almost unavoidable, 
because of the sheer number of obscure optional parameters in real-world APIs
(which, fortunately, are rarely needed to solve programmer's tasks).

Recall that the other failure mode of type mining is spuriously merging unrelated locations.
We did not observe any spurious merges among the randomly sampled methods,
but anecdotally we did encounter one such merge elsewhere in the Slack API: 
between \T{Channel.name} and \T{Message.name}.
Note that spurious merges might slow down the search and 
produce some ``semantically ill-typed'' solutions, 
but they do not prevent \tool from finding the desired solution. 

\medskip
\noindent
\framebox[\linewidth]{
  \parbox{.95\linewidth}{
    \centering
    \textbf{Takeaway:} Type mining increases the percentage of solved benchmarks from 12\% to 91\%.
    }
}

\begin{figure}[t]
  \includegraphics[width=\linewidth]{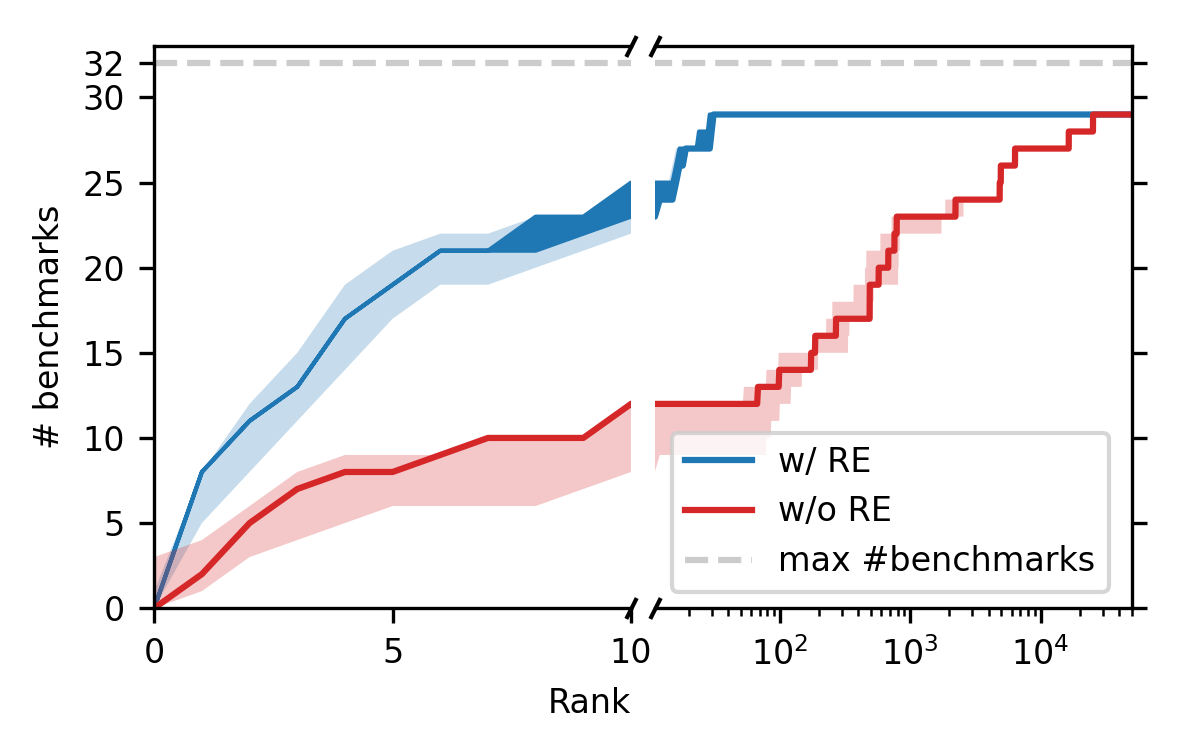}
  \caption{Number of benchmarks whose solution is reported within a given rank.
    The filled blue area is the range of ranks one might get
    depending on when they inspect the candidates.
    The shaded area is the 95\% confidence interval.}
  \label{fig:rank-cactus}
\end{figure}

\subsection{RQ3: Ranking}

To measure the effectiveness of RE-based ranking,
we compare the last three columns of \autoref{tab:bench}:
\rorig denotes the rank of the desired solution in the order it was generated by TTN search
(which is based on path length, and hence correlated with solution size);
\rre denotes the RE-based rank of the solution at the time it was generated,
and \rreto denotes its RE-based rank by the timeout
(which can be lower than \rre as other candidates generated later might end up being ranked higher).
We report both of these RE-based ranks
because we envision an \tool user inspecting the candidate solutions
some time between they are generated and the timeout,
and hence the relevant rank value is between \rre and \rreto.
We plot the number of benchmarks whose solutions lie at or below each rank in \autoref{fig:rank-cactus},
with the range between \rre and \rreto represented as a filled area.

As you can see from \autoref{fig:rank-cactus}, RE-based ranking
significantly increases the chances that the desired solution makes the short-list of candidates.
In particular, \emph{without RE-based ranking}
only \nTopFive benchmarks (28\% of solved) return the correct solution in top five, 
and only \nTopTen (41\%) return it in top ten;
in contrast, \emph{with RE-based ranking},
\nTopFiveRE (65\%) benchmarks return the correct solution in top five (after timeout), 
and \nTopTenRE (79\%) in top ten.
Moreover, as we can see from \autoref{tab:bench},
the solution's rank never gets worse after RE,
in all but two cases it strictly improves,
and for all long-running benchmarks it improves drastically
(the average rank improves from 2230.5 to 7.0).

A closer look at the six benchmarks that do not land in top ten after RE
reveals two main reasons for these suboptimal rankings.
In most cases the solution is simply large, 
and there are many smaller candidates that are still meaningful. 
For example, the query ``Delete all catalog items" (3.11) takes no arguments and 
returns an array of all deleted items; 
there are many valid and simple ways to construct an array of catalog items without deleting them.
In a few cases, APIphany fails to throw out meaningless programs 
due to the imprecision of retrospective execution.
For example, in 1.6 it reports a solution 
that posts an update to a given channel with a given timestamp, 
even though this timestamp might be invalid for this channel; 
APIphany instead thinks that this call always succeeds 
by relying on approximate matches during retrospective execution.

We also recorded the time \tool takes to compute the cost for all generated candidates
(which involves executing each candidate \nMultiplicity times).
Although \tool generates thousands of well-typed candidates for most benchmarks,
cost computation only takes about 1\% of total synthesis time.

\medskip
\noindent
\framebox[\linewidth]{
  \parbox{.95\linewidth}{
    \centering
  \textbf{Takeaway:} RE-based ranking takes a negligible amount of time
  and increases the percentage of correct solutions reported in top ten from 41\% to 79\%.
  }
}

\subsection{Discussion and Limitations}\label{sec:discussion}

\mypara{Witness generation} 
One threat to validity of our evaluation is that the results of type minings (and therefore synthesis)
depend heavily on the witness set.
In particular, if our benchmarks required methods that are not covered by the witness set,
\tool most likely would not be able to solve them,
since they would be ill-typed with inferred semantic types.
We ran our experiments using a particular witness set, 
which we collected using one methodology (described in 
\iflong
\autoref{sec:appendix:witness-collection}%
\else
the technical report~\cite{techreport}%
\fi
);
our findings might not generalize to using \tool with witness sets collected by other means.

\mypara{Effectful methods}
We observe that effectful methods in REST APIs have an interesting property:
they make the effect explicit in their response.
For example, the method for posting a message on \slack also returns the message object,
and the method for deleting a catalog item in \squareapi returns the ID of the deleted item
(instead of just returning \T{void}).
This property makes REST APIs particularly suitable for type-directed synthesis
and expressing user intent with types:
for example, the query ``Send a message to a user with a given email''
can be expressed as the type $\T{Profile.email} \to \T{Message}$
instead of a much less informative type $\T{Profile.email} \to \T{void}$.
The downside, of course, is that the return type of an effectful method might not be obvious to the user
(for example, does deleting a catalog item return an object or its ID?)
One way to overcome this limitation is to let the user specify the name of the last method they want to call
(\eg \T{catalog_object_delete}) instead of the output type;
this kind of specification is straightforward to integrate into TTN search.

\mypara{DSL restrictions}
In our search for benchmarks,
we encountered (very few) snippets that were inexpressible in our DSL because
they required functional transformations on primitive values,
as opposed to just structural transformations on objects and arrays,
for example: ``Get all members of a channel and \emph{concatenate} them together". 
We consider such functional transformations beyond the scope of \tool 
because its type-based specifications are too coarse to 
distinguish between different functional transformations.
This is also the reasoning behind our design decision to only support equality inside guards,
as opposed to more general predicates:
if the specification cannot distinguish between, say, $=$ and $\leq$,
there is little use in generating programs with both.
More generally, we view programs synthesized by \tool as a starting point, 
which helps the programmer figure out how to plumb data through a set of API calls; 
we envision the user building on top of those programs to add functional modifications 
and more expressive predicates. 
This interaction model motivates both our DSL restrictions and our type-based specifications.

\mypara{Value-based location merging}
Value-based merging works well for strings,
since their large domain makes it unlikely that two \T{String} locations share a value by chance.
It works less well for other primitive types, such as integers and booleans.
To reduce the risk of spurious merges, 
our implementation performs value-based merging 
only for strings and large integers ($> 1000$),
but not for booleans or small integers.
In the future, 
we plan to investigate more sophisticated approaches to location merging.
One idea is to use probabilistic reasoning to estimate the likelihood of two locations having the same type based on 
\begin{inparaenum}[(1)]
   \item how common a value is across locations and
   \item what proportion of values is shared between the two locations. 
\end{inparaenum}
%
%
Another approach is to cluster locations using NLP techniques, 
such as sentiment analysis of object and field names, 
as well as documentation.

\mypara{User interface}
Another important direction for future work is to investigate 
usable ways of specifying semantic type queries 
and comprehending synthesis results.
In particular, existing work from the HCI community~\cite{overcode,examplore} 
might help users quickly explore a large space of related candidate solutions, 
thereby mitigating the limitations of ranking.
\section{Related Work}\label{sec:related}

\tool is a component-based synthesizer 
and primarily compares with related work in this space.  
It also draws on techniques from specification mining 
and type inference. 

\mypara{Type-directed component-based synthesis}
The goal of component-based synthesis is to find a \emph{composition}
of components (library functions) that implements a given task. 
In \emph{type-directed} component-based synthesis
both the task and the components are specified using types.
%
The traditional approach to this problem based on proof search~\cite{djinn,Norell08,rehof16}
scales poorly with the size of the component library.
An alternative, more scalable \emph{graph-based approach}
was introduced in \prospector~\cite{Mandelin05} for unary components,
and generalized to $n$-ary components in \sypet~\cite{FengM0DR17},
by replacing graphs with Petri nets.
\tygar~\cite{GuoJJZWJP20} further extends \sypet's 
search to polymorphic components using the idea of \emph{abstract types},
which are inspired by \emph{succinct types} from another component-based synthesizer, 
\insynth \cite{GveroKKP13}.
\tool's program search phase is using the Petri net encoding from \sypet and \tygar
with minor adaptations (support for optional arguments and ILP encoding).
Our array-oblivious encoding is related to abstract and succinct types
in that it helps make the Petri net smaller,
but it is also substantially different
in that, unlike prior work, it can efficiently encode a certain class of higher-order programs
(array comprehensions) 
into the Petri net.
%

\mypara{API navigation}
Beyond type-directed synthesis,
other work focuses on smart auto-completion~\cite{Perelman12,Raychev14,Aroma}
but relies on static analysis and mining client code,
which \tool does not require.
Among tools that leverage dynamic analysis,
\tname{EdSynth}~\cite{Yang2018} uses test executions to generate
snippets that involve both API calls and control structures. 
\tname{MatchMaker}~\cite{Yessenov11} and \tname{DemoMatch}~\cite{Yessenov17}
are similar to \tool in that they rely on observed program traces
to suggest code that uses complex APIs
(the former from types and the latter from demonstrations).
All these techniques work in the context of Java,
and hence assume that sufficiently precise types are already present.

\mypara{SQL synthesis}
The problem of generating projections and filters is related to synthesis of SQL queries~\cite{scythe,sqlizer}.
%
Existing SQL synthesis techniques are not directly applicable to our problem domain, because
\begin{inparaenum}[(1)]
\item our programs also contain arbitrary API method invocations, and
\item we manipulate semi-structured data instead of relational data.
\end{inparaenum}

\mypara{API discovery and specification mining}
A complimentary approach to API navigation using program synthesis
is to infer specifications~\cite{specmining,ShohamYFP07,Mishne12} 
or example usages~\cite{Heydarnoori09,Buse12,Barnaby20}
to help the user understand the API better.
%
\tool's type mining is inspired by \citet{specmining}, where they build
probabilistic finite state automata representing data and temporal dependencies
between API methods.  
\tool implements a simpler form of their algorithm,
which discovers data flows (but not temporal dependencies),
but the novelty lies in using this information to drive program synthesis.

%

%

Type mining is also related to prior work on 
inferring type annotations for dynamically typed languages from executions%
~\cite{Chugh2011,DRuby,typedclojure}.
%
However, this work is for structural types,
whereas we infer domain-specific nominal types. 


\mypara{Simulated execution}
An alternative to our retrospective execution is to synthesize a \emph{model} of the API,
and evaluate program candidates against that model.
Previous work~\cite{Heule15,Jeon16} synthesizes models for complex frameworks and opaque code;
our retrospective execution is simpler: it skips the extra step of model synthesis.

\mypara{Ranking solutions}
Specifications in program synthesis are often ambiguous,
so synthesizers have to rank their candidate solutions
and return the top result(s).
Existing tools most commonly rely on hand-crafted~\cite{Gulwani11}
or learned~\cite{GveroKKP13,Raychev14,SinghG15} ranking functions
based on syntactic features of generated programs.
%
%
\hplus~\cite{JamesGWDPJP20} is most similar to \tool in that it
ranks programs based on the results of their \emph{execution},
using heuristics like whether the program always fails, 
and how similar it is to other candidates.


\begin{acks}
  The authors would like to thank the anonymous reviewers, 
  our shepherd Yuepeng Wang, 
  as well as Hila Peleg and Ilya Sergey
  for their valuable feedback on earlier drafts of this paper.
  This work was supported by the National Science Foundation under Grants
  No.~1943623, 1911149, and 2107397.
\end{acks}

\balance
\bibliography{main}

\iflong
\clearpage
\appendix
\section{Type Mining}\label{appendix:mining}

\begin{figure}
  \textbf{Location-Based Type Inference}\quad$\boxed{\jtinfer{\lloc}{\sema{t}}}$ 
  \begin{gather*}
    \inference[ObjStart]
    {\jtinferacc{o}{\many{l}}{\sema{t}}}
    {\jtinfer{o.\many{l}}{\sema{t}}}
    \quad
    \inference[FunStart]
    {\jtinferacc{f}{\many{l}}{\sema{t}}}
    {\jtinfer{f.\many{l}}{\sema{t}}}
    \\
    \inference[ObjBase]
    {  }
    {\jtinferacc{o}{[]}{o}}
    \quad
    \inference[PathBase]
    { \lloc \neq o }
    {\jtinferacc{\lloc}{[]}{\{\lloc\}}}
    \\
    \inference[ObjFollow]
    { \Lambda(\lloc.l_1) = o &
      \jtinferacc{o}{\many{l}}{\sema{t}}
    }
    {\jtinferacc{\lloc}{l_1.\many{l}}{\sema{t}}}
    \\
    \inference[Arr]
    { \Lambda(\lloc.l) = [t] &
      \jtinferacc{\lloc.l}{\lidx}{\sema{t}}
    }
    {\jtinferacc{\lloc}{l}{[\sema{t}]}}
    \\
    \inference[AdHoc]
    { \Lambda(\lloc.l) = \{\many{l_i : t_i}\} &
      \many{\jtinferacc{\lloc.l}{l_i}{\sema{t_i}}}
    }
    {\jtinferacc{\lloc}{l}{\{\many{l_i : \sema{t_i}}\}}}
    \\
    \inference[PathFollow]
    { \Lambda(\lloc.l_1) = \T{String} \vee (\Lambda(\lloc.l_1) \neq o \wedge \many{l} \neq []) \\
     \jtinferacc{\lloc.l_1}{\many{l}}{\sema{t}} }
    {\jtinferacc{\lloc}{l_1.\many{l}}{\sema{t}}}
  \end{gather*}
  \caption{Rules for location-based type inference.}
  \label{fig:type-inference}
\end{figure}

\mypara{Location-based type inference}
The location-based type inference judgement \jtinfer{\lloc}{\sema{t}} is defined in~\autoref{fig:type-inference}.
In this figure, the notation $\Lambda(\lloc) = t$ denotes looking up the syntactic type of location $\lloc$
in the library $\Lambda$,
\eg $\Lambda(\T{User.profile}) = \T{Profile}$
and $\Lambda(\T{c_members}.\lout.\lidx) = \T{String}$.
At the same time, $\Lambda(\T{User.profile.email})$ is undefined
because it does not directly appear in $\Lambda$ 
(instead we need to ask for \T{Profile.email}).
The full definition of
syntactic lookup is straightforward and therefore omitted.

The main complication during location-based type inference
is that we need to ``fold'' locations that denote named objects,
replacing them with object names;
\eg 
$$\jtinfer{\T{users_info}.\lout.\T{id}}{\{\T{User.id}\}}$$
To this end we introduce an auxiliary judgment \jtinferacc{\lloc}{\many{l}}{\sema{t}},
where intuitively $\lloc$ and $\many{l}$ correspond to a prefix and a suffix of the location of interest,
except that $\lloc$ is sufficiently ``folded''.
For example, to derive the judgment above, we will start with
$$\jtinferacc{\T{users_info}}{\lout.\T{id}}{\cdots}$$ 
but then use ObjFollow to fold the path and rewrite the judgment it into 
$$\jtinferacc{\T{User}}{\T{id}}{\cdots}$$
and then by PathBase we can establish 
$$\jtinferacc{\T{User}}{\T{id}}{\{\T{User.id}\}}$$
Note that because $\lloc$ in this judgment is always sufficiently folded,
all applications of $\Lambda(\lloc.l)$ in our rules are actually well-defined.

The rules Arr and AdHoc deal with inference of array types and record types.
The rule Arr applies when the syntactic type of the top-level location of interest is an array;
in this case we infer the semantic type $\sema{t}$ of the array element
and return $[\sema{t}]$ for the location.
Note that this rule is not used when an array-typed location occurs in the middle of a path and not at the top level:
in this case the default rule PathFollow applies.
The rule AdHoc takes care of locations that have \emph{ad-hoc} record types in $\Lambda$
(as opposed to names object types);
for example, we infer the type $\{\T{user}: \T{User.id}\}$ for the location $\T{users_info}.\lin$.
Just like Arr, AdHoc only applies when the top-level location has a record type,
and otherwise PathFollow suffices.

\section{Program Synthesis}\label{appendix:synthesis}

\begin{figure}
  \textbf{Expression Typing}\quad$\boxed{\jtyping{\Gamma}{e}{\sema{t}}}$\quad$\boxed{\jtypinglib{\sema{\Lambda}}{\prog}{\sema{s}}}$
  \begin{gather*}
  \inference[\textsc{T-Var}]
  {x:\sema{t}\in \Gamma}
  {\jtyping{\Gamma}{x}{\sema{t}} }
  \quad
  \inference[\textsc{T-Proj}]
  {\jtyping{\Gamma}{e}{\{l:\sema{t}, \ldots\}}}
  {\jtyping{\Gamma}{e.l}{\sema{t}} }
  \\
  \inference[\textsc{T-Ret}]
  {\jtyping{\Gamma}{e}{\sema{t}}}
  {\jtyping{\Gamma}{\eret{e}}{[\sema{t}]} }
  \\
  \inference[\textsc{T-Call}]
  {f: \{\many{\ell_j: \sema{t}_j}\}\to \sema{t}_o \in \sema{\Lambda} &
   \jtyping{\Gamma}{e_i}{\sema{t}_i} \\
   \forall j . \ell_j = l_j \Rightarrow \exists i . l_j = l_i \wedge \sema{t}_j = \sema{t}_i \\
   \forall j, i . \ell_j = ?l_j \wedge l_j = l_i \Rightarrow \sema{t}_j = \sema{t}_i}
  {\jtyping{\Gamma}{f(\many{l_i=e_i})}{\sema{t}_o} }
  \\
  \inference[\textsc{T-If}]
  {\jtyping{\Gamma}{e_x}{\{\many{\lloc}\}} & \jtyping{\Gamma}{e_y}{\{\many{\lloc}\}} &  \jtyping{\Gamma}{e}{[\sema{t}]}}
  {\jtyping{\Gamma}{\eseq{\eif{e_x}{e_y}}{e}}{[\sema{t}]} }
  \\
  \inference[\textsc{T-Let}]
  {\jtyping{\Gamma}{e_1}{\sema{t}_1} & \jtyping{\Gamma,x:\sema{t}_1}{e_2}{\sema{t}_2}}
  {\jtyping{\Gamma}{\eseq{\elet{x}{e_1}}{e_2}}{\sema{t}_2} }
  \\
  \inference[\textsc{T-Bind}]
  {\jtyping{\Gamma}{e_1}{[\sema{t}_1]} & \jtyping{\Gamma,x:\sema{t}_1}{e_2}{[\sema{t}_2]}}
  {\jtyping{\Gamma}{\eseq{\ebind{x}{e_1}}{e_2}}{[\sema{t}_2]} }
  \\
  \inference[\textsc{T-Obj}]
  {\jtyping{\Gamma}{e}{o} & o:\sema{t}\in \sema{\Lambda}}
  {\jtyping{\Gamma}{e}{\sema{t}}}
  \\
  \inference[\textsc{T-Top}]
  {\jtyping{\many{x_i:\sema{t}_i}}{e}{\sema{t}}}
  {\jtypinglib{\sema{\Lambda}}{\elam{\many{x_i}}{e}}{\{\many{x_i:\sema{t}_i}\} \to \sema{t}} }
  \end{gather*}
  \caption{\corelang: typing judgment.}\label{fig:lang}  
\end{figure}

\mypara{Expression typing}
The (semantic) typing rules for programs of \corelang are presented in \autoref{fig:lang}.
The typing judgement $\jtyping{\Gamma}{e}{\sema{t}}$ states
that a term $e$ has the semantic type $\sema{t}$
under context $\Gamma$ and the semantic library $\sema{\Lambda}$.
The rule \textsc{T-Call} check that
all required arguments are provided and
all provided arguments have correct types.
In a monadic binding $\eseq{\ebind{x}{e_1}}{e_2}$, both $e_1$ and $e_2$ must have array types (\textsc{T-Bind}).
In a guard $\eseq{\eif{e_1}{e_2}}{e}$, $e$ must have an array type,
while $e_1$ and $e_2$ cannot have an array or record type,
since equality is only supported over string values (\textsc{T-If}).
Finally, typing follows object definitions in $\sema{\Lambda}$:
if $e$ has type $o$ and $o:\sema{t} \in \Lambda$, then $e$ also has the type $\sema{t}$ (\textsc{T-Obj}).

\subsection{TTN Construction}\label{appendix:synthesis:construction}

%
A TTN $\net$ is a 4-tuple $(P, T, E, O)$, where $P$ is a set of \emph{places},
$T$ is a set of \emph{transitions}, and
$E: (P \times T) \cup (T \times P) \to \mathbb{N}$ is a matrix of \emph{edge multiplicities}.
$E(p,\tau)$ denotes how many required arguments of type $p$ component $\tau$ consumes,
and $E(\tau,p)$ denotes how many responses of type $p$ it produces.
$E(p,\tau) = 0$ means that there is no edge from $p$ to $\tau$ (and symmetrically for $E(\tau,p) = 0$).
To model optional arguments, in this work we augment the TTN 
with the matrix of \emph{optional multiplicities} $O: P \times T \to \mathbb{N}$,
which denotes the number of optional arguments of a given type.

A \emph{marking} is a mapping $M: P \to \mathbb{N}$ that
assigns a non-negative number of tokens to every place.
A \emph{transition firing} is a triple $M_1 \xrightarrow{\tau} M_2$, 
such that for all places $p$: 
\begin{inparaenum}[(a)]
  \item $M_1$ contains at least as many tokens as $\tau$'s incoming edges require: $M_1(p) \geq E(p, \tau)$, and
  \item $M_2$ loses tokens consumed by the incoming edge, but gains tokens produced by the outgoing edge: 
  $\exists c . E(p,\tau) \leq c \leq E(p,\tau) + O(p,\tau) \wedge M_2(p) = M_1(p) - c + E(\tau, p)$.
\end{inparaenum}
A \emph{path} between $M$ and $M'$ is a sequence of transitions $\tau_1, \ldots, \tau_n$
such that $M \xrightarrow{\tau_1} M_1 \xrightarrow{\tau_2} \ldots \xrightarrow{\tau_{n-1}} M_{n-1} \xrightarrow{\tau_n} M'$
is a sequence of transition firings.

\mypara{Array-oblivious encoding}
Procedure $\textsc{BuildTTN}(\sema{\Lambda})$ (line 2 in \autoref{alg:synthesis})
constructs a TTN $\net$ given a semantic library.
Intuitively, its goal is to add transitions for all methods in $\sema{\Lambda}$,
as well as other operations of \corelang, such as projections and filtering,
such that any well-typed \corelang program can be encoded as a path in $\net$.
There is one major issue, however:
\corelang programs also contain the higher-order monadic bind operations (aka ``flat maps''),
which cannot be easily encoded in a TTN;
moreover, the presence of array types in the TTN
increases the number of places and transitions, and slows down the search.
To combat this issue we propose the \emph{array-oblivious encoding}
of \corelang programs into the TTN,
which replaces array types with types of their elements
and monadic bindings with regular \T{let}-bindings
(which in the TTN corresponds to simple sequencing of transitions).
Formally, we define the \emph{downgrading} operation on semantic types $\downgrade{\sema{t}}$ as follows:
$$
    \downgrade{\sema{t}} =
    \begin{cases}
      \downgrade{\sema{t'}}\quad \text{if}\ \sema{t} = [\sema{t'}]\\
      \sema{t}\quad \text{otherwise}
    \end{cases}
$$

\begin{figure}
    \small
    \textbf{TTN Construction}\quad$\boxed{\construct{\sema{\Lambda}}{\net}{c}{\net'}}$ $\boxed{\construct{\sema{\Lambda}}{\net}{c:\sema{t}}{\net'}}$
    \begin{gather*}
        \inference[\textsc{C-Method}]
        {
            f: \{\many{\ell_i: \sema{t}_i}\} \to \sema{t}_o \in \sema{\Lambda} &
            \sema{t}_{in} = \{\many{\ell_i: \downgrade{\sema{t}_i}}\} \\
            O' = O[(p,f) \mapsto |\{l_i \mid ?l_i:p \in \sema{t}_{in} \}|] \\
            E' = E[(p,f) \mapsto |\{l_i \mid l_i:p \in \sema{t}_{in} \}|][(f, \downgrade{\sema{t}_o}) \mapsto 1]                
        }
        {
            \construct{\sema{\Lambda}}{(P,T,E,O)}{f}{(P \cup \{\many{\downgrade{\sema{t}_i}},\downgrade{\sema{t}_o}\}, T \cup \{f\}, E', O')}
        }
        \\
        \inference[\textsc{C-Object}]
        {
            o: \{\many{\ell_i: \sema{t}_i}\} \in \sema{\Lambda} &
            \construct{\sema{\Lambda}}{\net}{\proj_{o.l_i}:\downgrade{\sema{t}_i}}{\net'_i} \\
            \construct{\sema{\Lambda}}{\bigcup \net'_i}{\guard_{o.l_i}:\downgrade{\sema{t}_i}}{\net''_i}
        }
        {
            \construct{\sema{\Lambda}}{\net}{o}{\bigcup \net''_i}
        }
        \\
        \inference[\textsc{C-Proj}]
        {
            E' = E[(o, \proj_{o.l}) \mapsto 1, (\proj_{o.l}, \sema{t}) \mapsto 1] 
        }
        {
            \construct{\sema{\Lambda}}{(P,T,E,O)}{\proj_{o.l}:\sema{t}}{(P' \cup \{o, \sema{t}\}, T' \cup \{\proj_{o.l}\}, E',O)}
        }
        \\
        \inference[\textsc{C-Filter}]
        {
            \sema{t}\ \text{not obj. id}  \\
            E' = E[(o, \guard_{o\many{.l}}) \mapsto 1, (\sema{t}, \guard_{o\many{.l}}) \mapsto 1, (\guard_{o\many{.l}}, o) \mapsto 1]
        }
        {
            \construct{\Lambda}{(P,T,E,O)}{\guard_{o\many{.l}}:\sema{t}}{(P\cup\{o, \sema{t}\},T \cup\{\guard_{o\many{.l}}\},E',O)}
        }
        \\
        \inference[\textsc{C-Filter-Obj}]
        {
          o': \{\many{\ell_i: \sema{t}_i}\} \in \sema{\Lambda} &            
          \construct{\sema{\Lambda}}{\net}{\guard_{o\many{.l}.l_i}:\downgrade{\sema{t}_i}}{\net'_i}
        }
        {
            \construct{\Lambda}{\net}{\guard_{o\many{.l}}:o'}{\bigcup \net'_i}
        }
    \end{gather*}
    \caption{TTN construction from a semantic library.}
    \label{fig:ttn-construction}
\end{figure}

\mypara{Construction rules}
We formalize TTN construction using the step relation $\construct{\sema{\Lambda}}{\net}{c}{\net'}$ 
defined in \autoref{fig:ttn-construction}.
A construction step adds an encoding of a library component $c$ (method or object) to $\net$ and produces a new TTN $\net'$;
to encode the entire library $\sema{\Lambda}$,
we compose steps for all methods and objects in the library.
For example, \autoref{fig:ttn} depicts (a fragment of) the TTN built from the semantic library in \autoref{tab:library};
in this figure, all edges have multiplicity 1 
(since no method in $\sema{\Lambda}$ takes multiple arguments of the same type),
and there are no optional edges.
In general, there are three kinds of transitions in a TTN: 
method transitions, such as \T{conversations_members}, 
projection transitions, such as $\proj_{\texttt{User.id}}$, 
and filter transitions, such as $\guard_{\texttt{Channel.name}}$.
We now describe the construction rules for the three kinds of transitions in more detail.

The rule \textsc{C-Method} adds a transition for a method $f$
and connects it to the $f$'s (downgraded) input and output types.
For example, let $f = \T{conversations_members}$,
whose downgraded type is $\{\T{channel}: \T{Channel.id}\} \to \T{User.id}$.
In this case, the rule \textsc{C-Method} extends the TTN with a transition $f$
and places $\{\T{Channel.id}, \T{User.id}\}$;
it also sets $E[(\T{Channel.id},f)] = 1$, $E[(f,\T{User.id})] = 1$
(and all other incoming and outgoing edge multiplicities to $0$).
\autoref{fig:ttn-construction} uses the notation $E[k \mapsto v]$,
to denote a map that is equal to $E$ except for mapping $k$ to $v$.
Note that \textsc{C-Method} is the only rule that modifies optional multiplicities $O$,
since only methods can have optional arguments.

The rule \textsc{C-Object} adds projection and filter transitions
for every field $l$ of an object identifier $o$,
merging the resulting TTNs component-wise.
The rule \textsc{C-Proj} adds a projection transition $\proj_{o.l}$
between an on object identifier $o$ and the type of $l$.

Filter transitions are a little more involved,
as they do not mirror the structure of \corelang guards one-to-one.
Guards in their general form are higher-order operations,
and hence cannot be directly encoded in the TTN.
Fortunately, \tool only uses guards for one purpose:
filtering objects from an array $xs$ if their constituent is equal to some $y$
(in code: $\eseq{\ebind{x}{xs}}{\eseq{\eif{x.\many{l}}{y}}{\eret{x}}}$).
We encode this filtering operation in the TTN as a filter transition $\guard_{o.\many{l}}$,
which consumes $o$ (the downgraded type of $xs$) 
and the type of $y$, and produces $o$.
For example, the filter transition $\guard_{\texttt{Channel.name}}$ in \autoref{fig:ttn}
consumes a \T{Channel} and a \T{Channel.name} and produces a \T{Channel}.
When $o.\many{l}$ is an object, 
the rule \textsc{C-Filter-Obj} recursively creates filter transitions for all its fields
(recall that \corelang only supports guards on primitive values).
For example, for the object ID \T{User},
we will add a transition $\guard_{\texttt{User.profile.email}}$,
but not $\guard_{\texttt{User.profile}}$.

\subsection{TTN Search}\label{appendix:synthesis:search}

Once the TTN $\net$ has been constructed,
the algorithm \textsc{Synthesize} proceeds to enumerate paths 
from the initial marking $I$ to the final marking $F$ in $\net$.
$I$ and $F$ are constructed from the query type $\sema{t}_{in} \to \sema{t}_o$ as follows:
\begin{equation*}
  \begin{aligned}
  I(p) &= |\{l_i \mid l_i:p \in \downgrade{\sema{t}_{in}}\}| \\
  F(p) &= \mathsf{if}\ p =\downgrade{\sema{t}_o}\ \mathsf{then}\ 1\ \mathsf{else}\ 0
  \end{aligned}
\end{equation*}
For example, for the query $\T{Channel.name} \to [\T{Profile.email}]$,
the initial marking contains a single token in \T{Channel.name}
and the final marking contains a single token in \T{Profile.email}.
Hence, any valid path from $I$ to $F$ must consume all inputs and produce the output of the query type.
Note that the TTN as defined in \autoref{appendix:synthesis:construction} encodes a \emph{linear type system},
\ie can only generate programs where each input is used \emph{exactly once}.
Following prior work \cite{FengM0DR17,GuoJJZWJP20}, our implementation adds copy transitions to the TTN,
which results in a \emph{relevant type system}, 
\ie one where every input has to be used \emph{at least once}.
Prior work shows that this relevancy requirement is crucial to filtering out meaningless solutions during search.

\mypara{ILP encoding}
To find paths in the TTN, prior work has relied on a SAT/SMT encoding of TTN reachability.
We have found that although the SMT encoding from \cite{GuoJJZWJP20} works well to find a handful of paths,
in our domain we often need to enumerate thousands of paths,
which becomes very inefficient.
To address this problem, we instead use an \emph{integer linear programming} (ILP) solver,
which provides native functionality for computing all solutions to a constraint.

Given a TTN $\net = (P,T,E,O)$ with the initial marking $I$ and the final marking $F$, 
we show how to build an ILP formula that
encodes all valid paths of a given length $L$.
The overall search proceeds by iteratively increasing the path length $L$.
We encode the number of tokens in each place $p\in P$ at each time step $k\in [0,L]$
as a variable $\tok_{k}^{p}$.
We encode firing of transition $\tau \in T$ at time step $k \in [0,L-1]$
as a variable $\fire_{k}^{\tau} \in \{0,1\}$,
such that $\fire_{k}^{\tau} = 1$ indicates that $\tau$ is fired at time step $k$.
For any $\tau \in T$, we define the pre-image of $\tau$ as
$\pre(\tau) = \{p \in P \mid E(p, \tau) > 0 \vee O(p, \tau) > 0\}$ and the post-image of $\tau$ as
$\post(\tau) = \{p \in P \mid E(\tau, p) > 0\}$.
%

The formula for TTN reachability is a conjunction of the following constraints:
\begin{enumerate}[label=(\arabic*)]
    \item If a transition $\tau$ is fired at time step $k$, 
    then all required places $p \in \pre(\tau)$ have sufficiently many tokens:
    $\bigwedge_{k=0}^{L-1} \bigwedge_{\tau \in T} \bigwedge_{p \in \pre(\tau)} \tok_{k}^{p} \geq E(p, \tau) \times \fire_{k}^{\tau}$
    \item If a transition $\tau$ is fired at time step $k$, 
    then all places $p \in \pre(\tau) \cup \post(\tau)$ will have their marking updated at time step $k+1$:
    $\bigwedge_{k=0}^{L-1} \bigwedge_{\tau \in T} \bigwedge_{p \in \pre(\tau) \cup \post(\tau)} 
    \tok_{k}^{p} - (E(p, \tau) + O(p,\tau) - E(\tau, p)) \times \fire_{k}^{\tau} \leq \tok_{k+1}^{p} \leq \tok_{k}^{p} - (E(p, \tau) - E(\tau, p)) \times \fire_{k}^{\tau}$
    \item At each time step, only one transition is fired:\\ 
    $\bigwedge_{k=0}^{L-1} \sum_{\tau \in T} \fire_{k}^{\tau} = 1$
    \item Variable domains are respected: 
    $\bigwedge_{k=0}^{L} \bigwedge_{p \in P} 0\leq \tok_{k}^{p}$ and $\bigwedge_{k=0}^{L-1} \bigwedge_{\tau \in T} 0 \leq \fire_{k}^{\tau} \leq 1$
    \item The initial marking $I$ is valid: 
    $\bigwedge_{p \in P} \tok_{0}^{p} = I(p)$
    \item The final marking is valid: 
    $\bigwedge_{p \in P} \tok_{L}^{p} = F(p)$
\end{enumerate}

Note that constraint (2) \emph{approximates}
consumption of optional arguments as a range between
consuming all of them ($E(p, \tau) + O(p, \tau)$) and 
consuming none of them ($E(p, \tau)$).
This encoding is unsound when an optional argument has the same type as the output.
Consider firing a transition $\tau$ at step $k$ such that $\tok_{k}^{p} = 0, E(p, \tau) = 0, O(p, \tau) = 1, E(\tau, p) = 1$;
our encoding $0 \leq \tok_{k+1}^{p} \leq 1$,
whereas the TTN definition in \autoref{appendix:synthesis:construction} requires $\tok_{k+1}^{p}=1$ 
(the optional argument could not be consumed since there was no token to consume).
We use the approximate encoding because in our experience 
it is significantly more efficient than the exact alternatives,
which require additional variables and/or constraints.
In our evaluation, the unsoundness arises very rarely,
and when it does, the path is simply rejected by the type checker when converted into a program.

\subsection{Program Lifting}

\mypara{From paths to programs}
The function $\textsc{Progs}(\pi)$ (line 5 in \autoref{alg:synthesis}) converts a TTN path $\pi$
into a set of array-oblivious programs in A-Normal Form (ANF).
An ANF program is sequence of \emph{statements} $\sigma$ followed by a variable;
the syntax of ANF terms is given in \autoref{fig:lifting}.
$\textsc{Progs}$ converts each transition in $\pi$ into a sequence of statements with a dedicated output variable $x_o$:
a method transition becomes $\elet{x_o}{f(\many{l_i=x_i})}$,
a projection transition becomes $\elet{x_o}{x.l}$,
and a filter transition becomes $\elet{x_1}{x_o.l_1};\ldots\elet{x_n}{x_{n-1}.l_n};\eif{x_n}{y}$.
All emitted statements are then concatenated into an ANF term
$\sigma_1;\ldots;\sigma_n;x_n$, where $x_n$ is the output variable of the last transition.
\autoref{fig:lift-example} (left) shows the full array-oblivious program extracted from the bold path in \autoref{fig:ttn}.
The reason we use ANF as the intermediate representation instead of generating \corelang terms directly
is that ANF has a more direct correspondence with TTN paths
and also is more convenient to work with during lifting;
and ANF term can be translated into a regular \corelang term
by recursing through the sequence of statements and replacing sequential composition 
with \corelang \T{let} bindings, monadic bindings, or guard expressions.

\begin{figure}
  \centering
  \small
  \textbf{ANF Syntax}
    $$
    \begin{array}{lll}
    \sigma ::= & \elet{x}{f(\many{l_i=x_i})} \mid \elet{x}{x.l} & \text{Statements}\\
               & \mid \eif{x}{x} \mid \ebind{x}{x} \mid \elet{x}{\eret{x}} & \\
    a ::= & \eseq{\many{\sigma}}{x} &\text{ANF terms}
    \end{array}
    $$

  \textbf{Statement Lifting}\quad$\boxed{\jlift{\Gamma}{\sigma}{\many{\sigma'}}{\Gamma'}}$
  \begin{gather*}
      \inference[\textsc{L-Call}]
      {
        f:\{\many{l_i: \sema{t}_i}\} \to \sema{t}_o \in \sema{\Lambda} &        
        \jliftexp{\Gamma_{i - 1}}{x_i}{\sema{t}_i}{\many{\sigma}_i;x_i'}{\Gamma_i} \\
        \Gamma_0 = \Gamma &
        \many{\sigma'} \;=\; (\many{\sigma}_1;\ldots;\many{\sigma}_n;\elet{x}{f(\many{l_i = x_i'})})
      }
      {
        \jlift{\Gamma}{\elet{x}{f(\many{l_i = x_i})}}{\many{\sigma'}}{\Gamma_n,x:\sema{t}_o}
      }
      \\
      \inference[\textsc{L-Proj}]
      {
        \jtypinglib{\Gamma}{y}{\sema{t}} &
        \jliftexp{\Gamma}{y}{\downgrade{\sema{t}}}{\many{\sigma};y'}{\Gamma'} &
        \jtypinglib{\Gamma'}{y'.l}{\sema{t}'}
      }
      {        
        \jlift{\Gamma}{\elet{x}{y.l}}{\many{\sigma};\elet{x}{y'.l}}{\Gamma',x:\sema{t}'}         
      }
      \\
      \inference[\textsc{L-Guard}]
      {
        \jtypinglib{\Gamma}{x}{\sema{t}_x} & \jliftexp{\Gamma}{x}{\downgrade{\sema{t}_x}}{\many{\sigma_x};x'}{\Gamma'}  \\
        \jtypinglib{\Gamma}{y}{\sema{t}_y} & \jliftexp{\Gamma'}{y}{\downgrade{\sema{t}_y}}{\many{\sigma_y};y'}{\Gamma''}
      }
      {
        \jlift{\Gamma}{\eif{x}{y}}{\many{\sigma_x};\many{\sigma_y};\eif{x'}{y'}}{\Gamma''}
      }
  \end{gather*}  
  \textbf{ANF Term Lifting}\quad$\boxed{\jliftexp{\Gamma}{a}{\sema{t}}{a'}{\Gamma'}}$
  \begin{gather*}
      \inference[\textsc{L-Seq}]
      {
        \jlift{\Gamma}{\sigma_0}{\many{\sigma_0'}}{\Gamma'} &
        \jliftexp{\Gamma'}{\many{\sigma};x}{\sema{t}}{\many{\sigma'};x'}{\Gamma''}
      }
      {
        \jliftexp{\Gamma}{\sigma_0;\many{\sigma};x}{\sema{t}}{\many{\sigma_0'};\many{\sigma'};x'}{\Gamma''}
      }
      \\
      \inference[\textsc{L-Var}]
      {
        \jtypinglib{\Gamma}{x}{\sema{t}}
      }
      {
        \jliftexp{\Gamma}{x}{\sema{t}}{x}{\Gamma}
      }
      \\
      \inference[\textsc{L-Var-Down}]
      {
        \jtypinglib{\Gamma}{x}{[\sema{t}']} &
        \sema{t} \neq [\sema{t}'] &
        \_ :^x \sema{t}' \notin \Gamma &
        \mathsf{fresh}(x') \\
        \jliftexp{\Gamma,x' :^x \sema{t}'}{x'}{\sema{t}}{\many{\sigma};y}{\Gamma'}
      }
      {
        \jliftexp{\Gamma}{x}{\sema{t}}{\ebind{x'}{x};\many{\sigma};y}{\Gamma'}
      }
      \\
      \inference[\textsc{L-Var-Repeat}]
      {
        \jtypinglib{\Gamma}{x}{[\sema{t}']} &
        \sema{t} \neq [\sema{t}'] &
        x' :^x \sema{t}' \in \Gamma \\
        \jliftexp{\Gamma}{x'}{\sema{t}}{\many{\sigma};y}{\Gamma'}
      }
      {
        \jliftexp{\Gamma}{x}{\sema{t}}{\many{\sigma};y}{\Gamma'}
      }
      \\
      \inference[\textsc{L-Var-Up}]
      {
        \jtypinglib{\Gamma}{x}{\sema{t}'} &
        \sema{t}' \neq [\sema{t}] &
        \mathsf{fresh}(x') \\
        \jliftexp{\Gamma,x' : [\sema{t}']}{x'}{\sema{t}}{\many{\sigma};y}{\Gamma'}
      }
      {
        \jliftexp{\Gamma}{x}{[\sema{t}]}{\elet{x'}{\eret{x}};\many{\sigma};y}{\Gamma'}
      }
  \end{gather*}  
  \caption{Lifting rules. 
  }
  \label{fig:lifting}
\end{figure}

\mypara{Lifting rules}
We formalize lifting of ANF terms as a \emph{term lifting} judgement \jliftexp{\Gamma}{a}{\sema{t}}{a'}{\Gamma'},
defined in \autoref{fig:lifting}.
Here $a$ is the array-oblivious term to be lifted, whose free variables are defined in $\Gamma$,
$\sema{t}$ is the target type,
$a'$ is the lifted term, 
and $\Gamma'$ is $\Gamma$ extended with all the variables bound in $a'$.
The definition of term lifting relies on the auxiliary \emph{statement lifting} judgment \jlift{\Gamma}{\sigma}{\many{\sigma'}}{\Gamma'},
which lifts a single statement $\sigma$ that appears in the environment $\Gamma$
into a sequence of statements $\many{\sigma'}$; 
again $\Gamma'$ is $\Gamma$ extended with variables bound in $\many{\sigma'}$.
Both judgments implicitly rely on the semantic library $\sema{\Lambda}$,
which we omit for brevity since it is not modified.
For example, the statement in line 4 of \autoref{fig:lift-example} is lifted as follows:
\begin{multline*}
x_1:[\T{Channel}] \vdash \elet{x_2}{x_1.\T{name}} \leadsto \\
\ebind{x_1'}{x_1};\elet{x_2}{x_1'.\T{name}} \dashv \\ 
x_1:[\T{Channel}],x_1':^{x_1}\T{Channel},x_2:\T{Channel.name}
\end{multline*}
Note that the binding for $x_1'$ in $\Gamma'$ is annotated with $x_1$;
we introduce these annotations to keep track of \emph{mapping variables}: 
here $x_1'$ is the mapping variable for the array $x_1$,
and it will be reused later in the program whenever an element of $x_1$ is required again (see line 6).

Let us now describe the rules in \autoref{fig:lifting} in more detail.
The statement lifting rules enforce well-typing for the three kinds of statements that appear in array-oblivious programs:
\textsc{L-Call} enforces that the arguments of a method call agree with it definition in $\sema{\Lambda}$,
while \textsc{L-Proj} and \textsc{L-Guard} make sure their operands are scalars.
The heavy lifting is done by the rules \textsc{L-Var-Down} and \textsc{L-Var-Up},
both of which handle the case of lifting a variable whose type $\sema{t}'$ in $\Gamma$ does not match the target type $\sema{t}$.
\textsc{L-Var-Down} applies when $\sema{t}' = [..[\sema{t}]..]$:
in this case we generate a monadic binding $\ebind{x'}{x}$
(and then lift $x'$ again in case the array type was nested);
this rule is responsible for the monadic bindings in lines 3 and 8 of \autoref{fig:lift-example}.
Note that in the last premise we add the binding $x' :^x \sema{t}'$ to $\Gamma$ to record that $x'$ is the mapping variable for $x$.
If $\Gamma$ already has a mapping variable for $x$, we want to reuse that variable instead of generating a new binding;
this is accomplished by the rule \textsc{L-Var-Repeat}.
Finally, \textsc{L-Var-Up} applies when the opposite is true, \ie $\sema{t} = [..[\sema{t}']..]$;
in this case we simply need to wrap $x$ in a \T{return}.
This rule generates line 12 of \autoref{fig:lift-example}:
here the target return type of the program is $\sema{t} = [\T{Profile.email}]$
and the type of $x_7$ is $\sema{t}' = \T{Profile.email}$.

\section{Retrospective Execution}\label{appendix:re}

The full definition of the RE judgement is presented in \autoref{fig:re-full}.

\section{Witness Collection}\label{sec:appendix:witness-collection}
For each API, we manually created a test environment via the corresponding web interface
and filled in some arbitrary test data.
We then ran several operations in the test environment, 
using Google Chrome to record the web traffic in an HTTP Archive (HAR) file:
a JSON-formatted file that logs browser interactions with a server.
%
%
We then extracted initial witnesses $\witnesses_0$ from the HAR file and
ran the algorithm \textsc{MineTypes} (\autoref{fig:comp-inference}) on these witnesses,
resulting in an initial semantic library $\sema{\Lambda}_0$. 

Unfortunately, the set $\witnesses_0$ obtained this way is sparse,
which may prevent the mining algorithm from merging equivalent locations
or even inferring types for some methods altogether.
For example, if we only have initial witnesses as shown in \autoref{fig:witnesses},
\tool is unable to infer the semantic type of \T{users_lookupByEmail}
that is required by one of our benchmarks,
and hence fail to solve this task.
At the same time, generating useful tests for this method is challenging,
because it only succeeds on inputs that correspond to existing users' emails.
To improve the quality and coverage of inferred types,
\tool generates additional witnesses,
using a combination of type-directed random testing and a small amount of manual annotations.
Specifically,
\tool draws test inputs from the bank of values it has observed in the existing witnesses;
in our example, one of the strings that appear in \autoref{fig:witnesses} is \T{"xyz@gmail.com"}.
Calling \T{users_lookupByEmail} with this string succeeds and returns \T{"UJ5RHEG4S"}.
Based on the previously mined types of these two strings,
\tool infers the type $\T{Profile.email} \to \T{User.id}$ for the new method.

\begin{figure}
  \small
  \textbf{Retrospective Execution}\quad$\boxed{\stepenv{\Sigma}{e}{v}}$
  \begin{gather*}
  \inference[\textsc{E-Var}]
  {
    \Sigma(x) = v
  }
  {
    \stepenv{\Sigma}{x}{v}
  }
  \\
  \inference[\textsc{E-Var-Lazy}]
  {
    x \not\in \Sigma & 
    \jtypinglib{\Gamma}{x}{\sema{t}} &  
    v \in \witnesses(\sema{t})
  }
  {
    \stepenv{\Sigma}{x}{v}
  }
  \\
  \inference[\textsc{E-Bind-Pure}]
  {
    \stepenv{\Sigma}{e_1}{v} \\
    \stepenv{x \mapsto v, \Sigma}{e_2}{v'}}
  {
    \stepenv{\Sigma}{\eseq{\elet{x}{e_1}}{e_2}}{v'}
  }
  \\
  \inference[\textsc{E-Bind-Monad}]
  {
    \stepenv{\Sigma}{e_1}{[\many{v_i}]} \\
    \stepenv{x \mapsto v_i, \Sigma}{e_2}{v'_i} \\
    v' = \biguplus v'_i
  }
  {
    \stepenv{\Sigma}{\eseq{\ebind{x}{e_1}}{e_2}}{v'}
  }
  \\
  \inference[\textsc{E-Return}]
  {
    \stepenv{\Sigma}{x}{v}
  }
  {
    \stepenv{\Sigma}{\eret{x}}{[v]}
  }
  \\
  \inference[\textsc{E-Projection}]
  {
    \stepenv{\Sigma}{x}{v} &
    \mathsf{hasField}(v, l)}
  {
    \stepenv{\Sigma}{x.l}{v.l}
  }
  \\
    \inference[\textsc{E-If-True-L}]
    {
      x_1 \in \Sigma &
      x_2 \not\in \Sigma &
      \Sigma(x_1) = v_1 \\
      \stepenv{x_2 \mapsto v_1, \Sigma}{e}{v}
    }
    {
      \stepenv{\Sigma}{\eseq{\eif{x_1}{x_2}}{e}}{v}
    }
    \\
    \inference[\textsc{E-If-True-R}]
    {
      x_1 \not\in \Sigma &
      \stepenv{\Sigma}{x_2}{v_2} \\
      \stepenv{x_1 \mapsto v_2, x_2 \mapsto v_2, \Sigma}{e}{v}
    }
    {
      \stepenv{\Sigma}{\eseq{\eif{x_1}{x_2}}{e}}{v}
    }
    \\
    \inference[\textsc{E-If-True-LR}]
    {
      x_1 \in \Sigma &
      x_2 \in \Sigma &
      \Sigma(x_1) = \Sigma(x_2) \\
      \stepenv{\Sigma}{e}{v}
    }
    {
      \stepenv{\Sigma}{\eseq{\eif{x_1}{x_2}}{e}}{v}
    }
    \\
    \inference[\textsc{E-If-False}]
    {
      x_1 \in \Sigma &
      x_2 \in \Sigma \\
      \Sigma(x_1) = v_1 &
      \Sigma(x_2) = v_2 &
      v_1 \neq v_2
    }
    {
      \stepenv{\Sigma}{\eseq{\eif{x_1}{x_2}}{e}}{[]}
    }
  \\
  \inference[\textsc{E-Method}]
  {
    \stepenv{\Sigma}{x_i}{v_i} \\
    \stepenv{\Sigma}{f(\many{l_i=v_i})}{v_{out}}
  }
  {
    \stepenv{\Sigma}{f(\many{l_i=x_i})}{v_{out}}
  }
  \\
  \inference[\textsc{E-Method-val}]
  {
    (f, \many{l_i=v_i}, v_{out}) \in \witnesses
  }
  {
    \stepenv{\Sigma}{f(\many{l_i=v_i})}{v_{out}}
  }
  \\
  \inference[\textsc{E-Method-name}]
  {
    \forall (f, \many{l_i=v'_i}, v_{out}) \in \witnesses.~\exists i : v'_i \neq v_i \\
    (f, \many{l_i=v'_i}, v_{out}) \in \witnesses
  }
  {
    \stepenv{\Sigma}{f(\many{l_i=v_i})}{v_{out}}
  }
  \end{gather*}
\caption{Retrospective execution.}
\label{fig:re-full}
\end{figure}

The top-level API analysis algorithm of \tool is depicted in \autoref{alg:analyze} (top).
The algorithm alternates between calls to \textsc{MineTypes}
(to compute the best semantic library $\sema{\Lambda}$ it can mine from the current witnesses $\witnesses$)
and \text{GenerateTests}
(to augment $\witnesses$ using the current $\sema{\Lambda}$).
Analysis terminates either if $\sema{\Lambda}$ and $\witnesses$ reach a fixpoint,
or when a timeout is reached.
The algorithm returns both $\sema{\Lambda}$, which is used to build the TTN during the synthesis step,
and $\witnesses$, which is used for retrospective execution during the ranking step;
hence augmenting $\witnesses$ also improves the quality of ranking.

Before we introduce the algorithm \textsc{GenerateTests},
we augment the semantic library $\sema{\Lambda}$
with a \emph{value bank} $\bank$,
which is a mapping from semantic types to sets of values.
The value bank gets populated during type mining
and contains all values that appear in $\witnesses$;
for arrays and objects, it also contains their constituents. 
With the value bank at hand,
let us explain the algorithm \textsc{GenerateTests} presented in \autoref{alg:analyze} (bottom).
\textsc{GenerateTests} iterates through the method signatures in $\sema{\Lambda}$,
and for every method $f\colon \sema{t}_i \to \sema{t}_o$,
it samples a random input of type $\sema{t}_i$,
makes an API call to $f$ with that input,
and if the call succeeds, yields the corresponding witness.
It uses the value bank (denoted $\sema{\Lambda}.\bank$) at line 7
to randomly sample an input from all values stored at semantic type $\sema{t}$.
We also observe that many API methods have optional arguments
and behave differently depending on which subset of optional arguments is provided.
To cover a wide range of method behaviors,
\textsc{GenerateTests} partitions the record of all arguments $\sema{t}_i$
into two records, each containing either only required or only optional arguments (line 3).
The algorithm then iterates over all subsets of optional arguments (line 4),
attempting to make a call for each subset
(our implementation only iterates over subsets up to a pre-defined size).

Fully automatic test generation helps us bootstrap semantic type inference
for methods that already appear in $\witnesses_0$,
but it cannot add witnesses for methods that are missing entirely.
To address this problem,
we manually add consumer-producer annotations 
to those methods missing from $\witnesses_0$.

In the evaluation, we collect witnesses by running the algorithm \textsc{AnalyzeAPI} (\autoref{alg:analyze}) until it converges,
alternating test generation and type mining,
which resulted in the final set of witnesses $\witnesses$ and library $\sema{\Lambda}$.
%
%
The total running time for API analysis depends on the number of 
methods and their arguments, 
and ranged from several minutes to several hours in our experiments.

\begin{figure*}
  \centering
  \small
  \begin{minipage}{.45\textwidth}
    \begin{algorithmic}[1]
        \Require{Library $\Lambda$, initial witnesses $\witnesses_0$}
        \Ensure{Semantic library $\sema{\Lambda}$}
        \Ensure{Augmented witnesses $\witnesses$}
        \Function{\textsc{AnalyzeAPI}}{$\Lambda, \witnesses_0$}
            \Let{$\witnesses$}{$\witnesses_0$}
            \Repeat
              \Let{$\sema{\Lambda}$}{\Call{MineTypes}{$\Lambda$, $\witnesses$}}
              \Let{$\witnesses$}{$\witnesses \cup \textsc{GenerateTests}(\sema{\Lambda})$}
            \Until{fixpoint or timeout}
            \State \Return{$\sema{\Lambda}, \witnesses$}
            \Statex
        \EndFunction
    \end{algorithmic}
  \end{minipage}
  \centering
  \small
  \begin{minipage}{.45\textwidth}
    \begin{algorithmic}[1]
        \Require{Semantic library $\sema{\Lambda}$}
        \Ensure{Generated witnesses $\witnesses$}
        \Function{\textsc{GenerateTests}}{$\sema{\Lambda}$}
          \For{$f\colon \{\many{l_i:\sema{t}_i}, \many{?l_j:\sema{t}_j}\} \to \sema{t}_o \in \sema{\Lambda}$}
            \Let{$\sema{t}_{req}, \sema{t}_{opt}$}{$\{\many{l_i:\sema{t}_i}\}, \{\many{?l_j:\sema{t}_j}\}$}
            \For{$\sema{t}_{sub} \subset \sema{t}_{opt}$}
              \Let{$v_{in}$}{$\{ \}$}
              \For{$l:\sema{t} \in \sema{t}_{req}\cup\sema{t}_{sub}$}
                \Let{$v_{in}.l$}{$\mathsf{random}(\sema{\Lambda}.\bank[\sema{t}])$}              
              \EndFor
              \Let{$v_{out}$}{$\mathsf{call}(f, v_{in})$}
              \State{\textbf{if} $v_{out} \neq \bot$ \textbf{then} \textbf{yield} $\langle f, v_{in}, v_{out} \rangle$}
            \EndFor
          \EndFor
        \EndFunction
    \end{algorithmic}
  \end{minipage}
  \caption{Top-level API analysis algorithm and test generation.}
  \label{alg:analyze}
\end{figure*}

\section{Benchmarks and Solutions}\label{appendix:solutions}

\autoref{tab:bench-extended} contains benchmark descriptions and detailed results.
\autoref{tab:qualitative} contains the results of the qualitative analysis of mined semantic types.

The rest of this section includes type queries and ``gold standard'' solutions for all benchmarks. 
Note that the type queries used here correspond directly to the OpenAPI spec; 
earlier in the paper, type and method names were simplified for readability.

\begin{table*}
  \caption{
    Synthesis benchmarks and results.
    Benchmarks marked with $\dagger$ are effectful.
    For each benchmark we report the size of the desired solution:
    AST, $n_f$, $n_p$ and $n_g$ correspond to
    number of AST nodes, method calls,
    projections and guards, respectively.
    We also report the time spent on RE-based ranking $t_{\mathit{RE}}$ and 
    the total synthesis time $t_{\mathit{Total}}$ in seconds.
    In the last four columns,
    we report the rank before RE \rorig,
    the rank after RE among candidates that we find before the desired solution \rre,
    the total number of candidates we get within timeout,
    and the rank after RE among all candidates within timeout \rreto.
    `-' means no solution is found in \synTO.
  }\label{tab:bench-extended}
  \small
\begin{tabular}{l|lp{6.5cm}|rrrr|rr|rrrr}
        \toprule
        \multirow{2}{*}{API} & \multicolumn{2}{c|}{Benchmark} & \multicolumn{4}{c|}{Solution Size} & \multicolumn{2}{c|}{Timing} & \multicolumn{4}{c}{Rank} \\
        \cmidrule(lr){2-3} \cmidrule(lr){4-7} \cmidrule(lr){8-9} \cmidrule(lr){10-13} 
         & ID & Description & AST & $n_{f}$ & $n_{p}$ & $n_{g}$ & $t_{\mathit{Total}}$ & $t_{\mathit{RE}}$ & \rorig & \rre & $\#$ cands & \rreto \\
        \midrule
        \multirow{8}{*}{\rotatebox{90}{\slack}} & 1.1 & Retrieve emails of all members in a channel & 17 & 3 & 6 & 1 & 83.5 & 0.5 & 25230& 5& 38212 & 5  \\
        & 1.2$^{\dagger}$ & Send a message to a user given their email & 12 & 3 & 5 & 0 & 5.6 & 0.1 & 2224& 10& 30437 & 10  \\
        & 1.3 & Get the unread messages of a user & 16 & 3 & 7 & 0 & - & - & -& -& - & -  \\
        & 1.4 & Get all messages associated with a user & 14 & 2 & 4 & 1 & 1.3 & 0.2 & 489& 24& 28012 & 31  \\
        & 1.5$^{\dagger}$ & Create a channel and invite a list of users & 10 & 2 & 3 & 0 & 3.4 & 0.1 & 788& 5& 22426 & 5  \\
        & 1.6$^{\dagger}$ & Reply to a message and update it & 9 & 2 & 2 & 0 & 1.7 & 0.1 & 573& 8& 39276 & 19  \\
        & 1.7$^{\dagger}$ & Send a message to a channel with the given name & 12 & 2 & 4 & 1 & 1.3 & $<$0.1 & 757& 8& 39078 & 9  \\
        & 1.8 & Get the unread messages of a channel & 9 & 2 & 3 & 0 & 42.0 & 0.8 & 16438& 29& 50757 & 30  \\
        \hline
        \multirow{13}{*}{\rotatebox{90}{\stripe}} & 2.1$^{\dagger}$ & Subscribe to a product for a customer & 9 & 2 & 2 & 0 & 95.4 & 0.6 & 4952& 3& 6312 & 3  \\
        & 2.2$^{\dagger}$ & Subscribe to multiple items & 10 & 2 & 2 & 0 & 92.4 & 0.6 & 4854& 4& 6167 & 4  \\
        & 2.3$^{\dagger}$ & Create a product and invoice a customer & 12 & 3 & 2 & 0 & 121.2 & 2.6 & 6363& 1& 6644 & 1  \\
        & 2.4 & Retrieve a customer by email & 8 & 1 & 2 & 1 & 0.5 & $<$0.1 & 3& 1& 1751 & 1  \\
        & 2.5 & Get a list of receipts for a customer & 8 & 2 & 2 & 0 & 1.0 & $<$0.1 & 10& 4& 4548 & 4  \\
        & 2.6$^{\dagger}$ & Get a refund for a subscription & 9 & 3 & 2 & 0 & 12.2 & 0.1 & 270& 3& 4584 & 3  \\
        & 2.7 & Get the emails of all customers & 5 & 1 & 2 & 0 & 0.6 & $<$0.1 & 4& 2& 1382 & 2  \\
        & 2.8 & Get the emails of the subscribers of a product & 16 & 2 & 7 & 1 & 20.2 & 0.2 & 679& 17& 3407 & 17  \\
        & 2.9 & Get the last 4 digits of a customer's card & 6 & 1 & 2 & 0 & 0.5 & $<$0.1 & 2& 1& 1812 & 1  \\
        & 2.10$^{\dagger}$ & Update payment methods for a user's subscriptions & 10 & 2 & 3 & 0 & 7.8 & 0.1 & 187& 6& 3068 & 6  \\
        & 2.11$^{\dagger}$ & Delete the default payment source for a customer & 7 & 2 & 1 & 0 & 17.2 & 0.1 & 490& 6& 1373 & 6  \\
        & 2.12$^{\dagger}$ & Save a card during payment & 11 & 3 & 2 & 0 & - & - & -& -& - & -  \\
        & 2.13$^{\dagger}$ & Send an invoice to a customer & 10 & 3 & 2 & 0 & - & - & -& -& - & -  \\
        \hline
        \multirow{11}{*}{\rotatebox{90}{\squareapi}} & 3.1 & List invoices that match a location id & 4 & 1 & 1 & 0 & 0.2 & $<$0.1 & 2& 1& 9544 & 1  \\
        & 3.2 & List subscriptions by location, customer, and plan & 16 & 1 & 4 & 3 & 0.5 & $<$0.1 & 10& 4& 2526 & 4  \\
        & 3.3 & Get all items a tax applies to & 10 & 1 & 3 & 1 & 0.4 & $<$0.1 & 6& 1& 11039 & 1  \\
        & 3.4 & Get a list of discounts in the catalog & 5 & 1 & 2 & 0 & 0.7 & $<$0.1 & 2& 1& 11704 & 1  \\
        & 3.5$^{\dagger}$ & Add order details to order & 14 & 2 & 3 & 0 & 2.2 & $<$0.1 & 99& 2& 7222 & 2  \\
        & 3.6 & Get payment notes of a payment & 5 & 1 & 2 & 0 & 0.2 & $<$0.1 & 1& 1& 9590 & 1  \\
        & 3.7 & Get order ids of current user's transactions & 6 & 1 & 2 & 0 & 0.3 & $<$0.1 & 7& 4& 8669 & 4  \\
        & 3.8 & Get order names from a transaction id & 9 & 1 & 3 & 0 & 0.7 & $<$0.1 & 1& 1& 12323 & 1  \\
        & 3.9 & Find customers by name & 8 & 1 & 2 & 1 & 0.2 & $<$0.1 & 3& 2& 3177 & 2  \\
        & 3.10$^{\dagger}$ & Delete catalog items with names & 16 & 2 & 5 & 1 & 1.9 & $<$0.1 & 174& 10& 11336 & 12  \\
        & 3.11$^{\dagger}$ & Delete all catalog items & 8 & 2 & 3 & 0 & 1.0 & $<$0.1 & 68& 16& 7429 & 16  \\
        \bottomrule\end{tabular}
\end{table*}

\begin{table*}
  \caption{A sample of API methods, with their expected and inferred semantic types.
  Non-string parameters and responses are omitted since we do not perform inference for those.}\label{tab:qualitative}
  \begin{center}
\resizebox{\textwidth}{!}{
\begin{tabular}{c | c | c | c | c | c | c } 
 \toprule
API & Method & Field Type & Required & Name & Expected Type & Inferred Type \\ 
 \midrule

\multirow{11}{*}{\rotatebox{90}{\slack}} & \multirow{4}{*}{\T{/stars.add\_POST}} & \multirow{4}{*}{Parameter} & No & \T{channel} & \T{defs\_channel}   & \begin{tabular}{@{}c@{}} \T{defs\_group\_id, $\textbf{defs\_dm\_id}$} \end{tabular} \\\cline{4-7}
                            &   &  & No & \T{file} & \T{defs\_file\_id}  & \begin{tabular}{@{}c@{}} \T{/stars.add.in.file }\end{tabular} \\\cline{4-7}
                            &   &   & No & \T{file\_comment} &  \T{defs\_comment\_id} & \begin{tabular}{@{}c@{}} \T{/stars.add.in.file\_comment }\end{tabular} \\\cline{4-7}
                            &   &   & No & \T{timestamp} &  \T{defs\_ts} & \begin{tabular}{@{}c@{}} \T{/stars.add.in.timestamp }\end{tabular} \\\cline{2-7}
                            & \multirow{1}{*}{\T{/conversations.list\_GET}} & \multirow{1}{*}{Parameter} & No & \T{types} &  \T{objs\_conversation.types} & \begin{tabular}{@{}c@{}} \T{/conversations.list.in.types }\end{tabular} \\\cline{2-7}
                            & \multirow{2}{*}{\T{/users.profile.get\_GET}} & \multirow{1}{*}{Parameter}  & No & \T{user} &  \T{defs\_user\_id} & \begin{tabular}{@{}c@{}} \T{objs\_file.user, defs\_bot\_id, defs\_topic\_purpose\_creator,} \\ \T{$\textbf{defs\_user\_id}$ }\end{tabular} \\\cline{3-7}
                            &   & \multirow{1}{*}{Response} & - & \T{profile} & \T{objs\_user\_profile}  & \begin{tabular}{@{}c@{}} \T{$\textbf{objs\_user\_profile}$ }\end{tabular} \\\cline{2-7}
                            & \multirow{1}{*}{\T{/reminders.list\_GET}} & \multirow{1}{*}{Response} & - & \T{reminders} &  \T{objs\_reminder} & \begin{tabular}{@{}c@{}} \T{$\textbf{objs\_reminder}$ }\end{tabular} \\\cline{2-7}
                            & \multirow{2}{*}{\T{/users.conversations\_GET}} & \multirow{2}{*}{Parameter} & No & \T{user} &  \T{defs\_user\_id} & \begin{tabular}{@{}c@{}} \T{objs\_file.user, defs\_bot\_id, defs\_topic\_purpose\_creator,}\\ \T{$\textbf{defs\_user\_id}$ }\end{tabular} \\\cline{4-7}
                            &   &   & No & \T{types} &  \T{objs\_conversation.types} & \begin{tabular}{@{}c@{}} \T{/users.conversations.in.types }\end{tabular} \\
\hline

\multirow{44}{*}{\rotatebox{90}{\stripe}} & \multirow{10}{*}{\T{/v1/invoiceitems/\{invoiceitem\}\_POST}} & \multirow{11}{*}{Parameter} & Yes & \T{invoiceitem} &  \T{invoiceitem.id} & \begin{tabular}{@{}c@{}} \T{$\textbf{invoiceitem.id,}$ line\_item.invoice\_item }\end{tabular} \\\cline{4-7}
                             &   &   & No & \T{description} &  \T{invoiceitem.description} & \begin{tabular}{@{}c@{}} \T{product.name, $\textbf{invoiceitem.description,}$}\\ \T{credit\_note\_line\_item.description,}\\ \T{line\_item.description }\end{tabular} \\\cline{4-7}
                         &   &   & No & \T{discounts[0][coupon]} & \T{discounts.coupon}  & \begin{tabular}{@{}c@{}} \T{/v1/invoiceitems/\{invoiceitem\}.in.discounts.0.coupon }\end{tabular} \\\cline{4-7}
                         &   &   & No & \T{discounts[0][discount]} & \T{discounts.discount}  & \begin{tabular}{@{}c@{}} \T{/v1/invoiceitems/\{invoiceitem\}.in.discounts.0.discount }\end{tabular} \\\cline{4-7}
                         &   &   & No & \T{price} &  \T{price.id} & \begin{tabular}{@{}c@{}} \T{ $\textbf{price.id}$, plan.id }\end{tabular} \\\cline{4-7}
                         &   &   & No & \T{price\_data[currency]} & \T{invoice.lines.data.price.currency} & \begin{tabular}{@{}c@{}} \T{/v1/invoiceitems/\{invoiceitem\}.in.price\_data.currency }\end{tabular} \\\cline{4-7}
                         &   &   & No & \T{price\_data[product]} &   \T{invoice.lines.data.price.product} & \begin{tabular}{@{}c@{}} \T{/v1/invoiceitems/\{invoiceitem\}.in.price\_data.product }\end{tabular} \\\cline{4-7}
                         &   &   & No & \T{tax\_rates[0]} &  \T{tax\_rate.id} & \begin{tabular}{@{}c@{}} \T{/v1/invoiceitems/\{invoiceitem\}.in.tax\_rates.0 }\end{tabular} \\\cline{2-7}
                         & \multirow{1}{*}{\T{/v1/webhook\_endpoints\_GET}} &  \multirow{1}{*}{Response} & - & \T{object} &  \T{N/A} (Method returns constant string). & \begin{tabular}{@{}c@{}} \T{radar.value\_list.list\_items.object,}\\\T{credit\_note.lines.object, customer.sources.object,}\\ \T{subscription.items.object, payment\_intent.charges.object,}\\ \T{charge.refunds.object, file.links.object,}\\ \T{customer.subscriptions.object, invoice.lines.object }\end{tabular} \\\cline{2-7}
                         & \multirow{8}{*}{\T{/v1/transfers\_GET}} & \multirow{2}{*}{Parameter}  & No & \T{destination} &  \T{account.id} & \begin{tabular}{@{}c@{}} \T{/v1/transfers.in.destination }\end{tabular} \\\cline{4-7}
                         &   &   & No & \T{transfer\_group} & \T{transfer.transfer\_group}  & \begin{tabular}{@{}c@{}} \T{/v1/transfers.in.transfer\_group }\end{tabular} \\\cline{3-7}
                         &   & \multirow{1}{*}{Response} & - &  \T{object} &  \T{N/A} (Method returns constant string). & \begin{tabular}{@{}c@{}} \T{radar.value\_list.list\_items.object,}\\\T{credit\_note.lines.object, customer.sources.object,}\\\T{subscription.items.object, payment\_intent.charges.object,}\\\T{charge.refunds.object, file.links.object,}\\\T{customer.subscriptions.object, invoice.lines.object }\end{tabular} \\\cline{2-7}
                         & \multirow{6}{*}{\T{/v1/subscription\_items\_GET}} & \multirow{1}{*}{Parameter}  & Yes & \T{subscription} &  \T{subscription.id} & \begin{tabular}{@{}c@{}} \T{invoiceitem.subscription, invoice.subscription,}\\\T{discount.subscription, subscription\_item.subscription,}\\\T{line\_item.subscription, $\textbf{subscription.id}$ }\end{tabular} \\\cline{3-7}
                         &   & \multirow{1}{*}{Response}  & - & \T{object} &  \T{subscription.items.object} & \begin{tabular}{@{}c@{}} \T{radar.value\_list.list\_items.object,}\\\T{credit\_note.lines.object, customer.sources.object,}\\\T{$\textbf{subscription.items.object}$, payment\_intent.charges.object,}\\\T{charge.refunds.object, file.links.object,}\\\T{customer.subscriptions.object, invoice.lines.object }\end{tabular} \\\cline{2-7}
                         & \multirow{14}{*}{\T{/v1/tax\_rates/\{tax\_rate\}\_POST}} & \multirow{14}{*}{Parameter} & Yes & \T{tax\_rate} &  \T{tax\_rate.id} & \begin{tabular}{@{}c@{}} \T{$\textbf{tax\_rate.id}$, invoice\_tax\_amount.tax\_rate }\end{tabular} \\\cline{4-7}
                         &   &   & No & \T{country} &  \T{tax\_rate.country} & \begin{tabular}{@{}c@{}} \T{/v1/tax\_rates/\{tax\_rate\}.in.country }\end{tabular} \\\cline{4-7}
                         &   &   & No & \T{description} &  \T{tax\_rate.description} & \begin{tabular}{@{}c@{}} \T{$\textbf{tax\_rate.description}$ }\end{tabular} \\\cline{4-7}
                         &   &   & No & \T{display\_name} & \T{tax\_rate.display\_name}  & \begin{tabular}{@{}c@{}} \T{$\textbf{tax\_rate.display\_name}$ }\end{tabular} \\\cline{4-7}
                         &   &   & No & \T{jurisdiction} &  \T{tax\_rate.jurisdiction} & \begin{tabular}{@{}c@{}} \T{address.state, invoice.account\_country,}\\\T{card.country, $\textbf{tax\_rate.jurisdiction}$,}\\\T{tax\_rate.country, source\_type\_card.country,}\\\T{account.country, payment\_method\_card.country,}\\\T{payment\_method\_details\_card.country,}\\\T{country\_spec.supported\_transfer\_countries.[?],}\\\T{card.address\_state, country\_spec.id,}\\\T{address.country }\end{tabular} \\\cline{4-7}
                         &   &   & No & \T{state} &  \T{tax\_rate.state} & \begin{tabular}{@{}c@{}} \T{/v1/tax\_rates/\{tax\_rate\}.in.state }\end{tabular} \\\cline{2-7}
\hline

\multirow{20}{*}{\rotatebox{90}{\squareapi}} & \multirow{11}{*}{/v2/customers\_POST} & \multirow{11}{*}{Parameter} & No & \T{given\_name} & \T{Customer.given\_name } & \begin{tabular}{@{}c@{}} \T{$\textbf{Customer.given\_name}$, InvoiceRecipient.given\_name} \end{tabular} \\\cline{4-7}
                             &   &   & No & \T{family\_name} & \T{ Customer.famile\_name } & \begin{tabular}{@{}c@{}} \T{$\textbf{Customer.family\_name}$, InvoiceRecipient.family\_name} \end{tabular} \\\cline{4-7}
                             &   &   & No & \T{company\_name} & \T{ Customer.company\_name} & \begin{tabular}{@{}c@{}} \T{$\textbf{Customer.company\_name}$} \end{tabular} \\\cline{4-7}
                             &   &   & No & \T{nickname} & \T{Customer.nickname } & \begin{tabular}{@{}c@{}} \T{$\textbf{Customer.nickname}$ } \end{tabular} \\\cline{4-7}
  &   &   & No & \T{email\_address} & \T{ Customer.email\_address} & \begin{tabular}{@{}c@{}} \T{/v2/customers.in.email\_address} \end{tabular} \\\cline{4-7}
  &   &   & No & \T{address} & \T{ Address} & \begin{tabular}{@{}c@{}} \T{$\textbf{Address}$ } \end{tabular} \\\cline{4-7}
  &   &   & No & \T{phone\_number} & \T{ Customer.phone\_number} & \begin{tabular}{@{}c@{}} \T{$\textbf{Customer.phone\_number}$ } \end{tabular} \\\cline{4-7}
  &   &   & No & \T{reference\_id} & \T{Customer.reference\_id } & \begin{tabular}{@{}c@{}} \T{$\textbf{Customer.reference\_id}$ } \end{tabular} \\\cline{4-7}
  &   &   & No & \T{note} & \T{Customer.note } & \begin{tabular}{@{}c@{}} \T{$\textbf{Customer.note}$ } \end{tabular} \\\cline{4-7}
  &   &   & No & \T{birthday} & \T{Customer.birthday } & \begin{tabular}{@{}c@{}} \T{/v2/customers.in.birthday} \end{tabular} \\\cline{2-7}
  & \multirow{1}{*}{\T{/v2/orders/\{order\_id\}\_GET}} & \multirow{1}{*}{Parameter} & Yes & \T{order\_id} & \T{Order.id }  & \begin{tabular}{@{}c@{}} \T{Transaction.id, Payment.order\_id, $\textbf{Order.id}$,}\\\T{Transaction.order\_id, Invoice.order\_id,}\\\T{Tender.transaction\_id, OrderEntry.order\_id }\end{tabular} \\\cline{2-7}
  & \multirow{2}{*}{\T{/v2/catalog/list\_GET}} & \multirow{2}{*}{Parameter} & No & \T{types} &  \T{CatalogObject.type} & \begin{tabular}{@{}c@{}} \T{$\textbf{CatalogObject.type}$ }\end{tabular} \\\cline{4-7}
                         &   &   & No & \T{catalog\_version} &  \T{CatalogObject.version} & \begin{tabular}{@{}c@{}} \T{/v2/catalog/list.in.catalog\_version }\end{tabular} \\\cline{2-7}
                         & \multirow{1}{*}{\T{/v2/labor/break-types\_GET}} & \multirow{1}{*}{Parameter} & No & \T{location\_id} & \T{ Location.id} & \begin{tabular}{@{}c@{}} \T{/v2/labor/break-types.in.location\_id} \end{tabular} \\\cline{2-7}
                         & \multirow{4}{*}{\T{/v2/inventory/batch-retrieve-counts\_POST}} & \multirow{4}{*}{Parameter} & No & \T{catalog\_object\_ids[0]} & \T{CatalogObject.id} & \begin{tabular}{@{}c@{}} \T{/v2/inventory/batch-retrieve-counts.in.catalog\_object\_ids.0 }\end{tabular} \\\cline{4-7}
                         &   &   & No & \T{location\_ids[0]} & \T{Location.id}  & \begin{tabular}{@{}c@{}} \T{/v2/inventory/batch-retrieve-counts.in.location\_ids.0 }\end{tabular} \\\cline{4-7}
                         &   &   & No & \T{updated\_after} & \T{Counts.updated\_at}  & \begin{tabular}{@{}c@{}} \T{/v2/inventory/batch-retrieve-counts.in.updated\_after }\end{tabular} \\\cline{4-7}
                         &   &   & No & \T{states[0]} & \T{Counts.states}  & \begin{tabular}{@{}c@{}} \T{/v2/inventory/batch-retrieve-counts.in.states.0 }\end{tabular} \\
\bottomrule

\end{tabular}
}
\end{center}
  
\end{table*}
  
\newpage
\setlength{\parindent}{0pt}

\subsection{\slack}
\textbf{1.1. Retrieve emails of all members in a channel}

Type query: \begin{lstlisting}[style=dsl,basicstyle=\ttfamily\footnotesize,xleftmargin=5pt,breaklines=true,postbreak=\mbox{\textcolor{red}{$\hookrightarrow$}\space}]
{ channel_name: objs_conversation.name } -> [objs_user_profile.email]

\end{lstlisting}

Solution:
\begin{lstlisting}[style=dsl,basicstyle=\ttfamily\footnotesize,xleftmargin=5pt,breaklines=true,postbreak=\mbox{\textcolor{red}{$\hookrightarrow$}\space}]
\channel_name -> {
    let x0 = /conversations_list_GET()    
    x1 <- x0.channels    
    if x1.name = channel_name    
    let x2 = /conversations_members_GET(channel=x1.id)    
    x3 <- x2.members    
    let x4 = /users_profile_get_GET(user=x3)    
    return x4.profile.email
}
\end{lstlisting}

\vspace{-0.25em}{\footnotesize Source: \url{https://stackoverflow.com/questions/41564027/slack-api-retrieve-all-member-emails-from-a-slack-channel}

\vspace{1em}

\textbf{1.2. Send a message to a user given their email}

Type query: \begin{lstlisting}[style=dsl,basicstyle=\ttfamily\footnotesize,xleftmargin=5pt,breaklines=true,postbreak=\mbox{\textcolor{red}{$\hookrightarrow$}\space}]
{ email: objs_user_profile.email } -> objs_message

\end{lstlisting}

Solution:
\begin{lstlisting}[style=dsl,basicstyle=\ttfamily\footnotesize,xleftmargin=5pt,breaklines=true,postbreak=\mbox{\textcolor{red}{$\hookrightarrow$}\space}]
\email -> {
    let x0 = /users_lookupByEmail_GET(email=email)    
    let x1 = /conversations_open_POST(users=x0.user.id)    
    let x2 = /chat_postMessage_POST(channel=x1.channel.id)    
    return x2.message
}
\end{lstlisting}

\vspace{-0.25em}{\footnotesize Source: \url{https://stackoverflow.com/questions/43733375/slack-api-post-message-via-user-email}

\vspace{1em}

\textbf{1.3. Get the unread messages of a user}

Type query: \begin{lstlisting}[style=dsl,basicstyle=\ttfamily\footnotesize,xleftmargin=5pt,breaklines=true,postbreak=\mbox{\textcolor{red}{$\hookrightarrow$}\space}]
{ user_id: defs_user_id } -> [[objs_message]]

\end{lstlisting}

Solution:
\begin{lstlisting}[style=dsl,basicstyle=\ttfamily\footnotesize,xleftmargin=5pt,breaklines=true,postbreak=\mbox{\textcolor{red}{$\hookrightarrow$}\space}]
\user_id -> {
    let x0 = /users_conversations_GET(user=user_id)    
    x1 <- x0.channels    
    let x2 = /conversations_info_GET(channel=x1.id)    
    let x3 = /conversations_history_GET(channel=x2.channel.id, oldest=x2.channel.last_read)    
    return x3.messages
}
\end{lstlisting}

\vspace{-0.25em}{\footnotesize Source: \url{https://stackoverflow.com/questions/64561594/is-it-possible-to-know-the-number-of-unread-slack-messages-a-user-has-with-the-s}

\vspace{1em}

\textbf{1.4. Get all messages associated with a user}

Type query: \begin{lstlisting}[style=dsl,basicstyle=\ttfamily\footnotesize,xleftmargin=5pt,breaklines=true,postbreak=\mbox{\textcolor{red}{$\hookrightarrow$}\space}]
{ user_id: defs_user_id,
  ts: defs_ts 
} -> [objs_message]

\end{lstlisting}

Solution:
\begin{lstlisting}[style=dsl,basicstyle=\ttfamily\footnotesize,xleftmargin=5pt,breaklines=true,postbreak=\mbox{\textcolor{red}{$\hookrightarrow$}\space}]
\user_id ts -> {
    let x0 = /conversations_list_GET()    
    x1 <- x0.channels    
    let x2 = /conversations_history_GET(channel=x1.id, oldest=ts)    
    x3 <- x2.messages    
    if x3.user = user_id    
    return x3
}
\end{lstlisting}

\vspace{-0.25em}{\footnotesize Source: \url{https://github.com/hisabimbola/slack-history-export/blob/e53868d8820ba65e5e726bd5968c80d5eb54c0db/src/utils.js}

\vspace{1em}

\textbf{1.5. Create a channel and invite a list of users}

Type query: \begin{lstlisting}[style=dsl,basicstyle=\ttfamily\footnotesize,xleftmargin=5pt,breaklines=true,postbreak=\mbox{\textcolor{red}{$\hookrightarrow$}\space}]
{ user_ids: [defs_user_id],
  channel_name: objs_conversation.name 
} -> [objs_conversation]

\end{lstlisting}

Solution:
\begin{lstlisting}[style=dsl,basicstyle=\ttfamily\footnotesize,xleftmargin=5pt,breaklines=true,postbreak=\mbox{\textcolor{red}{$\hookrightarrow$}\space}]
\user_ids channel_name -> {
    let x0 = /conversations_create_POST(name=channel_name)    
    x1 <- user_ids    
    let x2 = /conversations_invite_POST(channel=x0.channel.id, users=x1)    
    return x2.channel
}
\end{lstlisting}

\vspace{-0.25em}{\footnotesize Source: \url{https://stackoverflow.com/questions/48328380/slack-api-channels-create-followed-by-channels-invite-info-returns-channel-not}

\vspace{1em}

\textbf{1.6. Reply to a message and update it}

Type query: \begin{lstlisting}[style=dsl,basicstyle=\ttfamily\footnotesize,xleftmargin=5pt,breaklines=true,postbreak=\mbox{\textcolor{red}{$\hookrightarrow$}\space}]
{ channel: defs_channel,
  ts: defs_ts 
} -> objs_message

\end{lstlisting}

Solution:
\begin{lstlisting}[style=dsl,basicstyle=\ttfamily\footnotesize,xleftmargin=5pt,breaklines=true,postbreak=\mbox{\textcolor{red}{$\hookrightarrow$}\space}]
\channel ts -> {
    let x1 = /chat_postMessage_POST(channel=channel, thread_ts=ts)    
    let x2 = /chat_update_POST(channel=channel, ts=x1.ts)    
    return x2.message
}
\end{lstlisting}

\vspace{1em}

\textbf{1.7. Send a message to a channel with the given name}

Type query: \begin{lstlisting}[style=dsl,basicstyle=\ttfamily\footnotesize,xleftmargin=5pt,breaklines=true,postbreak=\mbox{\textcolor{red}{$\hookrightarrow$}\space}]
{ channel: objs_conversation.name } -> objs_message

\end{lstlisting}

Solution:
\begin{lstlisting}[style=dsl,basicstyle=\ttfamily\footnotesize,xleftmargin=5pt,breaklines=true,postbreak=\mbox{\textcolor{red}{$\hookrightarrow$}\space}]
\channel -> {
    let x0 = /conversations_list_GET()    
    x1 <- x0.channels    
    if x1.name = channel    
    let x2 = /chat_postMessage_POST(channel=x1.id)    
    return x2.message
}
\end{lstlisting}

\vspace{-0.25em}{\footnotesize Source: \url{https://github.com/backspace/slack-statsbot/blob/primary/src/statsbot.js}

\vspace{1em}

\textbf{1.8. Get the unread messages of a channel}

Type query: \begin{lstlisting}[style=dsl,basicstyle=\ttfamily\footnotesize,xleftmargin=5pt,breaklines=true,postbreak=\mbox{\textcolor{red}{$\hookrightarrow$}\space}]
{ channel_id: defs_channel } -> [[objs_message]]

\end{lstlisting}

Solution:
\begin{lstlisting}[style=dsl,basicstyle=\ttfamily\footnotesize,xleftmargin=5pt,breaklines=true,postbreak=\mbox{\textcolor{red}{$\hookrightarrow$}\space}]
\channel_id -> {
    let x2 = /conversations_info_GET(channel=channel_id)    
    let x3 = /conversations_history_GET(channel=channel_id, oldest=x2.channel.last_read)    
    return x3.messages
}
\end{lstlisting}

\vspace{-0.25em}{\footnotesize Source: \url{https://stackoverflow.com/questions/64561594/is-it-possible-to-know-the-number-of-unread-slack-messages-a-user-has-with-the-s}

\vspace{1em}

\subsection{\stripe}
\textbf{2.1. Subscribe to a product for a customer}

Type query: \begin{lstlisting}[style=dsl,basicstyle=\ttfamily\footnotesize,xleftmargin=5pt,breaklines=true,postbreak=\mbox{\textcolor{red}{$\hookrightarrow$}\space}]
{ customer_id: customer.id,
  product_id: product.id 
} -> [subscription]

\end{lstlisting}

Solution:
\begin{lstlisting}[style=dsl,basicstyle=\ttfamily\footnotesize,xleftmargin=5pt,breaklines=true,postbreak=\mbox{\textcolor{red}{$\hookrightarrow$}\space}]
\customer_id product_id -> {
    let x1 = /v1/prices_GET(product=product_id)    
    x2 <- x1.data    
    let x3 = /v1/subscriptions_POST(customer=customer_id, items[0][price]=x2.id)    
    return x3
}
\end{lstlisting}

\vspace{-0.25em}{\footnotesize Source: \url{https://github.com/stripe-samples/charging-for-multiple-plan-subscriptions/blob/master/server/node/server.js}

\vspace{1em}

\textbf{2.2. Subscribe to multiple items}

Type query: \begin{lstlisting}[style=dsl,basicstyle=\ttfamily\footnotesize,xleftmargin=5pt,breaklines=true,postbreak=\mbox{\textcolor{red}{$\hookrightarrow$}\space}]
{ customer_id: customer.id,
  product_ids: [product.id] 
} -> [subscription]

\end{lstlisting}

Solution:
\begin{lstlisting}[style=dsl,basicstyle=\ttfamily\footnotesize,xleftmargin=5pt,breaklines=true,postbreak=\mbox{\textcolor{red}{$\hookrightarrow$}\space}]
\customer_id product_ids -> {
    x0 <- product_ids    
    let x1 = /v1/prices_GET(product=x0)    
    x2 <- x1.data    
    let x3 = /v1/subscriptions_POST(customer=customer_id, items[0][price]=x2.id)    
    return x3
}
\end{lstlisting}

\vspace{-0.25em}{\footnotesize Source: \url{https://github.com/stripe-samples/charging-for-multiple-plan-subscriptions/blob/master/server/node/server.js}

\vspace{1em}

\textbf{2.3. Create a product and invoice a customer}

Type query: \begin{lstlisting}[style=dsl,basicstyle=\ttfamily\footnotesize,xleftmargin=5pt,breaklines=true,postbreak=\mbox{\textcolor{red}{$\hookrightarrow$}\space}]
{ product_name: product.name,
  customer_id: customer.id,
  currency: fee.currency,
  unit_amount: plan.amount 
} -> invoiceitem

\end{lstlisting}

Solution:
\begin{lstlisting}[style=dsl,basicstyle=\ttfamily\footnotesize,xleftmargin=5pt,breaklines=true,postbreak=\mbox{\textcolor{red}{$\hookrightarrow$}\space}]
\product_name customer_id currency unit_amount -> {
    let x0 = /v1/products_POST(name=product_name)    
    let x1 = /v1/prices_POST(currency=currency, product=x0.id, unit_amount=unit_amount)    
    let x2 = /v1/invoiceitems_POST(customer=customer_id, price=x1.id)    
    return x2
}
\end{lstlisting}

\vspace{-0.25em}{\footnotesize Source: \url{https://stripe.com/docs/invoicing/prices-guide}

\vspace{1em}

\textbf{2.4. Retrieve a customer by email}

Type query: \begin{lstlisting}[style=dsl,basicstyle=\ttfamily\footnotesize,xleftmargin=5pt,breaklines=true,postbreak=\mbox{\textcolor{red}{$\hookrightarrow$}\space}]
{ email: customer.email } -> customer

\end{lstlisting}

Solution:
\begin{lstlisting}[style=dsl,basicstyle=\ttfamily\footnotesize,xleftmargin=5pt,breaklines=true,postbreak=\mbox{\textcolor{red}{$\hookrightarrow$}\space}]
\email -> {
    let x0 = /v1/customers_GET()    
    x1 <- x0.data    
    if x1.email = email    
    return x1
}
\end{lstlisting}

\vspace{-0.25em}{\footnotesize Source: \url{https://stackoverflow.com/questions/26767150/stripe-is-it-possible-to-search-a-customer-by-their-email}

\vspace{1em}

\textbf{2.5. Get a list of receipts for a customer}

Type query: \begin{lstlisting}[style=dsl,basicstyle=\ttfamily\footnotesize,xleftmargin=5pt,breaklines=true,postbreak=\mbox{\textcolor{red}{$\hookrightarrow$}\space}]
{ customer_id: customer.id } -> [charge]

\end{lstlisting}

Solution:
\begin{lstlisting}[style=dsl,basicstyle=\ttfamily\footnotesize,xleftmargin=5pt,breaklines=true,postbreak=\mbox{\textcolor{red}{$\hookrightarrow$}\space}]
\customer_id -> {
    let x1 = /v1/invoices_GET(customer=customer_id)    
    x2 <- x1.data    
    let x3 = /v1/charges/{charge}_GET(charge=x2.charge)    
    return x3
}
\end{lstlisting}

\vspace{-0.25em}{\footnotesize Source: \url{https://stackoverflow.com/questions/24335268/stripe-api-receipts-listing}

\vspace{1em}

\textbf{2.6. Get a refund for a subscription}

Type query: \begin{lstlisting}[style=dsl,basicstyle=\ttfamily\footnotesize,xleftmargin=5pt,breaklines=true,postbreak=\mbox{\textcolor{red}{$\hookrightarrow$}\space}]
{ subscription: subscription.id } -> refund

\end{lstlisting}

Solution:
\begin{lstlisting}[style=dsl,basicstyle=\ttfamily\footnotesize,xleftmargin=5pt,breaklines=true,postbreak=\mbox{\textcolor{red}{$\hookrightarrow$}\space}]
\subscription -> {
    let x0 = /v1/subscriptions/{subscription_exposed_id}_GET(subscription_exposed_id=subscription)    
    let x1 = /v1/invoices/{invoice}_GET(invoice=x0.latest_invoice)    
    let x2 = /v1/refunds_POST(charge=x1.charge)    
    return x2
}
\end{lstlisting}

\vspace{-0.25em}{\footnotesize Source: \url{https://stackoverflow.com/questions/62403075/stripe-api-get-upcoming-invoice-for-cancelled-subscription}

\vspace{1em}

\textbf{2.7. Get the emails of all customers}

Type query: \begin{lstlisting}[style=dsl,basicstyle=\ttfamily\footnotesize,xleftmargin=5pt,breaklines=true,postbreak=\mbox{\textcolor{red}{$\hookrightarrow$}\space}]
{  } -> [customer.email]

\end{lstlisting}

Solution:
\begin{lstlisting}[style=dsl,basicstyle=\ttfamily\footnotesize,xleftmargin=5pt,breaklines=true,postbreak=\mbox{\textcolor{red}{$\hookrightarrow$}\space}]
\ -> {
    let x0 = /v1/customers_GET()    
    x1 <- x0.data    
    return x1.email
}
\end{lstlisting}

\vspace{-0.25em}{\footnotesize Source: \url{https://stackoverflow.com/questions/65545997/python3-stripe-api-to-get-all-customer-email}

\vspace{1em}

\textbf{2.8. Get the emails of the subscribers of a product}

Type query: \begin{lstlisting}[style=dsl,basicstyle=\ttfamily\footnotesize,xleftmargin=5pt,breaklines=true,postbreak=\mbox{\textcolor{red}{$\hookrightarrow$}\space}]
{ product_id: product.id } -> [customer.email]

\end{lstlisting}

Solution:
\begin{lstlisting}[style=dsl,basicstyle=\ttfamily\footnotesize,xleftmargin=5pt,breaklines=true,postbreak=\mbox{\textcolor{red}{$\hookrightarrow$}\space}]
\product_id -> {
    let x1 = /v1/subscriptions_GET()    
    x2 <- x1.data    
    x3 <- x2.items.data    
    if x3.price.product = product_id    
    let x4 = /v1/customers/{customer}_GET(customer=x2.customer)    
    return x4.email
}
\end{lstlisting}

\vspace{-0.25em}{\footnotesize Source: \url{https://stackoverflow.com/questions/35882771/use-stripe-api-to-return-a-list-of-valid-subscribers}

\vspace{1em}

\textbf{2.9. Get the last 4 digits of a customer's card}

Type query: \begin{lstlisting}[style=dsl,basicstyle=\ttfamily\footnotesize,xleftmargin=5pt,breaklines=true,postbreak=\mbox{\textcolor{red}{$\hookrightarrow$}\space}]
{ customer_id: customer.id } -> bank_account.last4

\end{lstlisting}

Solution:
\begin{lstlisting}[style=dsl,basicstyle=\ttfamily\footnotesize,xleftmargin=5pt,breaklines=true,postbreak=\mbox{\textcolor{red}{$\hookrightarrow$}\space}]
\customer_id -> {
    let x0 = /v1/customers/{customer}/sources_GET(customer=customer_id)    
    x1 <- x0.data    
    return x1.last4
}
\end{lstlisting}

\vspace{-0.25em}{\footnotesize Source: \url{https://stackoverflow.com/questions/30447026/getting-last4-digits-of-card-using-customer-object-stripe-api-with-php}

\vspace{1em}

\textbf{2.10. Update payment methods for a user's subscriptions}

Type query: \begin{lstlisting}[style=dsl,basicstyle=\ttfamily\footnotesize,xleftmargin=5pt,breaklines=true,postbreak=\mbox{\textcolor{red}{$\hookrightarrow$}\space}]
{ payment_method: payment_method,
  customer_id: customer.id 
} -> [subscription]

\end{lstlisting}

Solution:
\begin{lstlisting}[style=dsl,basicstyle=\ttfamily\footnotesize,xleftmargin=5pt,breaklines=true,postbreak=\mbox{\textcolor{red}{$\hookrightarrow$}\space}]
\payment_method customer_id -> {
    let x0 = /v1/subscriptions_GET(customer=customer_id)    
    x1 <- x0.data    
    let x2 = /v1/subscriptions/{subscription_exposed_id}_POST(subscription_exposed_id=x1.id, default_payment_method=payment_method.id)    
    return x2
}
\end{lstlisting}

\vspace{-0.25em}{\footnotesize Source: \url{https://stackoverflow.com/questions/58270828/update-credit-card-details-of-user-for-all-subscriptions-in-stripe-using-api}

\vspace{1em}

\textbf{2.11. Delete the default payment source for a customer}

Type query: \begin{lstlisting}[style=dsl,basicstyle=\ttfamily\footnotesize,xleftmargin=5pt,breaklines=true,postbreak=\mbox{\textcolor{red}{$\hookrightarrow$}\space}]
{ customer_id: customer.id } -> payment_source

\end{lstlisting}

Solution:
\begin{lstlisting}[style=dsl,basicstyle=\ttfamily\footnotesize,xleftmargin=5pt,breaklines=true,postbreak=\mbox{\textcolor{red}{$\hookrightarrow$}\space}]
\customer_id -> {
    let x0 = /v1/customers/{customer}_GET(customer=customer_id)    
    let x1 = /v1/customers/{customer}/sources/{id}_DELETE(customer=customer_id, id=x0.default_source)    
    return x1
}
\end{lstlisting}

\vspace{-0.25em}{\footnotesize Source: \url{https://stackoverflow.com/questions/17807881/stripe-api-throwing-error-when-trying-to-delete-a-card}

\vspace{1em}

\textbf{2.12. Save a card during payment}

Type query: \begin{lstlisting}[style=dsl,basicstyle=\ttfamily\footnotesize,xleftmargin=5pt,breaklines=true,postbreak=\mbox{\textcolor{red}{$\hookrightarrow$}\space}]
{ cur: fee.currency,
  amt: plan.amount,
  pm: payment_method.id 
} -> payment_intent

\end{lstlisting}

Solution:
\begin{lstlisting}[style=dsl,basicstyle=\ttfamily\footnotesize,xleftmargin=5pt,breaklines=true,postbreak=\mbox{\textcolor{red}{$\hookrightarrow$}\space}]
\cur amt pm -> {
    let x1 = /v1/customers_POST()    
    let x2 = /v1/payment_intents_POST(customer=x1.id, payment_method=pm, currency=cur, amount=amt)    
    let x3 = /v1/payment_intents/{intent}/confirm_POST(intent=x2.id)    
    return x3
}
\end{lstlisting}

\vspace{-0.25em}{\footnotesize Source: \url{https://github.com/stripe-samples/saving-card-after-payment/blob/master/without-webhooks/server/node/server.js}

\vspace{1em}

\textbf{2.13. Send an invoice to a customer}

Type query: \begin{lstlisting}[style=dsl,basicstyle=\ttfamily\footnotesize,xleftmargin=5pt,breaklines=true,postbreak=\mbox{\textcolor{red}{$\hookrightarrow$}\space}]
{ customer_id: customer.id,
  price_id: plan.id 
} -> invoice

\end{lstlisting}

Solution:
\begin{lstlisting}[style=dsl,basicstyle=\ttfamily\footnotesize,xleftmargin=5pt,breaklines=true,postbreak=\mbox{\textcolor{red}{$\hookrightarrow$}\space}]
\customer_id price_id -> {
    let x1 = /v1/invoiceitems_POST(customer=customer_id, price=price_id)    
    let x2 = /v1/invoices_POST(customer=x1.customer)    
    let x3 = /v1/invoices/{invoice}/send_POST(invoice=x2.id)    
    return x3
}
\end{lstlisting}

\vspace{-0.25em}{\footnotesize Source: \url{https://stripe.com/docs/invoicing/integration#send-invoice}

\vspace{1em}

\subsection{\squareapi}
\textbf{3.1. List invoices that match a location id}

Type query: \begin{lstlisting}[style=dsl,basicstyle=\ttfamily\footnotesize,xleftmargin=5pt,breaklines=true,postbreak=\mbox{\textcolor{red}{$\hookrightarrow$}\space}]
{ location_id: Location.id } -> [Invoice]

\end{lstlisting}

Solution:
\begin{lstlisting}[style=dsl,basicstyle=\ttfamily\footnotesize,xleftmargin=5pt,breaklines=true,postbreak=\mbox{\textcolor{red}{$\hookrightarrow$}\space}]
\location_id -> {
    let x0 = /v2/invoices_GET(location_id=location_id)    
    return x0.invoices
}
\end{lstlisting}

\vspace{-0.25em}{\footnotesize Source: \url{https://github.com/square/connect-api-examples/blob/4283ac967c31b75dc17aceebd84f649093477e9a/connect-examples/v2/node_invoices/routes/management.js}

\vspace{1em}

\textbf{3.2. List subscriptions by location, customer, and plan}

Type query: \begin{lstlisting}[style=dsl,basicstyle=\ttfamily\footnotesize,xleftmargin=5pt,breaklines=true,postbreak=\mbox{\textcolor{red}{$\hookrightarrow$}\space}]
{ customer_id: Customer.id,
  location_id: Location.id,
  plan_id: CatalogObject.id 
} -> [Subscription]

\end{lstlisting}

Solution:
\begin{lstlisting}[style=dsl,basicstyle=\ttfamily\footnotesize,xleftmargin=5pt,breaklines=true,postbreak=\mbox{\textcolor{red}{$\hookrightarrow$}\space}]
\customer_id location_id plan_id -> {
    let x0 = /v2/subscriptions/search_POST()    
    x1 <- x0.subscriptions    
    if x1.customer_id = customer_id    
    if x1.location_id = location_id    
    if x1.plan_id = plan_id    
    return x1
}
\end{lstlisting}

\vspace{-0.25em}{\footnotesize Source: \url{https://github.com/square/connect-api-examples/blob/4283ac967c31b75dc17aceebd84f649093477e9a/connect-examples/v2/node_subscription/routes/subscription.js}

\vspace{1em}

\textbf{3.3. Get all items a tax applies to}

Type query: \begin{lstlisting}[style=dsl,basicstyle=\ttfamily\footnotesize,xleftmargin=5pt,breaklines=true,postbreak=\mbox{\textcolor{red}{$\hookrightarrow$}\space}]
{ tax_id: CatalogObject.id } -> [CatalogObject]

\end{lstlisting}

Solution:
\begin{lstlisting}[style=dsl,basicstyle=\ttfamily\footnotesize,xleftmargin=5pt,breaklines=true,postbreak=\mbox{\textcolor{red}{$\hookrightarrow$}\space}]
\tax_id -> {
    let x0 = /v2/catalog/search_POST()    
    x1 <- x0.objects    
    x2 <- x1.item_data.tax_ids    
    if x2 = tax_id    
    return x1
}
\end{lstlisting}

\vspace{-0.25em}{\footnotesize Source: \url{https://github.com/square/catalog-api-demo/blob/85b6754c90fa7b66fc5e605ee7a344314537eade/src/main/java/com/squareup/catalog/demo/example/ApplyTaxToAllIItemsExample.java}

\vspace{1em}

\textbf{3.4. Get a list of discounts in the catalog}

Type query: \begin{lstlisting}[style=dsl,basicstyle=\ttfamily\footnotesize,xleftmargin=5pt,breaklines=true,postbreak=\mbox{\textcolor{red}{$\hookrightarrow$}\space}]
{  } -> [CatalogDiscount]

\end{lstlisting}

Solution:
\begin{lstlisting}[style=dsl,basicstyle=\ttfamily\footnotesize,xleftmargin=5pt,breaklines=true,postbreak=\mbox{\textcolor{red}{$\hookrightarrow$}\space}]
\ -> {
    let x0 = /v2/catalog/list_GET()    
    x1 <- x0.objects    
    return x1.discount_data
}
\end{lstlisting}

\vspace{-0.25em}{\footnotesize Source: \url{https://github.com/square/catalog-api-demo/blob/master/src/main/java/com/squareup/catalog/demo/example/ListDiscountsExample.java}

\vspace{1em}

\textbf{3.5. Add order details to order}

Type query: \begin{lstlisting}[style=dsl,basicstyle=\ttfamily\footnotesize,xleftmargin=5pt,breaklines=true,postbreak=\mbox{\textcolor{red}{$\hookrightarrow$}\space}]
{ location_id: Location.id,
  order_ids: [Order.id],
  updates: [OrderFulfillment] 
} -> [Order]

\end{lstlisting}

Solution:
\begin{lstlisting}[style=dsl,basicstyle=\ttfamily\footnotesize,xleftmargin=5pt,breaklines=true,postbreak=\mbox{\textcolor{red}{$\hookrightarrow$}\space}]
\location_id order_ids updates -> {
    x0 <- order_ids    
    let x1 = /v2/orders/batch-retrieve_POST(location_id=location_id, order_ids[0]=x0)    
    x2 <- x1.orders    
    let x3 = {fulfillments=updates}    
    let x4 = /v2/orders/{order_id}_PUT(order_id=x2.id, order=x3)    
    return x4.order
}
\end{lstlisting}

\vspace{-0.25em}{\footnotesize Source: \url{https://github.com/square/connect-api-examples/blob/4283ac967c31b75dc17aceebd84f649093477e9a/connect-examples/v2/node_orders-payments/routes/checkout.js}

\vspace{1em}

\textbf{3.6. Get payment notes of a payment}

Type query: \begin{lstlisting}[style=dsl,basicstyle=\ttfamily\footnotesize,xleftmargin=5pt,breaklines=true,postbreak=\mbox{\textcolor{red}{$\hookrightarrow$}\space}]
{  } -> [Payment.note]

\end{lstlisting}

Solution:
\begin{lstlisting}[style=dsl,basicstyle=\ttfamily\footnotesize,xleftmargin=5pt,breaklines=true,postbreak=\mbox{\textcolor{red}{$\hookrightarrow$}\space}]
\ -> {
    let x0 = /v2/payments_GET()    
    x1 <- x0.payments    
    return x1.note
}
\end{lstlisting}

\vspace{-0.25em}{\footnotesize Source: \url{https://stackoverflow.com/questions/23252751/square-connect-api-list-payments-endpoint-not-showing-description}

\vspace{1em}

\textbf{3.7. Get order ids of current user's transactions}

Type query: \begin{lstlisting}[style=dsl,basicstyle=\ttfamily\footnotesize,xleftmargin=5pt,breaklines=true,postbreak=\mbox{\textcolor{red}{$\hookrightarrow$}\space}]
{ location_id: Location.id } -> [Order.id]

\end{lstlisting}

Solution:
\begin{lstlisting}[style=dsl,basicstyle=\ttfamily\footnotesize,xleftmargin=5pt,breaklines=true,postbreak=\mbox{\textcolor{red}{$\hookrightarrow$}\space}]
\location_id -> {
    let x0 = /v2/locations/{location_id}/transactions_GET(location_id=location_id)    
    x1 <- x0.transactions    
    return x1.order_id
}
\end{lstlisting}

\vspace{-0.25em}{\footnotesize Source: \url{https://stackoverflow.com/questions/46910044/getting-compact-information-from-square-connect-api}

\vspace{1em}

\textbf{3.8. Get order names from a transaction id}

Type query: \begin{lstlisting}[style=dsl,basicstyle=\ttfamily\footnotesize,xleftmargin=5pt,breaklines=true,postbreak=\mbox{\textcolor{red}{$\hookrightarrow$}\space}]
{ location_id: Location.id,
  transaction_id: Order.id 
} -> [Invoice.title]

\end{lstlisting}

Solution:
\begin{lstlisting}[style=dsl,basicstyle=\ttfamily\footnotesize,xleftmargin=5pt,breaklines=true,postbreak=\mbox{\textcolor{red}{$\hookrightarrow$}\space}]
\location_id transaction_id -> {
    let x0 = /v2/orders/batch-retrieve_POST(location_id=location_id, order_ids[0]=transaction_id)    
    x1 <- x0.orders    
    x2 <- x1.line_items    
    return x2.name
}
\end{lstlisting}

\vspace{-0.25em}{\footnotesize Source: \url{https://stackoverflow.com/questions/58047894/square-connect-how-to-retrieve-product-information-from-transaction-id}

\vspace{1em}

\textbf{3.9. Find customers by name}

Type query: \begin{lstlisting}[style=dsl,basicstyle=\ttfamily\footnotesize,xleftmargin=5pt,breaklines=true,postbreak=\mbox{\textcolor{red}{$\hookrightarrow$}\space}]
{ name: Customer.given_name } -> Customer

\end{lstlisting}

Solution:
\begin{lstlisting}[style=dsl,basicstyle=\ttfamily\footnotesize,xleftmargin=5pt,breaklines=true,postbreak=\mbox{\textcolor{red}{$\hookrightarrow$}\space}]
\name -> {
    let x0 = /v2/customers_GET()    
    x1 <- x0.customers    
    if x1.given_name = name    
    return x1
}
\end{lstlisting}

\vspace{-0.25em}{\footnotesize Source: \url{https://developer.squareup.com/forums/t/search-customers-by-name/1567}

\vspace{1em}

\textbf{3.10. Delete catalog items with names}

Type query: \begin{lstlisting}[style=dsl,basicstyle=\ttfamily\footnotesize,xleftmargin=5pt,breaklines=true,postbreak=\mbox{\textcolor{red}{$\hookrightarrow$}\space}]
{ item_type: CatalogObject.type,
  names: [CatalogItem.name] 
} -> [CatalogObject.id]

\end{lstlisting}

Solution:
\begin{lstlisting}[style=dsl,basicstyle=\ttfamily\footnotesize,xleftmargin=5pt,breaklines=true,postbreak=\mbox{\textcolor{red}{$\hookrightarrow$}\space}]
\item_type names -> {
    let x0 = /v2/catalog/search_POST(object_types[0]=item_type)    
    x1 <- x0.objects    
    x2 <- names    
    if x1.item_data.name = x2    
    let x3 = /v2/catalog/object/{object_id}_DELETE(object_id=x1.id)    
    x4 <- x3.deleted_object_ids    
    return x4
}
\end{lstlisting}

\vspace{-0.25em}{\footnotesize Source: \url{https://github.com/square/catalog-api-demo/blob/85b6754c90fa7b66fc5e605ee7a344314537eade/src/main/java/com/squareup/catalog/demo/example/DeleteCategoryExample.java}

\vspace{1em}

\textbf{3.11. Delete all catalog items}

Type query: \begin{lstlisting}[style=dsl,basicstyle=\ttfamily\footnotesize,xleftmargin=5pt,breaklines=true,postbreak=\mbox{\textcolor{red}{$\hookrightarrow$}\space}]
{  } -> [CatalogObject.id]

\end{lstlisting}

Solution:
\begin{lstlisting}[style=dsl,basicstyle=\ttfamily\footnotesize,xleftmargin=5pt,breaklines=true,postbreak=\mbox{\textcolor{red}{$\hookrightarrow$}\space}]
\ -> {
    let x0 = /v2/catalog/list_GET()    
    x1 <- x0.objects    
    let x2 = /v2/catalog/object/{object_id}_DELETE(object_id=x1.id)    
    return x2.deleted_object_ids
}
\end{lstlisting}

\vspace{-0.25em}{\footnotesize Source: \url{https://github.com/square/catalog-api-demo/blob/85b6754c90fa7b66fc5e605ee7a344314537eade/src/main/java/com/squareup/catalog/demo/example/DeleteAllItemsExample.java}

\vspace{1em}

\fi

\end{document}